\documentclass[prd,twocolumn,nofootinbib,showpacs,amsmath,amssymb,floatfix,eqsecnum]{revtex4-1}

\usepackage{amsmath}
\usepackage{amssymb}
\usepackage{amsthm}
\usepackage{amsfonts}

\usepackage{graphicx,color,framed}
\usepackage{hyperref}
\usepackage{times}
\usepackage{enumerate}
\usepackage{lipsum}
\usepackage{slashed}
\usepackage{url}

\hypersetup{
    colorlinks=true, 
    linktoc=all,     
    linkcolor=blue,  
}

\def \beq {\begin{equation}}
\def \eeq {\end{equation}}
\def \beqa {\begin{eqnarray}}
\def \eeqa {\end{eqnarray}}
\def \bseq {\begin{subequations}}
\def \eseq {\end{subequations}}

\newcommand \up {\uparrow}
\newcommand \down {\downarrow}

\newcommand{\<}{\langle}
\renewcommand{\>}{\rangle}

\begin{document}

\title{Microscopic definitions of anyon data}

\author{Kyle Kawagoe}
\author{Michael Levin}
\affiliation{Department of Physics, Kadanoff Center for Theoretical Physics, University of Chicago, Chicago, Illinois 60637,  USA}

\begin{abstract}
We present microscopic definitions of both the $F$-symbol and $R$-symbol -- two pieces of algebraic data that characterize anyon excitations in (2+1)-dimensional systems. An important feature of our definitions is that they are operational; that is, they provide concrete procedures for computing these quantities from microscopic models. In fact, our definitions, together with known results, provide a way to extract a \emph{complete} set of anyon data from a microscopic model, at least in principle. We illustrate our definitions by computing the $F$-symbol and $R$-symbol in several exactly solvable lattice models and edge theories. We also show that our definitions of the $F$-symbol and $R$-symbol satisfy the pentagon and hexagon equations, thereby providing a microscopic derivation of these fundamental constraints.
\end{abstract}

\maketitle


\section{Introduction}

It is generally believed that every (2+1)-dimensional many-body system with local interactions and an energy gap can be associated with a corresponding anyon theory (also known as a unitary braided fusion category). This anyon theory consists of a collection of algebraic data characterizing the anyon excitations of the many-body system. More specifically, an anyon theory consists of three pieces of data: (i) a set of anyon types $\{a,b,c,...\}$; (ii) a collection of ``fusion rules'' describing the outcomes of fusing pairs of anyons $a, b$; and (iii) an ``$F$-symbol'' and an ``$R$-symbol'', which can be thought of as collections of complex numbers that describe fusion and braiding properties of anyons~\cite{kitaevlongpaper, preskilltqc, frohlichbraidstat, bakalovkirillov}.

The mapping between gapped many-body systems and anyon theories has proven to be a powerful tool in the theory of topological phases of matter and it has been applied successfully in many different contexts~\cite{arovaswilczek, wenreview, kitaevtc, mooreread, ivanov}. Nevertheless, this mapping is still missing an important ingredient, namely a systematic procedure for extracting anyon data from a microscopic model. Another way to say this is that we are lacking \emph{microscopic} definitions of the anyon data, i.e. definitions that express each piece of data in terms of the underlying quantum many-body system. The goal of this paper is to find such definitions.

Our main results are microscopic definitions of both the $F$-symbol and the $R$-symbol. These definitions, together with the known definitions of anyon types and fusion rules\footnote{See Secs.~\ref{sec:abel_review} and \ref{sec:nonabel_review}.}, provide a way to extract a \emph{complete} anyon theory from a microscopic model, at least in principle. To illustrate our definitions, we compute the $F$-symbol and $R$-symbol for several exactly solvable lattice models and edge theories. We also show that our definitions of the $F$-symbol and $R$-symbol have all the expected properties. In particular, we show that these quantities obey the pentagon and hexagon equations -- algebraic relations that hold in any consistent anyon theory. 

\begin{figure}[tb]
\centering
\includegraphics[width=.99\columnwidth]{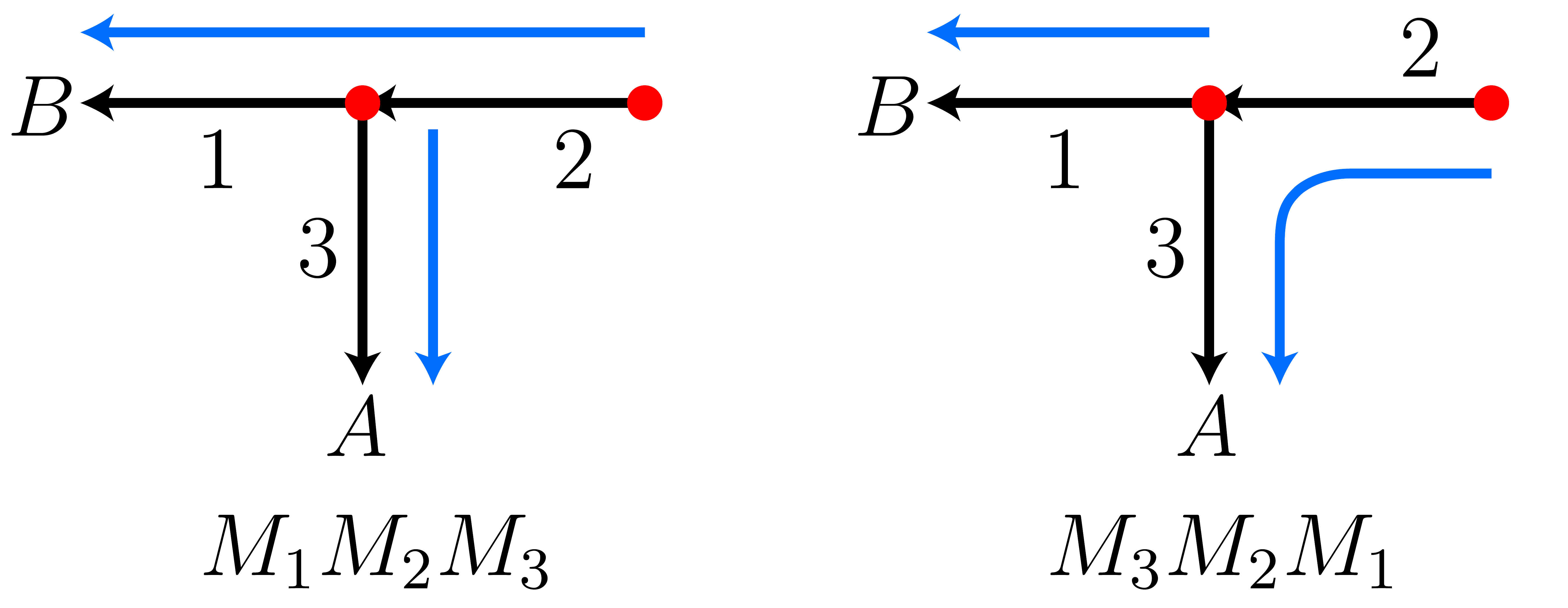}
\caption{Two processes involving two identical Abelian anyons moving along three paths $1,2,3$. In the first process ($M_1 M_2 M_3$), the anyon on the left travels to point $A$ and the anyon on the right travels to point $B$, and vice-versa in the second process ($M_3 M_2 M_1$). The final states of the two processes differ by $R(a,a)$, the exchange statistics of the anyons.}
\label{braid-stat}
\end{figure}

To get a sense of the problem that we address, consider one of the simplest pieces of anyon data: the exchange statistics of an Abelian anyon, $a$. Naively, one might try to define the exchange statistics, which we denote by $R(a,a)$, by considering an adiabatic process in which two identical $a$ particles are exchanged with one another in the counterclockwise direction. One could then define $R(a,a)$ as the Berry phase accumulated during this exchange process. The problem with this definition is that the Berry phase for such a process will generally include a \emph{geometric} phase, which depends on the details of the paths that the particles traverse, in addition to the statistical phase of interest. For example, if the anyon $a$ is charged and there is an external magnetic field, then the Berry phase will contain a contribution coming from the Aharonov-Bohm effect. 

To deal with this problem, Ref.~\onlinecite{levinwenferm} proposed a more careful definition which ensures that all geometric phases cancel. The idea is to start with an initial state, $|i\>$, with two identical $a$ particles, and then apply three ``movement operators'' $M_1, M_2, M_3$ that move the anyons along the three paths, labeled $1,2,3$ in Fig.~\ref{braid-stat}. By applying these operators to $|i\>$ in two different orders, namely $M_1 M_2 M_3$ vs. $M_3 M_2 M_1$, one obtains two final states that differ from one another by an exchange. The relative phase between $M_1 M_2 M_3 |i\>$ and $M_3 M_2 M_1 |i\>$ defines the exchange statistics $R(a,a)$:
\begin{align}
M_1M_2M_3|i\> = R(a,a) \cdot M_3M_2M_1|i\>
\label{exstat}
\end{align}
Note that this definition is manifestly independent of the phases of the three movement operators since each operator appears on both the left and right hand side of (\ref{exstat}). Thus, this definition succeeds in isolating the exchange statistics, $R(a,a)$, from extraneous geometric phases. In this paper, we present microscopic definitions of $F$ and $R$ that are similar in spirit to the above example.

Previous work on the problem of defining anyon data can be divided into two categories depending on whether the authors considered Abelian or non-Abelian anyons. In the Abelian case, most work has focused on computing the exchange statistics and mutual statistics of anyon excitations -- i.e., the statistical phases associated with exchanging two identical anyons or braiding one anyon around another. This line of work has been very successful: exchange statistics and mutual statistics have been computed for many microscopic models (see e.g. Refs.~\onlinecite{arovaswilczek, kitaevtc}), and a general procedure for computing/defining these quantities was presented in Ref.~\onlinecite{levinwenferm}. Moreover, all other anyon data is determined by the exchange and mutual statistics (Proposition 2.5.1 of Ref.~\onlinecite{Quinn}) so one could argue that the problem of computing Abelian anyon data has already been solved. This case is further supported by Refs.~\onlinecite{NaaijkensAQFT,ChaAQFT}, which showed how to \emph{rigorously} define and compute a complete set of Abelian anyon data.   

The situation for non-Abelian anyons is different, however. In the non-Abelian case, most work has focused on the ``topological $S$-matrix'' and the ``topological $T$-matrix''\cite{kitaevlongpaper} -- two pieces of data that reduce to the mutual statistics and exchange statistics in the Abelian case. Several methods have been proposed that allow one to extract $S$ or $T$ directly from ground state wave functions~\cite{wenmodulartrans, zhangvishwanath, zhangvishwanath2, zaletelpolarization, qipolarization, Haah2016}. The problem is that $S$ and $T$ carry some, but not all, of the physical information in the $F$ and $R$ symbols. Therefore, these methods do not provide a way to compute the complete set of anyon data. This paper fills in this gap in the literature by showing how to compute the $F$ and $R$ symbols (and hence all other data) in the general, non-Abelian case. 

Although the generality of our approach is one of its most important features, we will first present our definition of $F$ and $R$ in the context of Abelian anyons and only later discuss non-Abelian anyons. The reason that we organize the paper in this way is that the main subtlety in defining $F$ and $R$ has to do with the $U(1)$ \emph{phases} of these quantities, and this subtlety is the same in the Abelian and non-Abelian cases. Indeed, we will see that the generalization from the Abelian to the non-Abelian case is straightforward. 

This paper is organized as follows. In Sec.~\ref{sec:abel_review}, we review the basic data in Abelian anyon theories. Section~\ref{sec:abelian_F_def} presents our microscopic definition of the $F$-symbol in the case of Abelian anyons. In Sec.~\ref{sec:abelian_F_ex}, we illustrate our definition by computing the $F$-symbol in several lattice models and edge theories. This leads into Sec.~\ref{sec:R}, where we present our microscopic definition of the $R$-symbol in the case of Abelian anyons. We extend both definitions to the non-Abelian case in Sec.~\ref{sec:non-abelian}. We present our conclusions in Sec.~\ref{conclusion}. Technical details are contained in the appendices.

\section{Review: Abelian anyon data}\label{sec:abel_review}
We begin by reviewing some basic aspects of Abelian anyon theories~\cite{kitaevlongpaper, preskilltqc}. These theories consist of four pieces of data:
\begin{enumerate}
\item{{\bf Set of anyon types}: a finite set of anyon types,\\$\mathcal{A} = \{a, b, c, ...\}$.}
\item{{\bf Fusion product}: an associative and commutative multiplication law on $\mathcal{A}$, denoted $a \times b$ or $ab$.}
\item{{\bf $F$-symbol}: a function $F: \mathcal{A} \times \mathcal{A} \times \mathcal{A} \rightarrow U(1)$, denoted $F(a,b,c)$.}
\item{{\bf $R$-symbol}: a function $R: \mathcal{A} \times \mathcal{A} \rightarrow U(1)$, denoted $R(a,b)$.}
\end{enumerate}
We now explain the physical meaning of this data in the context of two-dimensional many-body systems with local interactions and an energy gap.

We begin with the idea of ``anyon types.'' At an intuitive level, the set of anyon types $\mathcal{A}$ is simply the set of topologically distinct particle-like excitations of the many-body system. More precisely, to define the anyon types associated with some many-body Hamiltonian $H$, consider all Hamiltonians of the form $H+V$ that have a unique gapped ground state, where $H$ is defined on an infinite plane geometry and $V$ is supported on a finite region of the plane. (We can think of $V$ as a trapping potential for particle-like excitations.) Denote the set of all ground states generated in this way as $\{|\Psi\>\}$. We define an equivalence relation on the set $\{|\Psi\>\}$ as follows: $|\Psi'\> \sim |\Psi\>$ if there exists a unitary operator, $U$, supported in a finite region of the plane with $|\Psi'\> = U |\Psi\>$. Under this equivalence relation, the collection of states $\{|\Psi\>\}$ breaks up into equivalence classes. These equivalence classes define the set of anyon types $\mathcal{A}=\{a, b, c, ...\}$. The infinitely many states contained within each equivalence class describe the infinitely many ways to realize an anyon excitation of type $a$, type $b$, type $c$, etc.

This discussion leads naturally to the idea of the ``fusion product.'' Let $a, b$ be any pair of (Abelian) anyons. We say that $a \times b = c$ if a pair of anyons $a, b$ can be converted into $c$ and vice-versa, by applying a unitary operator supported in a finite region around $a, b$. It is clear from this definition that the fusion product is both associative and commutative, as mentioned above.

To complete the picture, we need to explain the physical meaning of the $F$ and $R$ symbols. This is the main subject of this paper, so we will say much more about this below. For now, we only mention that the $F$-symbol describes $U(1)$ phases associated with fusing three anyons in different orders, while the $R$-symbol describes $U(1)$ phases associated with braiding or exchanging two anyons. In particular, the quantity $R(a,a)$ can be interpreted as the exchange statistics of anyon $a$, while $R(a,b) R(b,a)$ is the mutual statistics of $a$ and $b$ -- that is, the phase associated with braiding anyon $a$ around anyon $b$.

\section{Defining $F$ for Abelian anyons}
\label{sec:abelian_F_def}

\subsection{Abstract definition of $F$}
\label{sec:abelian_abstract}

\begin{figure}[tb]
\centering
\includegraphics[width=.99\columnwidth]{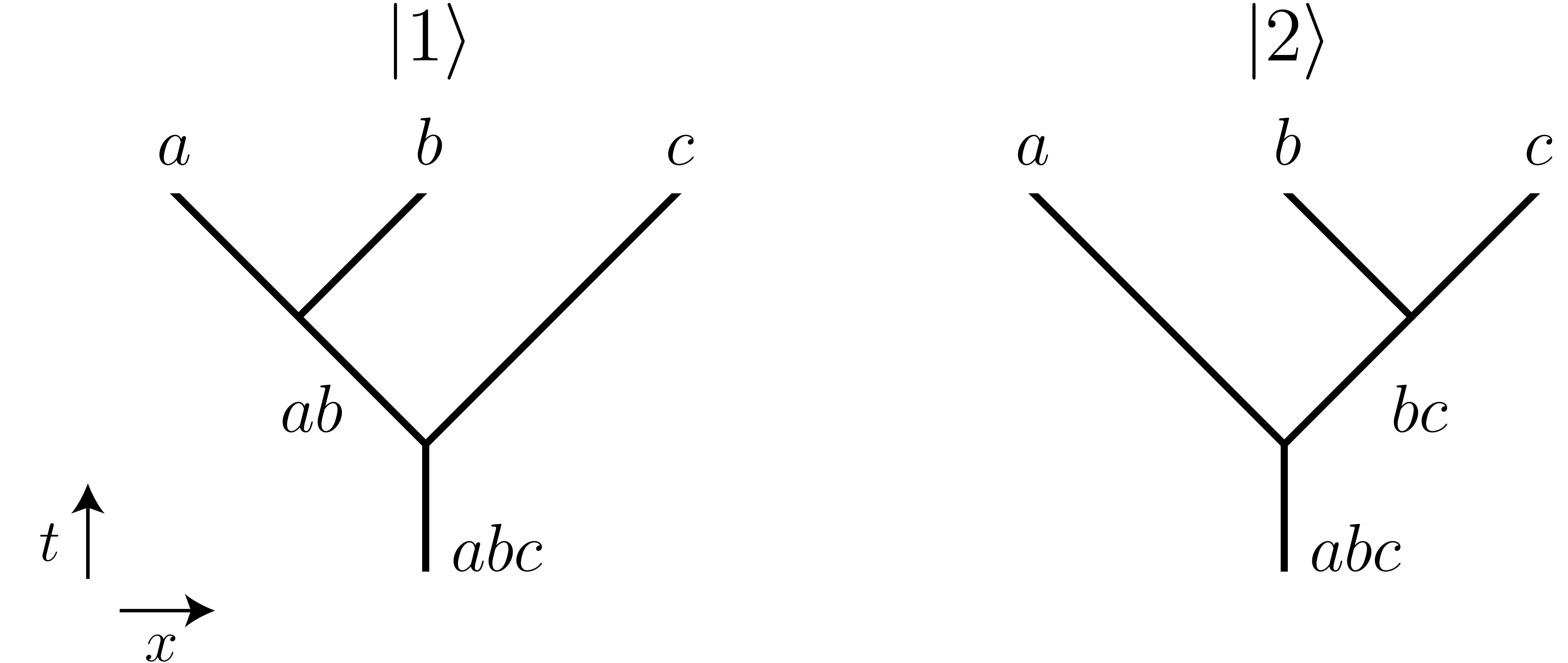}
\caption{Spacetime diagrams for two processes in which an anyon $abc$ splits into anyons $a,b,c$. The horizontal and vertical axes denote the space and time directions while the lines denote anyon worldlines. The final states produced by these processes, $|1\>, |2\>$, are equal up to the $U(1)$ phase $F(a,b,c)$.}
\label{fig:sketch}
\end{figure}

Before explaining our microscopic definition of the $F$-symbol, we first review the \emph{abstract} definition~\cite{kitaevlongpaper, preskilltqc}. We say that this definition is ``abstract'' because it captures the mathematical properties of the $F$-symbol but it is not obvious, a priori, how to make sense of it in a microscopic lattice model.

The basic idea is to consider two different physical processes in which an anyon of type $abc$ splits into three anyons of types $a, b, c$ (Fig. \ref{fig:sketch}). In one process, $abc$ splits into $ab$ and $c$, and then $ab$ splits into $a$ and $b$; in the other process, $abc$ splits into $a$ and $bc$ and then $bc$ splits into $b$ and $c$. By construction, the final states $|1\>, |2\>$ produced by these processes contain the same anyons $a, b, c$, at the same three positions. Therefore, the two final states $|1\>, |2\>$ must be the same up to a phase. The $F$-symbol $F(a,b,c)$ is defined to be the phase difference between the two states: 
\begin{align}
|1\> = F(a,b,c) |2\>.
\label{Fdefab}
\end{align}

The abstract definition of the $F$-symbol has two important implications. First, the $F$-symbol has an inherent ambiguity: it is only well-defined up to transformations of the form
\begin{equation}
F(a,b,c)\rightarrow F(a,b,c)\frac{e^{i\nu(ab,c)}e^{i\nu(a,b)}}{e^{i\nu(a,bc)}e^{i\nu(b,c)}}
\label{gaugetrans}
\end{equation} 
where $\nu(a,b) \in \mathbb{R}$. To understand where this ambiguity comes from, it is helpful to think about the physical processes in Fig.~\ref{fig:sketch} as being implemented by a sequence of two ``splitting operators'' applied to an initial state, $|abc\>$. The key point is that the phases of these splitting operators are arbitrary. If we multiply each of the four splitting operators by a corresponding phase, namely, $e^{i\nu(ab,c)}, e^{i\nu(a,b)}$ for the two splitting operators in the first process and $e^{i\nu(a,bc)}, e^{i\nu(b,c)}$ in the second, $F$ changes by exactly the above transformation (\ref{gaugetrans}). We will call the transformations in (\ref{gaugetrans}) ``gauge transformations.'' 

\begin{figure}[tb]
\centering
\includegraphics[width=1.0\columnwidth]{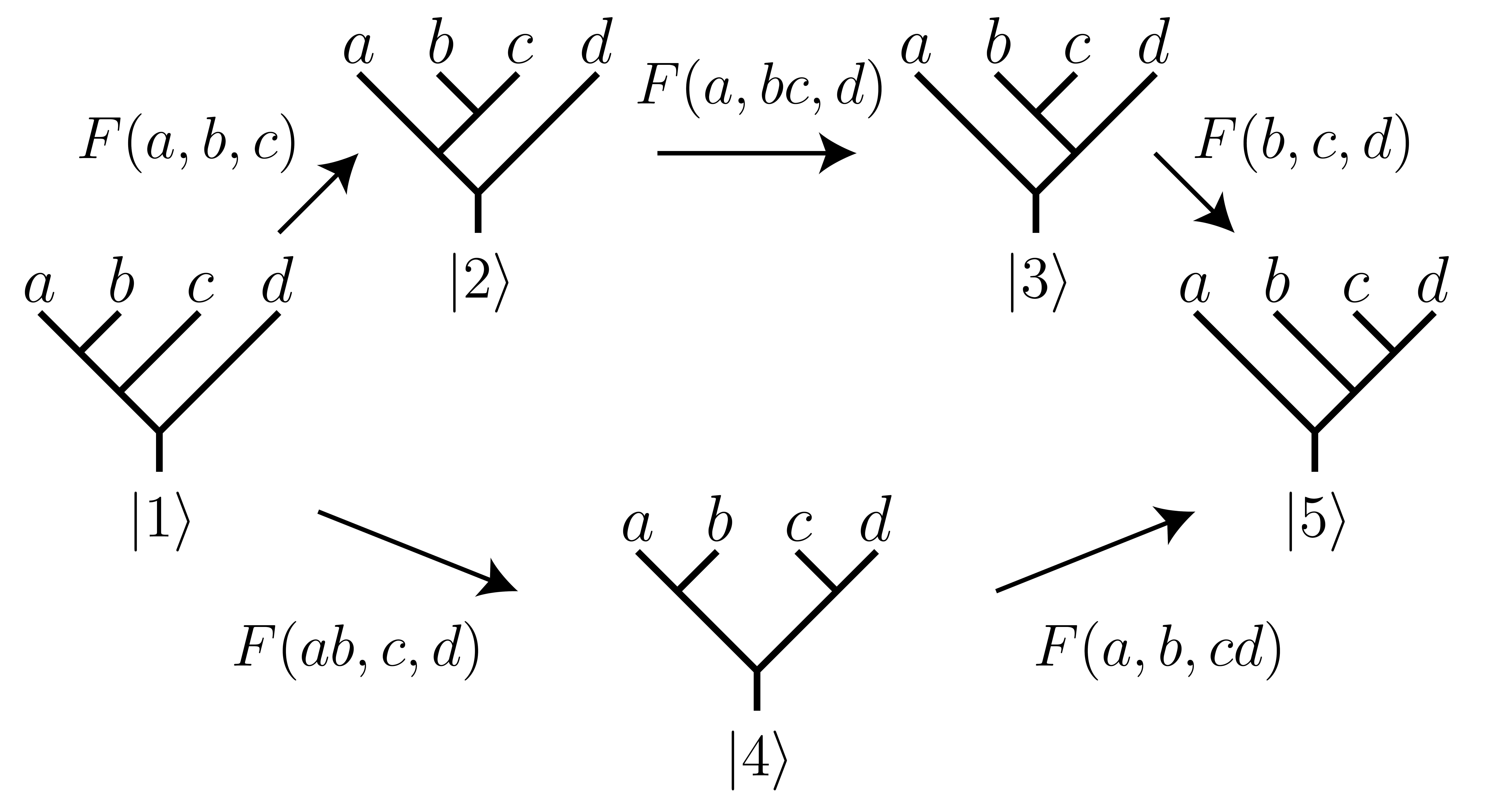}
\caption{The pentagon identity: consistency requires that the product of the three $F$-symbols on the upper path is equal to the product of the two $F$-symbols on the lower path.}
\label{fig:pentagon0}
\end{figure}

The second implication of the above definition is that $F$ must satisfy a non-trivial constraint known as the ``pentagon identity'':
\begin{equation}
F(a,b,c) F(a,bc,d) F(b,c,d) = F(ab,c,d) F(a,b,cd)
\label{pentid}
\end{equation}
To derive this identity, consider the $5$ processes shown in Fig. \ref{fig:pentagon0}. The first step is to note that the final states produced by these processes, namely $\{|1\>,...,|5\>\}$, are all the same up to a phase. Next, we compute the phase difference between states $|1\>$ and $|5\>$ in two different ways. In the first way, we compute the relative phases between $(|1\>, |2\>)$, $(|2\>, |3\>)$, and $(|3\>, |5\>)$ using (\ref{Fdefab}); in the second way, we compute the relative phases between $(|1\>, |4\>)$ and $(|4\>, |5\>)$. Demanding consistency between the two calculations gives the pentagon identity (\ref{pentid}). Below, we will show that our microscopic definition of $F$ also obeys the pentagon equation.

\subsection{Microscopic definition of $F$}
\label{sec-micro-F-abel}
The main problem with applying the above definition to a microscopic lattice model is that some of the anyon splittings in Fig.~\ref{fig:sketch} occur at different points in space. This means that the corresponding splitting operators are distinct operators which can be multiplied by \emph{independent} phases. For example, the two splitting operators in the first process in Fig. \ref{fig:sketch} can be multiplied by two independent phases $e^{i\nu_1(ab,c)}, e^{i\nu_2(a,b)}$. This leads to a larger ambiguity in $F(a,b,c)$ than (\ref{gaugetrans}). To solve this problem, we need to perform all anyon splittings using the same set of splitting operators, defined at the same location in space. However, this introduces another problem: to do this, we need to \emph{move} anyons in addition to splitting them. The operators that move anyons have arbitrary phases, just like the splitting operators; so, unless we are careful, the arbitrary choice of these phases will introduce additional ambiguities into $F$, beyond the ones in Eq. (\ref{gaugetrans}). In this section, we overcome this problem by constructing two microscopic processes in which these extraneous phases cancel out.

To begin, we introduce some notation for labeling anyon states. The first step is to fix a line in the 2D plane, say the $x$-axis. We will only consider states with anyons living along this line. Next, for each anyon type $a$ and for each point $x$ on the line, we let $|a_x\>$ denote a state containing a single\footnote{Although anyons can only be created in pairs, it is possible to have a state containing a \emph{single} anyon where the partner is located far away (i.e. ``at infinity''); this is the setup envisioned here.} anyon $a$ located at point $x$.\footnote{This step involves an arbitrary choice since there are infinitely many states that contain an anyon $a$ at point $x$, differing in their microscopic details; we will check that our definition of $F$ does not depend on these choices.} Similarly, we let 
\begin{align*}
|a_{x_1}, b_{x_2}, c_{x_3},...\>
\end{align*}
be the (normalized) multi-anyon state with the anyon $a$ at point $x_1$, the anyon $b$ at point $x_2$, and so on. Here we assume that the points are ordered left to right as $x_1 < x_2 <\cdots$. We also assume that $x_1, x_2, ...$ are \emph{well-separated}: every pair of neighboring anyons is separated by a distance that is much larger than the correlation length $\xi$ of the ground state. Given this assumption, we will neglect finite size corrections that are exponentially small in the distance between the anyons.

For reasons that will become clear below, it is useful to have a precise definition of multi-anyon states in terms of the single anyon states. To this end, we define $|a_{x_1}, b_{x_2}, c_{x_3},...\>$ to be the unique\footnote{In this discussion, multi-anyon states are \emph{unique} because the anyons are assumed to be Abelian. In the non-Abelian case, multi-anyon states come in multiplets, as discussed in Sec.~\ref{sec:non-abelian}.} state that has the same expectation values as the ground state for local operators supported away from all the anyons, the same expectation values as $|a_{x_1}\>$ for operators supported near $x_1$, the same expectation values as $|b_{x_2}\>$ for operators supported near $x_2$, and so on. In other words, multi-anyon states are characterized by the fact that 
\begin{align}
\<...,a_x,...|O|...,a_x,...\> = \<a_x|O|a_x\>
\label{locprop}
\end{align}
for every operator $O$ supported in the neighborhood of a single anyon $a_x$.

Having fixed our definitions of anyon states, the next step is to define \emph{movement} operators. For any anyon, $a$, and any pair of points, $x, x'$, we say that $M^a_{x'x}$ is a movement operator if it satisfies two conditions: (i) $M^a_{x'x}$ obeys
\begin{equation}
M^a_{x'x} |a_x\> \propto |a_{x'}\>
\label{mdef1}
\end{equation}
where the proportionality constant is a $U(1)$ phase; (ii) $M^a_{x'x}$ is \emph{local} in the sense that it is supported in the neighborhood of the interval containing $x$, $x'$. 

The locality condition on $M^a_{x'x}$ is important because it guarantees that $M^a_{x'x}$ acts the same way on any multi-anyon state of the form $|...,a_x,...\>$:  that is,
\begin{equation}
M^a_{x'x} |...,a_x,...\> \propto |...,a_{x'},...\>
\label{mdef2}
\end{equation}
as long as the other anyons in $|...,a_x,...\>$ are well-separated from the interval containing $x$ and $x'$. Indeed, to derive Eq.~\ref{mdef2} from Eq.~\ref{mdef1}, consider the expectation value of any operator, $O$, supported in the neighborhood of $a_{x'}$, in the two states $|...,a_{x'},...\>$ and $M^a_{x'x} |...,a_x,...\>$. Using (\ref{locprop}) and (\ref{mdef1}), we can see that $O$ has the same expectation value in the two states:
\begin{align}
\<...,a_{x'},...| O &|...,a_{x'},...\> = \<a_{x'}| O |a_{x'}\> \nonumber \\
&= \<a_{x}| (M^a_{x'x})^\dagger O M^a_{x'x}  |a_x\> \nonumber \\
&=\<...,a_{x},...| (M^a_{x'x})^\dagger O M^a_{x'x}  |...,a_x,...\>
\end{align}
This is also true for operators supported near any other anyon, since $M^a_{x'x}$ is supported away from those anyons. It follows that these two states must be the same, up to a phase.

In addition to the movement operators, we also define \emph{splitting} operators for our anyons. Fix two well-separated points on the line, which we will call `$1$' and `$2$'. For any pair of anyons $a,b$, we say that $S(a,b)$ is a splitting operator if it satisfies two conditions: (i) $S(a,b)$ satisfies
\begin{equation}
S(a,b)|ab_1\> \propto|a_1, b_2\>
\end{equation}
where the proportionality constant is a $U(1)$ phase; (ii) $S(a,b)$ is supported in the neighborhood of the interval $[1,2]$. Again, the second condition guarantees that the splitting operators can be applied to any multi-anyon state of the form $|...,ab_1,...\>$ provided that the other anyons are located far from the interval $[1,2]$:
\begin{equation}
S(a,b)|...,ab_1,...\> \propto|...,a_1, b_2,...\>
\label{smultiprop}
\end{equation}
where the proportionality constant is a $U(1)$ phase. Note that, unlike the movement operators, we only define splitting operators that act at a \emph{single} location `$1$' on the $x$-axis.
 
\begin{figure}[tb]
\centering
\includegraphics[width=0.9\columnwidth]{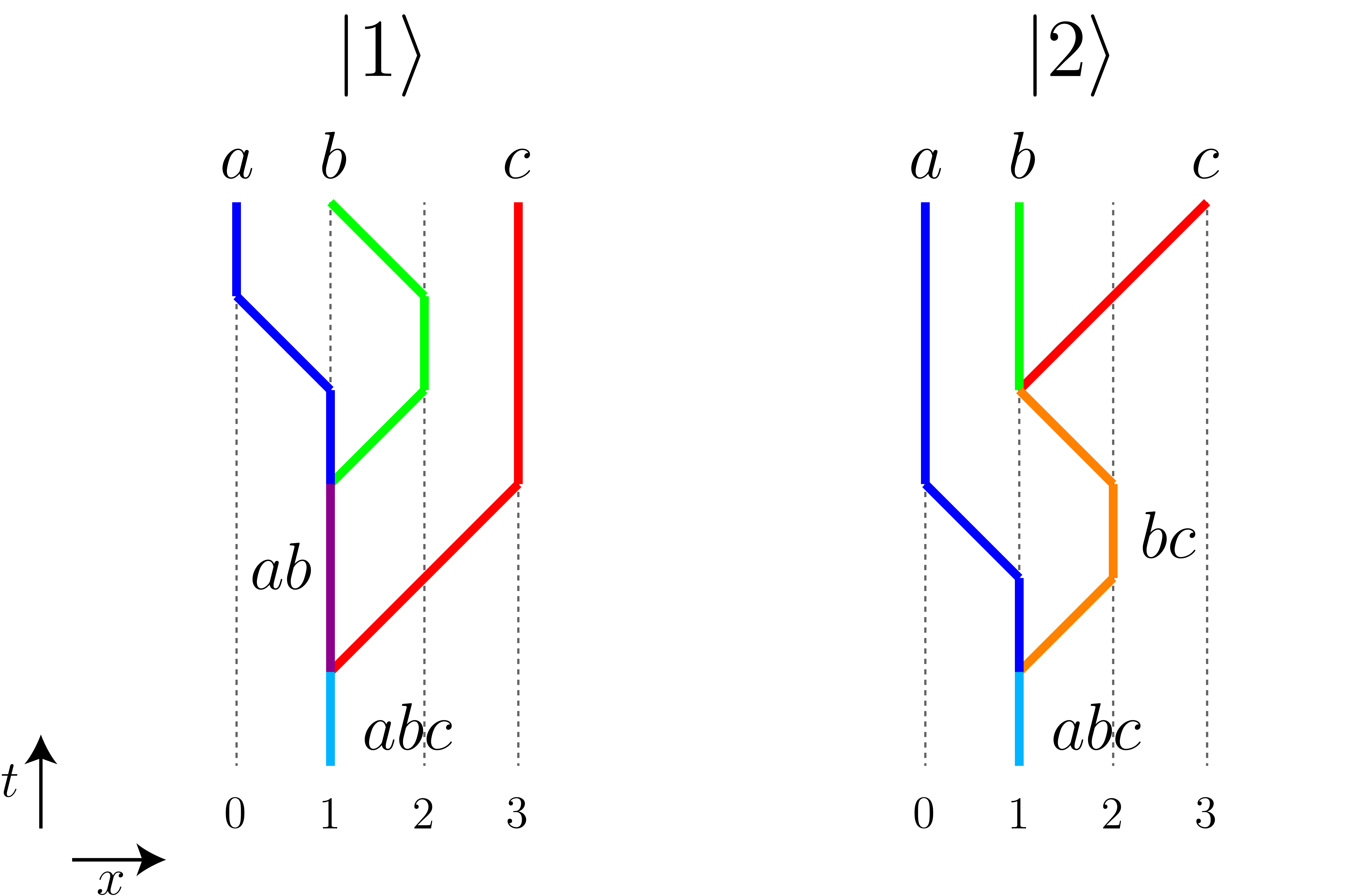}
\caption{The two processes that are compared in the microscopic definition of the $F$-symbol. 
}
\label{fig:two}
\end{figure}

With this setup, we are now ready to define the $F$-symbol. The first step is to fix some choice of anyon states $|a_x\>$ and some choice of movement and splitting operators $M^a_{x'x}, S(a,b)$. Next, consider the initial state $|abc_1\>$, i.e. the state with a single anyon $abc$ at position $1$. We then apply two different sequences of movement and splitting operators to $|abc_1\>$, denoting the final states by $|1\>$ and $|2\>$:
\begin{align}
|1\>&= M^b_{12} M^a_{01} S(a,b) M^c_{32} S(ab,c) |abc_1\> \nonumber \\
|2\>&= M^c_{32} S(b,c) M^{bc}_{12} M^a_{01} S(a,bc) |abc_1\> 
\label{fdef1}
\end{align}
These two processes are shown in Figure \ref{fig:two}. By construction, the final states $|1\>, |2\>$ produced by these processes both contain anyons $a, b, c$ at positions $0, 1, 3$, respectively. In particular, this means that $|1\>, |2\>$ are the same up to a phase. We define the $F$-symbol $F(a,b,c)$ to be this phase difference:
\begin{equation}
F(a,b,c)=\<2|1\>
\label{fdef2}
\end{equation}

\subsection{Checking the microscopic definition}
\label{Finvab}
To show that our microscopic definition is sensible, we need to establish two properties of $F$: (i) $F$ is well-defined in the sense that different choices of anyon states and movement and splitting operators give the same $F$ up to a gauge transformation (\ref{gaugetrans}); and (ii) $F$ obeys the pentagon identity (\ref{pentid}). We prove property (ii) in Appendix~\ref{pentagon}; the goal of this section is to prove property (i). 

As a warm-up, let us see how $F$ transforms if we change the \emph{phase} of the movement and splitting operators. That is, suppose we replace
\begin{align}
M^a_{x'x} \rightarrow e^{i\theta_{x'x}(a)} M^a_{x'x}, \quad S(a,b) \rightarrow e^{i\phi(a,b)} S(a,b) 
\end{align}
for some real-valued $\theta, \phi$. Substituting these transformations into (\ref{fdef1}-\ref{fdef2}) gives
\begin{align}
F(a,b,c) \rightarrow F(a,b,c) \frac{e^{i\phi(ab,c)}e^{i\phi(a,b)} e^{i\theta_{12}(b)}}{e^{i\phi(a,bc)}e^{i\phi(b,c)}e^{i\theta_{12}(bc)}}
\end{align}
Crucially, this transformation is identical to a gauge transformation (\ref{gaugetrans}) with
\begin{equation}
\nu(a,b)=\phi(a,b)+\theta_{12}(b)
\label{nuphitheta}
\end{equation}
This is exactly what we want: different phase choices lead to the same $F$, up to a gauge transformation. Although the above calculation is very simple, we would like to emphasize that this is actually a non-trivial test of our definition of $F$. In fact, the two processes in (\ref{fdef1}) were designed specifically to pass this test.

With this warm-up, we are now ready to consider the general case where we change the movement and splitting operators in an arbitrary way (for a fixed choice of anyon states $|a_x\>$). Denoting the new movement and splitting operators by
\begin{align}
M^a_{x'x} \rightarrow \tilde{M}^a_{x'x}, \quad \quad S(a,b) \rightarrow \tilde{S}(a,b),
\end{align}
it follows from properties (\ref{mdef2}), (\ref{smultiprop}) that
\begin{align}
\tilde{M}^a_{x'x} |...,a_x,...\> &= \omega_1 \cdot M^a_{x'x} |...,a_{x},...\> \label{maphase} \\
\tilde{S}(a,b)|...,ab_1,...\> &= \omega_2 \cdot S(a,b)|...,ab_1,...\>  
\end{align}
where $\omega_1, \omega_2$ are $U(1)$ phases. These $U(1)$ phases are highly constrained: taking the inner product of the two sides of (\ref{maphase}) with $M^a_{x'x} |...,a_{x},...\>$ and using property (\ref{locprop}), we can see that $\omega_1$ can only depend on $a, x, x'$. By the same reasoning, $\omega_2$ can only depend on $a, b$. Hence, we must have
\begin{align}
\tilde{M}^a_{x'x} |...,a_x,...\> &= e^{i\theta_{x'x}(a)} M^a_{x'x} |...,a_{x},...\> \nonumber \\
\tilde{S}(a,b)|...,ab_1,...\> &= e^{i\phi(a,b)} S(a,b)|...,ab_1,...\>
\end{align}
for some real-valued $\theta, \phi$. Substituting these relations into (\ref{fdef1}), we again see that $F$ changes by a gauge transformation with $\nu$ given by (\ref{nuphitheta}).

At this point, we have shown that different choices of movement and splitting operators lead to the same $F$, up to a gauge transformation. To complete the proof of property (i), we need to check that different choices of representative anyon states $|a_x\>$ also lead to the same $F$, up to a gauge transformation. In order to investigate this issue, we make a physical assumption which is inspired by the definition of anyon types given in Sec.~\ref{sec:abel_review}: we assume that different choices of anyon states are related to one another by a local unitary transformation, $U$. (By a `local unitary transformation', we mean a unitary operator generated by the finite time evolution of a local Hamiltonian that is supported near the $x$-axis~\cite{chenlocalunitary}). Given this assumption, our task is to understand how $F$ changes if we replace
\begin{align}
|a_{x_1}, b_{x_2}, c_{x_3}, ...\> \rightarrow |a_{x_1}, b_{x_2}, c_{x_3}, ...\>'
\end{align}
where 
\begin{align}
|a_{x_1}, b_{x_2}, c_{x_3}, ...\>' = U |a_{x_1}, b_{x_2}, c_{x_3}, ...\>
\end{align}
for some local unitary transformation $U$. To understand the effect on $F$, observe that we can choose movement and splitting operators for the states $\{|a_{x_1}, b_{x_2}, c_{x_3}, ...\>'\}$ however we like, changing $F$ by at most a gauge transformation. The simplest choice is
\begin{align}
(M^a_{x'x})' = U M^a_{x'x} U^\dagger, \quad \quad S'(a,b) = U S(a,b) U^\dagger
\end{align}
where $M^a_{x'x}$ and $S(a,b)$ are movement and splitting operators for $\{|a_{x_1}, b_{x_2}, c_{x_3}, ...\>\}$. With this choice, it is clear that $|1'\>= U|1\>$, and $|2'\> = U|2\>$. It follows that $F' = \<2'|1'\> = \<2|1\> = F$. Thus, we conclude that $F$ is invariant under a replacement of the anyon states, $|a_{x_1}, b_{x_2}, c_{x_3}, ...\> \rightarrow |a_{x_1}, b_{x_2}, c_{x_3}, ...\>'$. This completes the proof of property (i) above.


\section{Examples of $F$-symbol calculations}
In this section, we illustrate our definition by computing the $F$-symbol for several lattice models and edge theories.
\label{sec:abelian_F_ex}
\subsection{Toric code model}
\label{tcmodel}
\subsubsection{Hamiltonian}
\begin{figure}[tb]
\centering
\includegraphics[width=0.9\columnwidth]{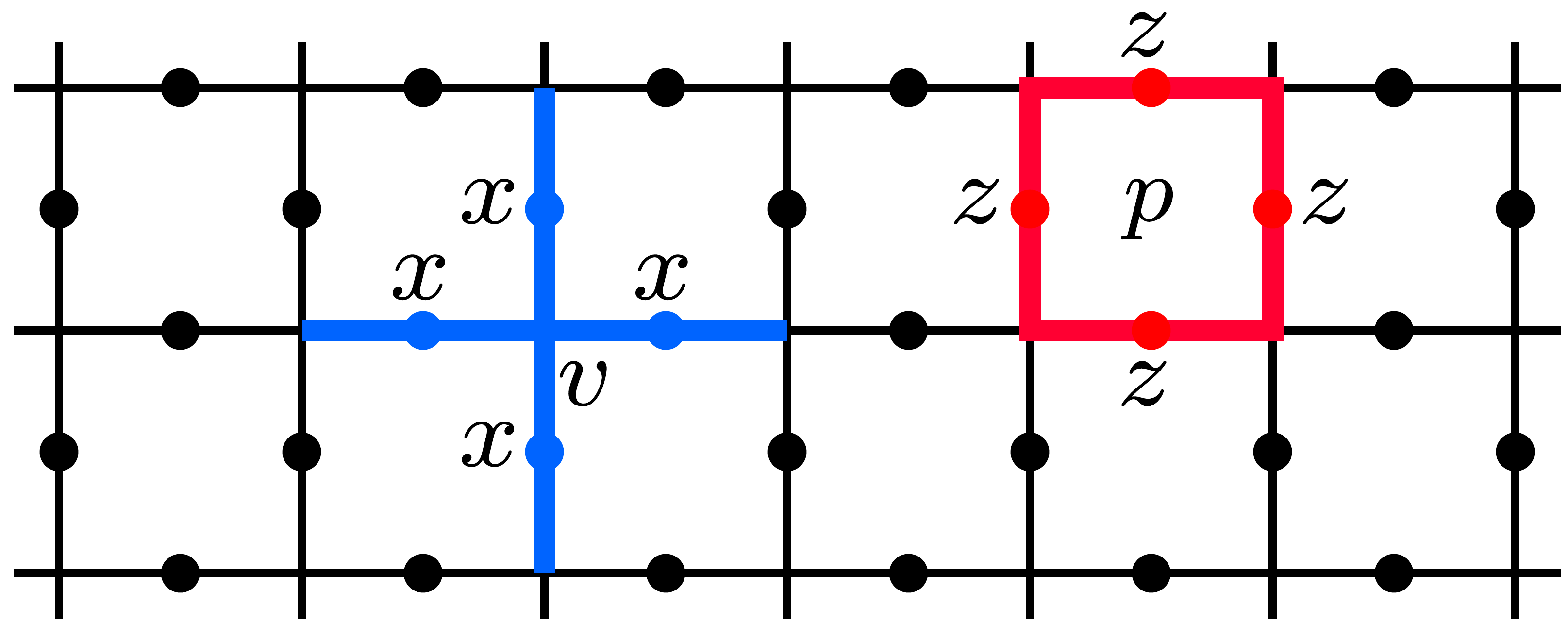}
\caption{The two operators $A_v$ and $B_p$ in the toric code Hamiltonian. 
}
\label{fig:tcham}
\end{figure}

The toric code model is an exactly solvable spin-$1/2$ model where the spins live on the links of the square lattice~\cite{kitaevtc}. The Hamiltonian is
\beq
H=-\sum_{v}A_v-\sum_{p}B_p
\eeq
where the two sums run over the vertices, $v$, and plaquettes, $p$, of the square lattice. The two operators $A_v, B_p$ are defined by
\begin{align}
A_v = \prod_{\ell\in v}\sigma^x_\ell, \quad \quad B_p =\prod_{\ell\in \partial p}\sigma^z_\ell
\end{align}
where the first product runs over the four links, $\ell$, that are adjacent to the vertex $v$, and the second product runs over the four links, $\ell$, that belong to the boundary of the plaquette $p$ (Fig.~\ref{fig:tcham}).

To find the ground state of this model in an infinite plane geometry, notice that the $A_v, B_p$ operators commute with one another and have eigenvalues $\pm 1$. From this, one can deduce that the state $|A_v = B_p = 1\>$ is the ground state since this state minimizes the energies of all the terms in the Hamiltonian.

As for excitations, the toric code model supports four different (anyon) types, which we denote by $\{1,e,m,em\}$. All four anyon types have simple realizations on the lattice: the `$e$' anyon corresponds to an excitation where $A_v = -1$ for a single vertex $v$; the `$m$' anyon corresponds to an excitation where $B_p = -1$ for a single plaquette $p$; the `$em$' anyon corresponds to an excitation where $A_v = -1$ and $B_p = -1$ for both a vertex $v$ and nearby plaquette $p$; finally the `$1$' anyon (also known as the ``trivial'' anyon) corresponds to having no excitation at all. The fusion rules for these anyons are 
\begin{align}
e\times e = m\times m= 1, \quad \quad e\times m = em
\end{align}

\subsubsection{$F$-symbol calculation}
\label{tcF}
\begin{figure}[tb]
\centering
\includegraphics[width=0.8\columnwidth]{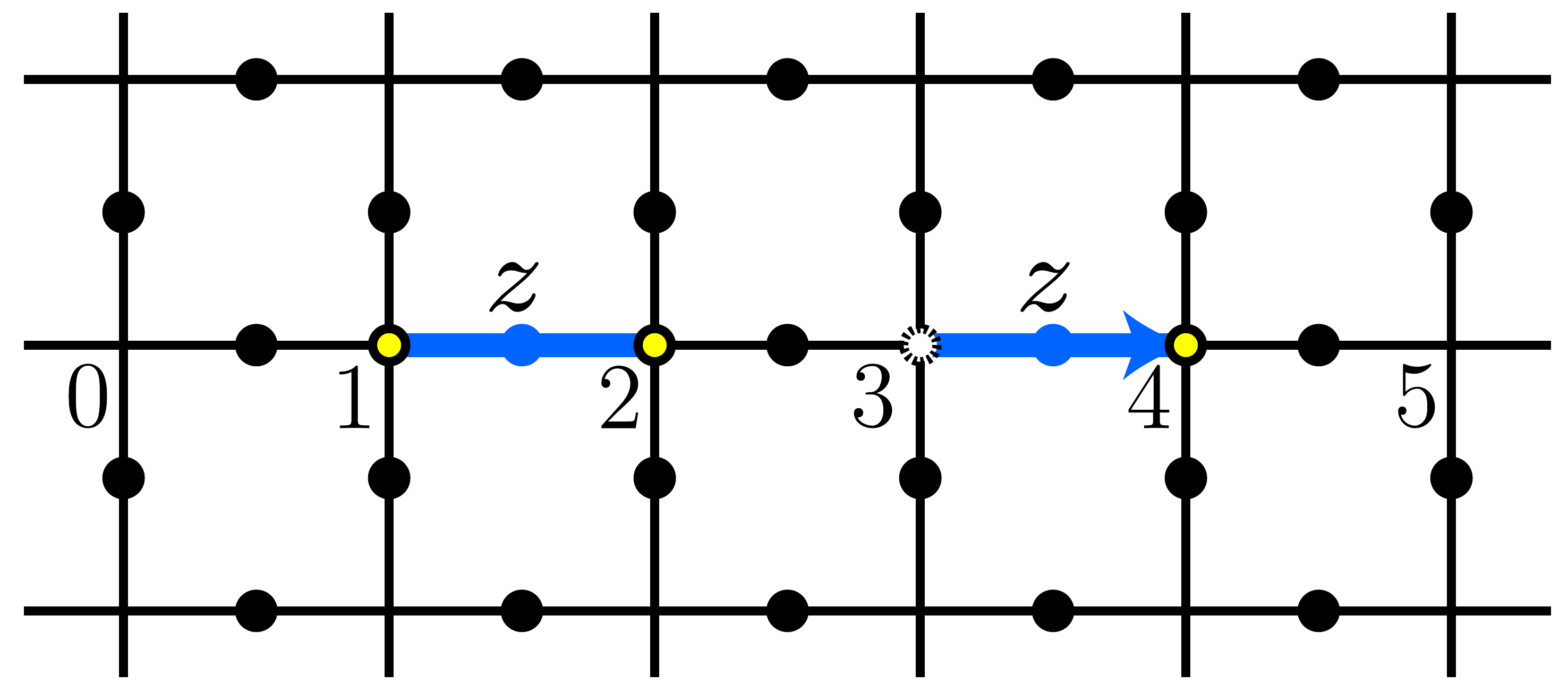}
\caption{Movement and splitting operators for $e$ anyons. The splitting operator for $e$ is $S(e,e)=\sigma^z_{1,2}$. The movement operator that moves $e$ from vertex $3$ to $4$ is $M^e_{4,3}=\sigma^z_{3,4}$.}
\label{fig:tcsplit}
\end{figure}

We now illustrate our definition of $F$ by computing $F(e,e,e)$. First, we define some notation for labeling anyon states. Following the procedure outlined in Sec.~\ref{sec-micro-F-abel}, we focus on anyon states where the anyons live along a line, specifically the $x$-axis. We then label the vertices along the $x$-axis by integers, $n = 0, \pm 1, \pm 2,...$ (Fig.~\ref{fig:tcsplit}), and we define $|e_n\>$ to be the state with $A_{v=n} = -1$, and all other $A_v$'s and $B_p$'s equal to $+1$. 

Next, we need to construct movement and splitting operators for the $e$ anyons. We define the movement operator between neighboring vertices $n$ and $n+1$ by
\beq
M^e_{(n+1)n}= \sigma^z_{n,n+1}
\eeq
Here $\sigma^z_{n,n+1}$ denotes the $\sigma^z$ operator on the link $\ell =\<n(n+1)\>$ (Fig.~\ref{fig:tcsplit}). It is easy to check that this movement operator obeys the required property, $M^e_{(n+1)n} |e_n\> \propto |e_{n+1}\>$ using the fact that $\sigma^z_{n,n+1}$ anticommutes with the $A_v$ operators on vertex $n$ and vertex $n+1$.

We define the reverse movement operator in the same way:
\beq
M^e_{n(n+1)} = \sigma^z_{n,n+1}
\eeq
Also, we define the splitting operator for the $e$ anyon to be (Fig.~\ref{fig:tcsplit})
\beq
S(e,e)=\sigma^z_{1,2}
\eeq
Again, one can check that $S(e,e)$ has the required property, $S(e,e)|1_1\> \propto |e_1, e_2\>$, using the fact that $\sigma^z_{1,2}$ anticommutes with $A_v$ on vertex $1$ and vertex $2$.

We will also need the splitting operator $S(1,e)$, which we define as
\beq
S(1,e)=\sigma^z_{1,2}
\eeq
Finally, we will need the movement operator $M^1_{n'n}$ and the splitting operator $S(e,1)$, which we take to be the identity operator:
\beq
 M^1_{n'n} = S(e,1) = 1
\eeq
(The reason why we can take $S(e,1) = 1$ is that $S(e,1)$ is defined by the condition $S(e,1)|e_1\> \propto |e_1, 1_2\>$, which is satisfied by the identity operator $S(e,1) = 1$.)

With these definitions, we are now equipped to calculate $F(e,e,e)$. Following Eq. (\ref{fdef1}), we have
\begin{align}
|1\> &= M^e_{12} M^e_{01}S(e,e)M^e_{32}S(1,e)|e_1\> \nonumber \\
&= \sigma^z_{1,2} \sigma^z_{0,1} \sigma^z_{1,2} \sigma^z_{2,3}\sigma^z_{1,2}|e_1\>
\end{align}
Similarly,
\begin{align}
|2\> &=M^e_{32}S(e,e)M^{1}_{12}M^e_{01}S(e,1)|e_1\> \nonumber\\
&= \sigma^z_{2,3}\sigma^z_{1,2}\sigma^z_{0,1}|e_1\>
\end{align}
Comparing the two expressions, we see that $|1\> = |2\>$, so that
\beq
F(e,e,e)=\<2|1\>=1
\eeq
More generally, one can check that all the $F$-symbols are $1$ for this model, i.e.
\begin{equation}
F(a,b,c) = 1
\end{equation}
for all $a,b,c \in \{1,e,m,em\}$ for an appropriate choice of movement and splitting operators. 

\subsection{Doubled semion model}
\begin{figure}[tb]
\centering
\includegraphics[width=1.0\columnwidth]{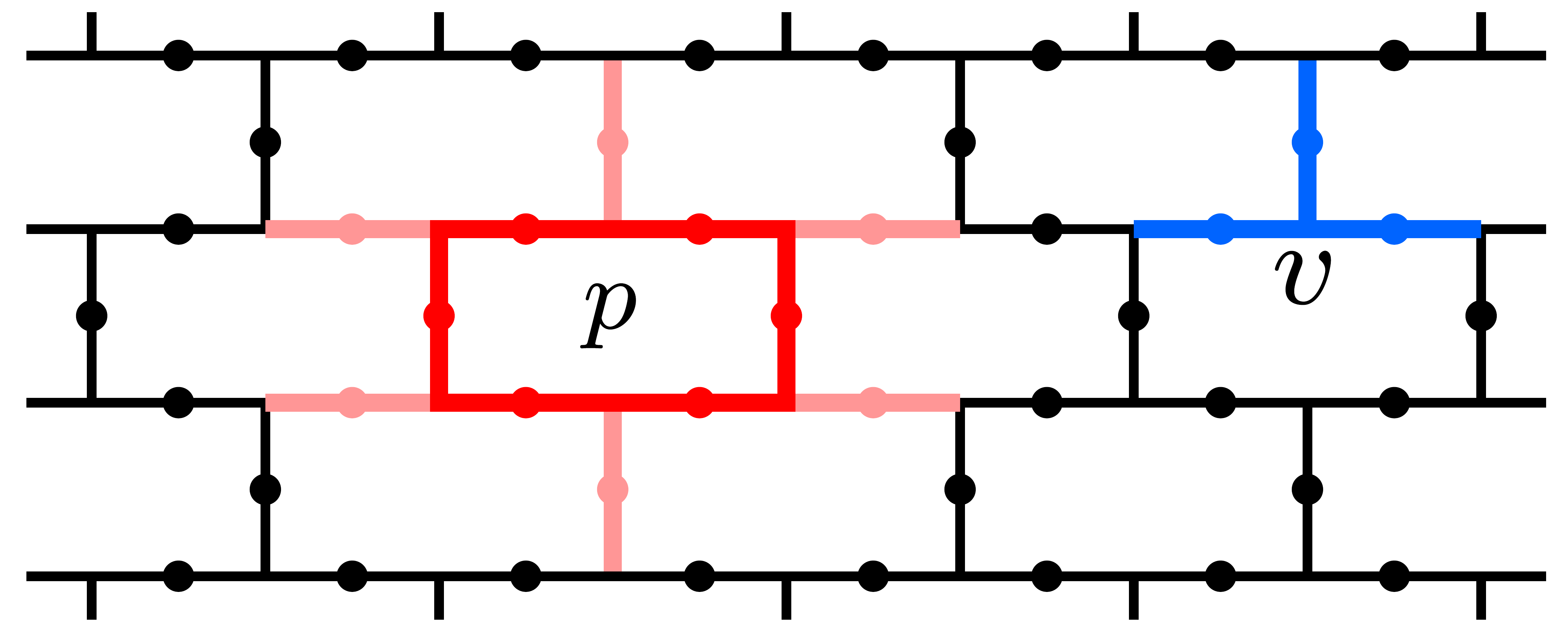}
\caption{The two operators $Q_v$ and $B_p$ in the doubled semion Hamiltonian: $Q_v$ is a product of $\sigma^x$ operators acting on the three blue links adjacent to the vertex $v$, while $B_p$ involves a product of $\sigma^z$ operators on the six red links adjacent to the plaquette $p$ and a product of $i^{(1-\sigma^x)/2}$ on the six pink ``legs'' of the plaquette $p$.}
\label{fig:semion}
\end{figure}

\subsubsection{Hamiltonian}
The doubled semion model is an exactly solvable spin-$1/2$ model where the spins live on the links of the honeycomb lattice~\cite{levinwenstrnet}. The Hamiltonian is 
\begin{equation}
H=-\sum_v Q_v-\sum_p B_p
\end{equation}
where the two sums run over the vertices, $v$, and plaquettes, $p$, of the honeycomb lattice. The operator $Q_v$ is defined by
\begin{equation}
Q_v= \frac{1}{2}\left(1 + \prod_{\ell\in v}\sigma^x_\ell \right)
\end{equation}
Likewise, $B_p$ is defined by
\begin{equation}
B_p=\frac{1}{2} \left(1 - \prod_{\ell\in \partial p}\sigma^z_\ell\prod_{\ell\in\text{legs of }p}i^{\frac{1-\sigma^x_\ell}{2}} \right) P_p
\end{equation}
where the second product runs over the six links, $\ell$, that form the ``legs'' of the plaquette $p$ (Fig.~\ref{fig:semion}) and $P_p$ is the projector
\begin{equation}
P_p=\prod_{v\in \partial p}Q_v
\end{equation}
Like the toric code model, the $Q_v, B_p$ operators commute with one another. Also, one can check that $Q_v$ and $B_p$ have eigenvalues $0$ and $1$ so that the ground state in an infinite plane geometry is the state $|Q_v = B_p = 1\>$. 

As for excitations, it is known that the doubled semion model supports four types of anyons, which we denote by $\{1,s,\bar{s},s\bar{s}\}$. The fusion rules for these anyons are
\begin{align}
s\times s= \bar{s} \times \bar{s} = 1, \quad \quad s \times \bar{s} = s\bar{s}
\end{align}
Similarly to the toric code model, these anyon excitations correspond to defects where $Q_v =0$ or $B_p = 0$ for some collection of vertices and plaquettes. However, unlike the toric code model, the excitations are not completely characterized by their $Q_v$ and $B_p$ eigenvalues due to degeneracies in the simultaneous eigenspaces of $\{Q_v, B_p\}$. Therefore, we cannot define these excitations unambiguously in terms of their $Q_v$ and $B_p$ eigenvalues; instead, we will define/construct them by writing down explicit (string-like) creation operators.

\subsubsection{$F$-symbol calculation}

\begin{figure}[tb]
\centering
\includegraphics[width=0.9\columnwidth]{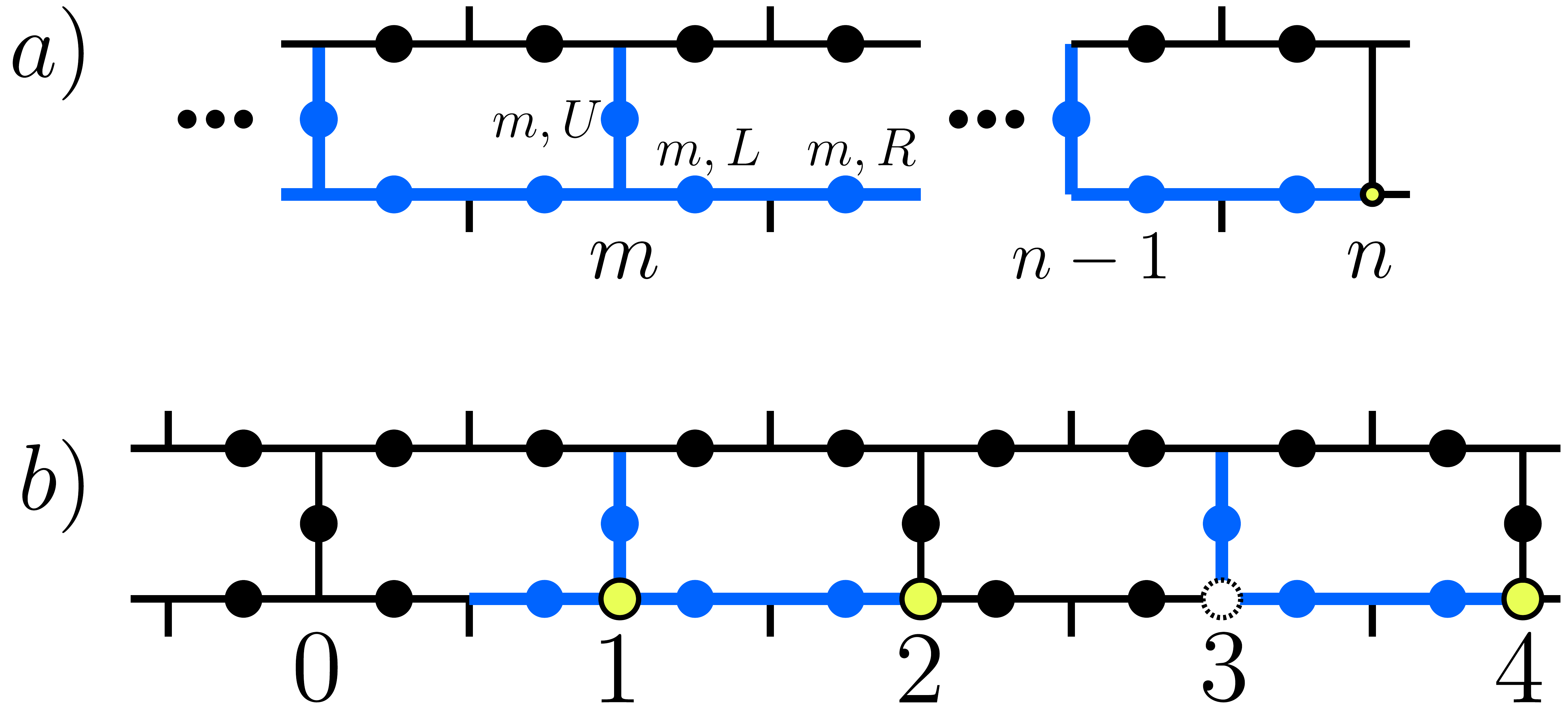}
\caption{(a) Definition of semion creation operator $a^\dagger_n$: the operator $a^\dagger_n$ acts on all $(m,U)$, $(m,L)$ and $(m,R)$ links with $m < n$. (b) Movement and splitting operators for semions: the splitting operator $S(s,s)$ acts on the blue links near $1$ and $2$ according to Eq.~\ref{ssemdef}, while the movement operator $M^s_{43}$ acts on the blue links near $3$ and $4$ according to Eq.~\ref{msemdef}.} 
\label{fig:move_semi}
\end{figure}

To illustrate our definition, we now compute the $F$-symbol $F(s,s,s)$. As before, we focus on anyon states where the $s$ anyons live along the $x$-axis. More specifically, we focus on states where the $s$ anyons live at vertices of type `$\perp$' along the $x$-axis (Fig.~\ref{fig:move_semi}b). We label these vertices by integers $n = 0, \pm 1, \pm 2,\cdots$. We then define $|s_n\>$ to be a particular state with an $s$ anyon at vertex $n$. More specifically, we define
\begin{align}
|s_n\> = a_n^\dagger |0\>
\end{align}
where $|0\> \equiv |Q_v = B_p = 1\>$ denotes the ground state, and where $a_n^\dagger$ is the following string-like creation operator for $s$ anyons:
\begin{align}
a^\dagger_n &= \prod_{m<n}\sigma^z_{m,L} \sigma^z_{m,R} \cdot \prod_{m < n}(-1)^{\frac{1}{4}(1+\sigma^x_{m,L})(1-\sigma^x_{m,R})} \nonumber \\
&\cdot \prod_{m < n}i^{\frac{1-\sigma^x_{m,U}}{2}}
\label{semstring}
\end{align}
Here, we have introduced some notation: we denote the three links near vertex $m$ by the labels $(m,U)$, $(m,L)$, $(m,R)$  (see Fig.~\ref{fig:move_semi}a). The justification for this string operator comes from Ref.~\onlinecite{levinwenstrnet}, where it was argued that string operators of the form (\ref{semstring}) create semion excitations at their endpoints. 
 
Next we need to construct movement and splitting operators for $s$. We define the movement operator between vertices $n$ and $n+1$ as (see Fig. \ref{fig:move_semi}b):
\begin{equation}
M^s_{(n+1)n}=\sigma^z_{n,L}\sigma^z_{n,R}(-1)^{\frac{1}{4}(1+\sigma^x_{n,L})(1-\sigma^x_{n,R})}i^{\frac{1-\sigma^x_{n,U}}{2}}
\label{msemdef}
\end{equation}
It is clear that $M^s_{(n+1)n}$ obeys the required property $M^s_{(n+1)n} |s_n\> \propto |s_{n+1}\>$ since $M^s_{(n+1)n} a^\dagger_n \propto a^\dagger_{n+1}$. Likewise, we define the reverse movement operator by
\begin{equation}
M^s_{n(n+1)} = (M^s_{(n+1)n})^{-1}
\end{equation}

As for the splitting operator, we define 
\begin{equation}
S(s,s)=\sigma^x_{0,R}\sigma^z_{1,L}\sigma^z_{1,R}(-1)^{\frac{1}{4}(1+\sigma^x_{1,L})(1-\sigma^x_{1,R})}i^{\frac{1-\sigma^x_{1,U}}{2}}
\label{ssemdef}
\end{equation}
(see Fig.~\ref{fig:move_semi}b). Notice that 
\begin{align}
S(s,s) = M^s_{21} \sigma^x_{0,R} 
\label{smdsemion}
\end{align}
The extra factor of $\sigma^x_{0,R}$ is necessary to make sure that the two excitations created by $S$ are the same as those created by $a_n^\dagger$, i.e.
\begin{equation}
S(s,s)|0\> \propto a^\dagger_1 a^\dagger_2 |0\>
\label{Sidentity_dsem}
\end{equation}
(see Appendix~\ref{semapp} for a derivation). Equation (\ref{Sidentity_dsem}) guarantees that $S(s,s)$ has the required property for a splitting operator: $S(s,s)|1_1\> \propto |s_1, s_2\>$. 

We will also need the splitting operator $S(1,s)$, which we define as
\begin{align}
S(1,s) = M^s_{21}
\end{align}
Finally, we will need the movement operator $M^1_{n'n}$ and the splitting operator $S(s,1)$, which we define as
\beq
M^1_{n'n} = S(s,1) = 1
\eeq

We are now ready to compute the $F$-symbol $F(s,s,s)$. Using Eq. (\ref{fdef1}), we have:
\begin{align}
|1\> &= M^s_{12} M^s_{01}S(s,s)M^s_{32}M^s_{21}|s_1\> \nonumber \\
|2\> &= M^s_{32}S(s,s)M^s_{01}|s_1\>
\end{align}
To proceed further, we note that all movement operators commute with one another since they act on non-overlapping spins. We also note that the splitting operator $S(s,s)$ \emph{anticommutes} with $M^s_{01}$ and $M^s_{10}$ but commutes with every other movement operator, as one can see from (\ref{smdsemion}). Using these facts, we can simplify $|1\>$ to
\begin{align}
|1\> &= -M^s_{32} S(s,s) M^s_{01} |s_1\> 
\end{align}
Hence, $|1\> = -|2\>$ so
\begin{align}
F(s,s,s) = \<2|1\> = -1
\end{align}

In fact, using the same movement and splitting operators, one can show that all the other $F$-symbols involving $1$ and $s$ are trivial:
\begin{align}
F(1,1,1) &= F(1,1,s) = F(1,s,1) = F(1,s,s)=1 \nonumber \\
F(s,1,1) &= F(s,1,s) = F(s,s,1) = 1 \nonumber
\end{align}
In particular, this means that
\begin{align}
F(s,s,s) F(s,1,s) = -1
\end{align}
The latter result is significant because one can check that $F(s,s,s) F(s,1,s)$ is \emph{invariant} under the gauge transformation (\ref{gaugetrans}). Therefore we have shown that the $F$-symbol for the doubled semion model is not gauge equivalent to $F(a,b,c) = 1$. A corollary of this result is that the toric code model and the doubled semion model must belong to different phases since they have different $F$-symbols.


\subsection{Semion edge theory}
\label{sec:semionfield}

We now show how to compute the $F$-symbol using an \emph{edge} theory of a topological phase. Specifically, we consider the simplest non-trivial topological phase, namely the Laughlin $\nu = 1/2$ bosonic fractional quantum Hall state. This topological phase supports two types of anyon excitations, which we denote by $\{1,s\}$. Here, $s$ can be thought of as the charge $1/2$ quasihole (which happens to be a semion), while $1$ denotes the trivial anyon. The anyons obey the following fusion rule: $s \times s = 1$.

Our goal is to compute the $F$-symbol for this topological phase using its edge theory. The edge theory consists of a single chiral boson field $\phi$ obeying commutation relations~\cite{wenreview}
\begin{equation}
[\phi(x), \partial_y \phi(y)] = \pi i \delta(x-y)
\end{equation}
The Hamiltonian for the edge theory is
\begin{equation}
H=\int dx\frac{v}{4\pi}(\partial_x\phi)^2
\end{equation}
where $v$ describes the velocity of the chiral boson edge mode. 

To fully define the edge theory, it is important to specify the set of local operators. For the above edge theory, the fundamental local operators are $e^{\pm i 2 \phi}$, which can be thought of as the creation/annihilation operators for the (charge $1$) boson on the edge, respectively. All other local operators can be constructed by taking derivatives and products of $e^{\pm i 2 \phi}$.

We now compute the $F$-symbol $F(s,s,s)$. The first step is to define the anyon states that we will manipulate. Following the standard edge theory formalism~\cite{wenreview}, we define the anyon state $|s_x\>$ by
\begin{align}
|s_x\> = a_x^\dagger |0\>
\end{align}
where $|0\>$ denotes the ground state of the edge theory, and $a_x^\dagger$ is defined by
\begin{align}
a_x^\dagger = e^{i\int_{-\infty}^x\partial_{y}\phi(y)dy}
\end{align}
Here, $a_x$ can be thought of as a (string-like) creation operator for $s$ at position $x$ along the edge. 

Next we define movement and splitting operators for $s$. We define the movement operator by
\begin{equation}
M^s_{x'x}=e^{i\int_x^{x'}\partial_y\phi(y)dy}
\end{equation}
Notice that $M^s_{x'x} a^\dagger_x \propto a^\dagger_{x'}$  which guarantees that $M^s_{x'x}$ obeys the required property $M^s_{x'x} |s_x\> \propto |s_{x'}\>$.

Likewise, we define the splitting operator for $s$ by
\begin{align}
S(s,s)&= M^s_{21} e^{i2\phi(1)} \nonumber \\
&= e^{i\int_{1}^2\partial_x\phi(x)dx}e^{i2\phi(1)}
\end{align}
Here the extra factor of $e^{2i\phi(1)}$ is necessary to ensure that
\begin{align}
S(s,s) |0\> \propto a_1^\dagger a_2^\dagger |0\>
\end{align}

The other splitting and movement operators that we need are $S(1,s)$, which we define by
\begin{align}
S(1,s) = M^s_{21}
\end{align}
and $M^1_{x'x}$ and $S(s,1)$ which we define as
\beq
M^1_{x'x} = S(s,1) = 1
\eeq

We are now ready to compute the $F$-symbol $F(s,s,s)$. Following the definition (\ref{fdef1}), we have
\begin{align}
|1\> &= M^s_{12} M^s_{01}M^s_{21}e^{i2\phi(1)}M^s_{32}M^s_{21}|s_1\> \nonumber \\
|2\> &=M^s_{32}M^s_{21}e^{i2\phi(1)}M^s_{01}|s_1\> 
\end{align}
To compare these two expressions we need to rearrange the order of these operators. We can do this with the help of the following relations, which can be derived from the Baker-Campbell-Hausdorff formula:
\begin{align}
M^s_{x''x'} M^s_{x'x}  &= e^{i \pi/2} M^s_{x'x} M^s_{x''x'}, \quad  x < x' < x'' \nonumber \\
e^{i2\phi(1)} M^s_{1x} &= -M^s_{1x} e^{i2\phi(1)},\quad x\neq 1
\end{align}
With these formulas and the identity $M^s_{xx'}=(M^s_{x'x})^{-1}$, we can rewrite $|1\>$ as:
\begin{align}
|1\> &= - M^s_{32}M^s_{21}e^{i2\phi(1)}M^s_{01}|s_1\>
\end{align}
Hence,
\begin{align}
F(s,s,s) = \<2|1\> = -1
\end{align}
In exactly the same way, one can show that $F(a,b,c) = 1$ for every other choice of $a,b,c \in \{1,s\}$. 

Before concluding, we need to address a question that may worry some readers: our microscopic definition of the $F$-symbol assumes a finite correlation length $\xi$, so what is the justification for applying it to a gapless edge theory? Our answer is that, in some edge theories, the anyon states in the bulk can be transformed into anyon states at the edge by a local unitary transformation. If this is the case, an edge calculation is guaranteed to give the same answer as a bulk calculation where our definition is on firmer ground. We believe that the edge theory analyzed above (and in the next section) has this property.


\subsection{General chiral boson edge theory}
\label{chibosonF}
We now extend the calculation of the previous section to a general bosonic Abelian topological phase. According to the $K$-matrix formalism, every bosonic Abelian topological phase can be described by a multi-component $U(1)$ Chern-Simons theory of the form~\cite{wenreview}
\begin{align}
L = \sum_{ij\mu\nu} \frac{K_{ij}}{4\pi} \epsilon^{\lambda \mu \nu} a_{\lambda i} \partial_\mu a_{\nu j}
\end{align}
where $K_{ij}$ is a non-degenerate $N \times N$ integer symmetric matrix with even elements on the diagonal. Our goal will be to compute the $F$-symbol corresponding to any given $K_{ij}$. (The example in the previous section corresponds to the case $K = 2$).

A word about notation: in the standard $K$-matrix formalism, anyon excitations are parameterized by equivalence classes of $N$-component integer vectors, where the equivalence relation is defined by $l\sim m$ if $l-m = K \Lambda$ for some integer vector $\Lambda$. Here, we use a slightly different notation: instead of working with equivalence classes of integer vectors, we choose a single \emph{representative} from each equivalence class, and we label each anyon by the corresponding representative $l$. In this notation, the set of anyons is given by a finite collection of integer vectors $\{l,m,n,...\}$. The fusion rules are given by 
\begin{align}
l \times m = [l+m]
\end{align}
where $[l+m]$ denotes the unique representative that belongs to the same equivalence class as $l+m$.

As in the previous section, we will compute the $F$-symbol using an edge theory. In particular, we will use the standard chiral boson edge theory consisting of $N$ chiral boson fields, $\Phi_1,. . .,\Phi_N$, obeying the commutation relations~\cite{wenreview}
\begin{equation}
[\Phi_i(x), \Phi_j(y)] = \pi i K^{-1}_{ij} \text{sgn}(y-x) + \pi i (K^{-1} X K^{-1})_{ij}
\label{phicommgen}
\end{equation}
with an edge Hamiltonian of the form
\begin{equation*}
H=\int dx\sum_{ij}\frac{V_{ij}}{4\pi}\partial_x\Phi_i\partial_x\Phi_j
\end{equation*}
In the above commutation relation, $X$ is (any) skew-symmetric integer matrix with $X \equiv K \pmod{2}$. The $K^{-1} X K^{-1}$ term in (\ref{phicommgen}) is not included in standard treatments of chiral boson edge theories~\cite{wenreview} but it plays the same role as the more well-known Klein factors\cite{senechalreview}: it guarantees that non-overlapping local operators commute with one another (see Eq. \ref{loc_op_K} below).

To complete the edge theory, we need to specify the set of local operators. In this case, the fundamental local operators are those of the form 
\begin{align}
\exp(i\Lambda^T K \Phi) \equiv \exp\left(i \sum_{ij} \Lambda_i K_{ij} \Phi_j \right)
\label{loc_op_K}
\end{align}
where $\Lambda$ is an $N$ component integer column vector. All other local operators can be constructed by taking derivatives and products of $e^{i \Lambda^T K \Phi}$. Physically, $\exp(i\Lambda^T K \Phi)$ can be thought of as a product of boson creation/annihilation operators acting on the different edge modes.    

We now compute the $F$-symbol $F(l,m,n)$. The first step is to define the anyon states that we will manipulate. For each anyon type, $l$, we define a state with anyon $l$ at position $x$, by
\begin{equation}
|l_x\> = (a^{l}_x)^\dagger |0\>
\end{equation}
where $|0\>$ denotes the ground state of the edge theory and
\begin{equation}
(a^{l}_x)^\dagger=e^{il^T\int_{-\infty}^x\partial_{y}\Phi(y)dy}
\end{equation}
As in the previous example, $(a^l_x)^\dagger$ can be thought of as the string-like creation operator for anyon $l$ at position $x$.

Next we define movement and splitting operators. We define the movement operator for $l$ by
\begin{equation}
M^{l}_{x'x}=e^{il^T\int_{x}^{x'}\partial_{y}\Phi(y)dy}
\label{Mdefchi}
\end{equation}
Likewise, we define the splitting operator by
\begin{align}
S(l,m) &= M^{m}_{21} \cdot C(l,m)  
\label{Sdefchi}
\end{align}
where $C(l,m)$ denotes the operator $C(l,m) = e^{i (l+m-[l+m])^T\Phi(1)}$. As in the previous example, the extra factor of $e^{i (l+m-[l+m])^T\Phi(1)}$ is necessary to ensure that
\begin{align}
S(l,m)|[l+m]_1\> \propto (a^{l}_1)^\dagger (a^{m}_2)^\dagger |GS\>
\end{align}

We are now ready to compute the $F$-symbol $F(l,m,n)$. From the definition (\ref{fdef1}) we have:
\begin{align}
|1\>&= M^m_{12} M^l_{01}M^m_{21} C(l,m) M^{n}_{32}M^n_{21}\nonumber \\
&\cdot C([l+m],n)|[l+m+n]_1\> \nonumber \\
|2\>&=M^n_{32}M^n_{21} C(m,n)M^{[m+n]}_{12}M^l_{01}M^{[m+n]}_{21} \nonumber \\
&\cdot C(l, [m+n])|[l+m+n]_1\> 
\end{align}
To simplify these expressions we use the following commutation relations:
\begin{align}
M^l_{x''x'} M^m_{x'x}  &= e^{i \alpha(l,m)} \cdot M^m_{x'x} M^l_{x''x'}, \quad x < x' < x'' \nonumber \\
M^l_{x1} e^{i \Lambda^T K \Phi(1)} &= e^{i \alpha(l, K\Lambda)} \cdot e^{i \Lambda^T K \Phi(1)} M^l_{x1}\quad x\neq 1
\label{McommK}
\end{align}
where 
\begin{align}
\alpha(l,m) = \pi l^T K^{-1} m. 
\label{alpha}
\end{align}
We also use the relation
\begin{align}
e^{i\Lambda^T K \Phi(1)} e^{i \Xi^T K \Phi(1)} &= e^{-i \beta(K\Lambda, K\Xi)} \cdot e^{i (\Lambda + \Xi)^T K \Phi(1)}
\end{align}
where 
\begin{align}
\beta(\Lambda, \Xi) = \frac{\pi}{2} \Lambda^T (K^{-1} X K^{-1}) \Xi. 
\end{align}
All of these relations follow from Eq.~\ref{phicommgen}, together with the Baker-Campbell-Hausdorff formula. 

Using the above relations, together with the identity $M^l_{xx'} = (M^l_{x'x})^{-1}$, we simplify $|1\>, |2\>$ to:
\begin{align}
|1\> &= e^{i \alpha_1} M^l_{01} M^{n}_{32}M^n_{21} e^{i(l+m+n-[l+m+n])^T\Phi(1)} \nonumber \\
&\cdot |[l+m+n]_1\> \nonumber \\
|2\> &= e^{i \alpha_2} M^l_{01} M^n_{32}M^n_{21} e^{i(l+m+n-[l+m+n])^T\Phi(1)}\nonumber \\
&\cdot |[l+m+n]_1\> 
\end{align}
where
\begin{align}
\alpha_1 &= \alpha(l,m) +\alpha (n, l+m-[l+m]) \nonumber \\
&- \beta( l+m-[l+m], [l+m] +n - [l+m+n]) \nonumber \\
\alpha_2 &= \alpha(l,m) \nonumber \\
&-\beta(m+n-[m+n],l+[m+n]-[l+m+n])
\end{align}
We conclude that
\begin{align}
F(l,m,n) = \<2|1\> = \exp[i (\alpha_1 - \alpha_2)] 
\end{align}
To complete the calculation, we simplify the expression for $F$ by making a gauge transformation (\ref{gaugetrans}) with
\begin{align}
\nu(l,m) = \beta(l+m - [l+m],[l+m])  +\beta(m,l) + \alpha(m,l)
\label{gaugetransK}
\end{align}
After this gauge transformation, we obtain
\begin{equation}
F(l,m,n)=\exp[i\pi l^T  K^{-1}(K-X)K^{-1}(m+n-[m+n])]
\label{FsymK}
\end{equation}


\section{Defining the $R$-Symbol for Abelian anyons}
\label{sec:R}

\subsection{Abstract definition of $R$}
\label{R-abs-sec}
\begin{figure}[tb]
\centering
\includegraphics[width=.5\columnwidth]{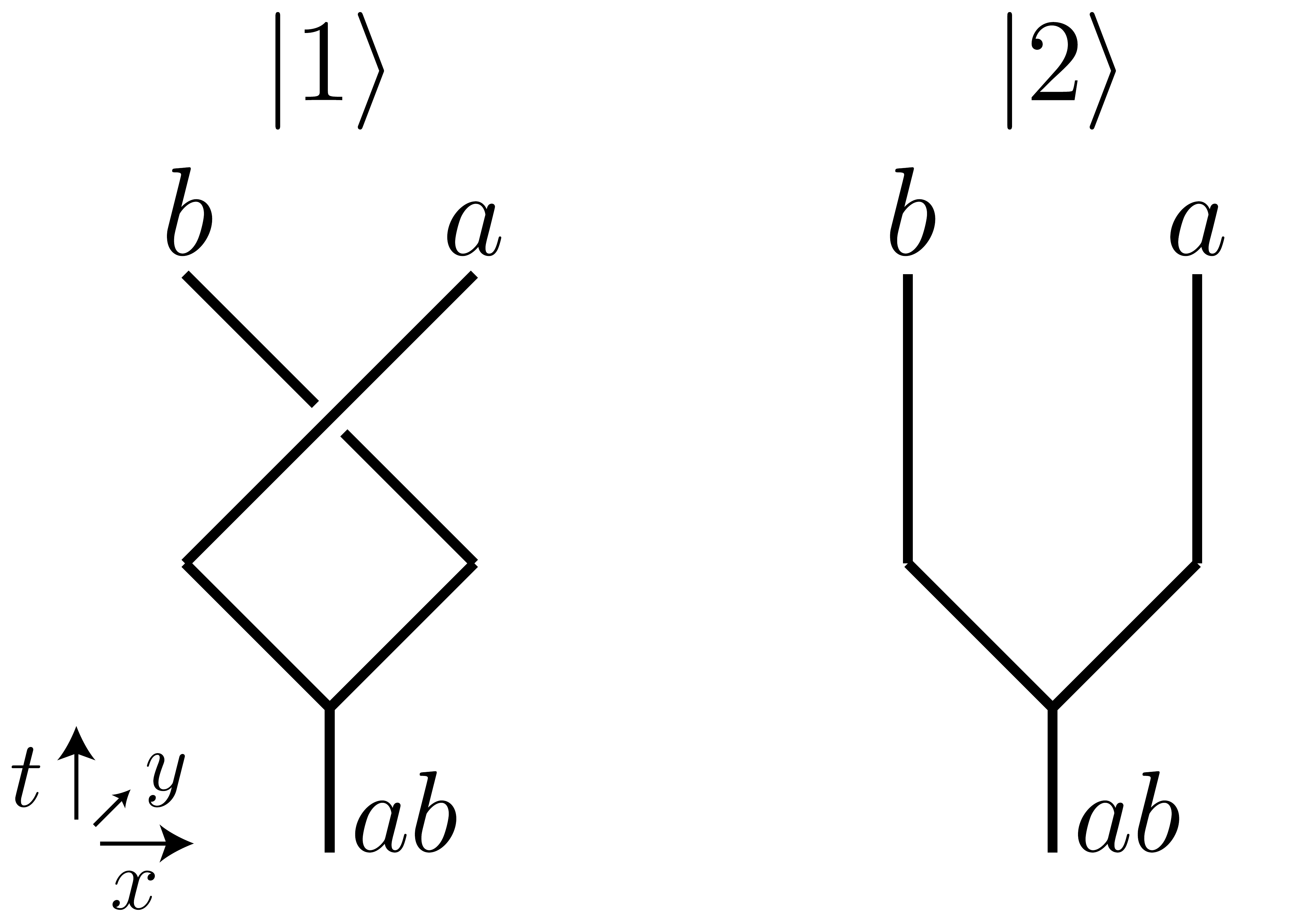}
\caption{Spacetime diagrams for two processes in which an anyon $ab$ splits into anyons $a, b$. The final states $|1\>, |2\>$ are equal up to a $U(1)$ phase, $R(a,b)$.}
\label{fig:R-abstract}
\end{figure}

We begin by reviewing the \emph{abstract} definition of the $R$-symbol in the case of Abelian anyons~\cite{kitaevlongpaper, preskilltqc}. According to this definition, $R$ is defined by comparing two different physical processes (Fig.~\ref{fig:R-abstract}). In one process, an anyon of type $ab$ splits into two anyons of type $a, b$, with $a$ on the left and $b$ on the right, and then $a, b$ are exchanged in the counterclockwise direction; the other process, $ab$ splits into $a,b$ with $b$ on the left and $a$ on the right, without any subsequent exchange. By construction, the final states $|1\>, |2\>$ produced by these processes contain the same anyons $a, b$ in the same positions. Therefore, $|1\>, |2\>$ are the same up to a phase. The $R$-symbol $R(a,b)$ is defined to be the phase difference between the two states:
\begin{align}
|1\> = R(a,b) |2\>
\label{Rdefab}
\end{align}

Two properties of the $R$-symbol follow from this definition. First, $R$ has an inherent ambiguity: it is only well-defined up to gauge transformations of the form
\begin{equation}
R(a,b)\rightarrow R(a,b)\frac{e^{i\nu(a,b)}}{e^{i\nu(b,a)}}
\label{gaugetransR}
\end{equation} 
where $\nu(a,b) \in \mathbb{R}$. As with $F$, this ambiguity comes about because the phases of splitting operators can be varied arbitrarily. In particular, if we multiply the two splitting operators in Fig.~\ref{fig:R-abstract} by two phases, $e^{i\nu(a,b)}, e^{i\nu(b,a)}$, this changes $R$ by exactly the above transformation (\ref{gaugetransR}). An important point is that the $\nu(a,b)$ in (\ref{gaugetransR}) is the same as the $\nu(a,b)$ that appears in the gauge transformation (\ref{gaugetrans}): that is, $R$ and $F$ transform under the same gauge transformation $\nu$. 

\begin{figure}[tb]
\centering
\includegraphics[width=1\columnwidth]{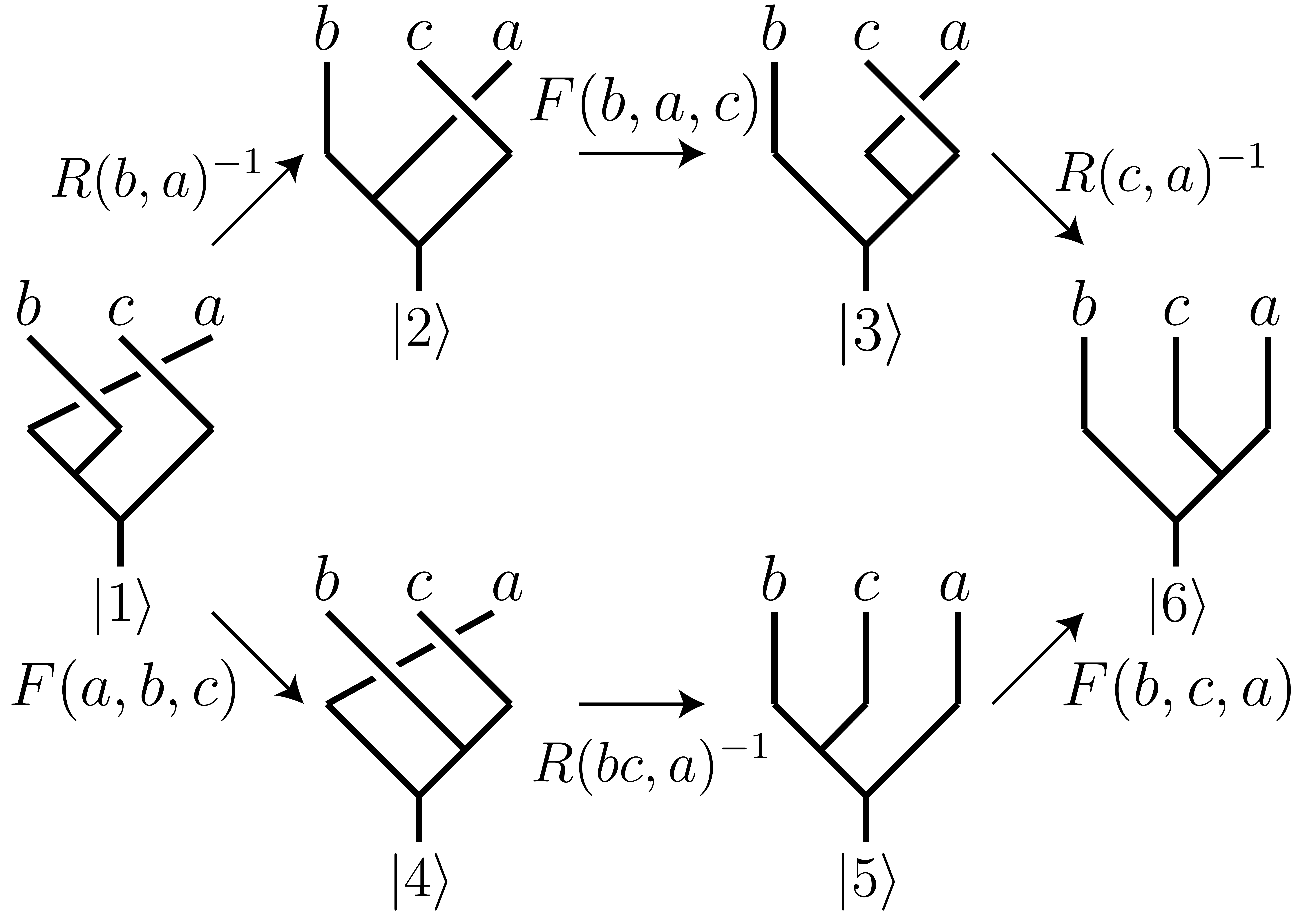}
\caption{The second hexagon equation (\ref{hexeq}). Consistency requires that the product of the $U(1)$ phases along the upper path is equal to the product along the lower path.}
\label{fig:hex2}
\end{figure}

The second property of the $R$-symbol is that it obeys two constraints, known as the hexagon equations:
\begin{align}
R(a,b) F(b,a,c) R(a,c) = F(a,b,c) &R(a,bc) F(b,c,a) \nonumber \\
R(b,a)^{-1} F(b,a,c) R(c,a)^{-1} = F(a,b,c) &R(bc,a)^{-1} \nonumber \\
& \cdot F(b,c,a)
\label{hexeq}
\end{align}
To derive the second equation, consider the $6$ processes shown in Fig.~\ref{fig:hex2}. Denote the final states produced by these processes by $\{|1\>,...,|6\>\}$. We can compute the phase difference between states $|1\>$ and $|6\>$ using either the upper or lower path of Fig.~\ref{fig:hex2}. Demanding consistency between the two calculations gives the second hexagon equation. The first hexagon equation follows in the same way by considering the $6$ processes shown in Fig.~\ref{fig:hex1}. In Appendix~\ref{hexeqapp}, we will show that our microscopic definitions of $F$ and $R$ also obey the hexagon equations.

\subsection{Microscopic definition of $R$}
\label{R-micro-sec}

\begin{figure}[tb]
\centering
\includegraphics[width=1.0\columnwidth]{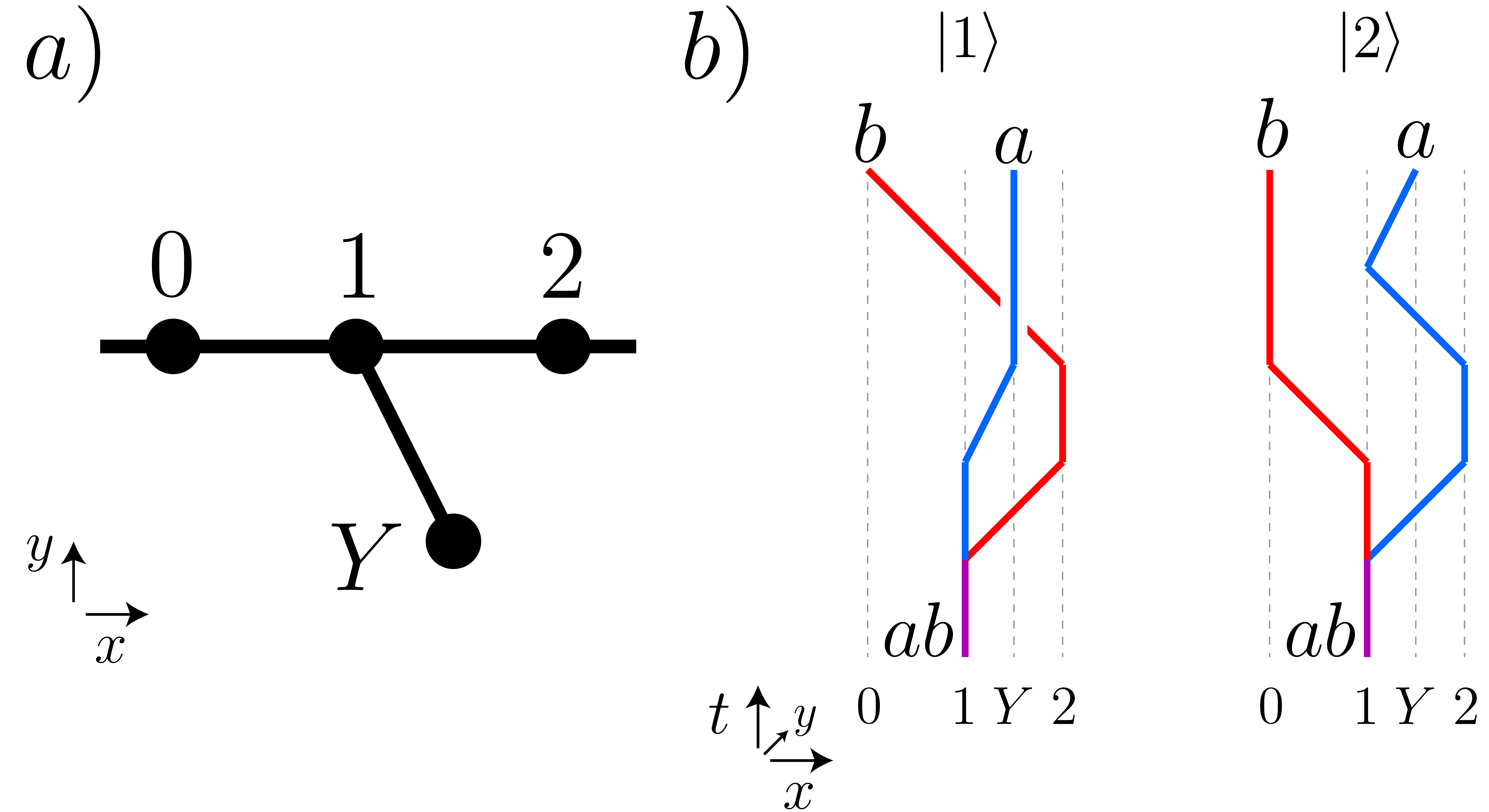}
\caption{(a) To define $R$, we include an additional set of anyon states $\{|a_Y\>\}$ located at a point $Y$ that is not on $x$-axis. We assume that $Y$ has a negative $y$-coordinate and an $x$-coordinate between $1$ and $2$. (b) The two processes that are compared in the microscopic definition of the $R$-symbol. }
\label{fig:R-micro}
\end{figure}

As with the $F$-symbol, the first step in the microscopic definition of the $R$-symbol is to choose a representative anyon state, $|a_x\>$, for each anyon type $a$ and each point $x$. The only new element here is that we need to consider a larger set of positions for $x$: specifically, in addition to the anyon states $\{|a_x\>\}$, where the position $x$ is located on the $x$-axis, we also include the anyon states $|a_Y\>$ for some point, labeled $Y$, that is \emph{not} on the $x$-axis. For concreteness, we will use a convention where the $y$-coordinate of $Y$ is \emph{negative} and larger in magnitude than the correlation length of the ground state, $\xi$, while the $x$-coordinate of $Y$ lies between the two points `$1$' and `$2$' where our splitting operators are defined (Fig.~\ref{fig:R-micro}a).

Moving on to \emph{multi}-anyon states, we use the same notation as in our definition of $F$: we label these states as $|a_{x_1}, b_{x_2}, c_{x_3},...\>$, where $a, b, c,...$ are the different anyons in the state and $x_1,x_2,x_3...$ are their positions. Like before, we assume that $x_1, x_2, x_3,...$ are ordered according to their $x$-coordinate and that every pair of $x_i$'s is separated by a distance of at least $\xi$. 

Likewise, we use the same movement and splitting operators as before, but with one addition: we include two additional movement operators $M^a_{Y1}$ and $M^a_{1Y}$ which move an anyon, $a$, from point $1$ to point $Y$ and vice-versa. Formally, these operators are defined by the condition
\begin{align}
M^a_{Y1} |a_1\> \propto |a_Y\>, \quad \quad M^a_{1Y} |a_Y\> \propto |a_1\>
\end{align}
where both proportionality constants are $U(1)$ phases. Note that we do \emph{not} include any movement operators between $Y$ and any other point on the $x$-axis: all anyon movements occur within the $T$-junction geometry shown in Fig.~\ref{fig:R-micro}a.

With this setup we are now ready to give our definition of $R$. Consider the initial state $|ab_1\>$, i.e. the state with a single anyon of type $ab$ at position $1$. Starting with this state, we apply two different sequences of movement and splitting operators, denoting the final states by $|1\>$ and $|2\>$:
\begin{align}
|1\> &= M^b_{01} M^b_{12} M^a_{Y1} S(a,b) |ab_1\> \nonumber \\
|2\> &= M^a_{Y1} M^a_{12} M^b_{01} S(b,a) |ab_1\>
\label{Rdef1}
\end{align}
These two processes are shown in Figure~\ref{fig:R-micro}b. By construction, the final states $|1\>, |2\>$ produced by these processes both contain anyons $a, b$ at positions $Y, 0$, respectively. In particular, this means that $|1\>, |2\>$ are the same up to a phase. We define $R(a,b)$ to be this phase difference:
\begin{equation}
R(a,b)=\<2|1\>
\label{Rdef2}
\end{equation}
At this point, one can check that our definition of $R(a,a)$ matches the definition from Ref.~\onlinecite{levinwenferm} reviewed in the introduction.

\subsection{Checking the microscopic definition}
\label{Rinvab}
To show that the microscopic definition is sensible, we need to establish two properties of $R$: (i) $R$ is well-defined in the sense that different choices of anyon states and movement and splitting operators give the same $R$ up to a gauge transformation; and (ii) $R$ obeys the hexagon equations (\ref{hexeq}). We prove property (ii) in Appendix~\ref{hexeqapp}; the goal of this section is to prove property (i).

To start, we analyze how $R$ transforms if we change the phase of the movement and splitting operators:
\begin{align}
M^a_{x'x} \rightarrow e^{i\theta_{x'x}(a)} M^a_{x'x}, \quad S(a,b) \rightarrow e^{i\phi(a,b)} S(a,b) 
\end{align}
Substituting these transformations into (\ref{Rdef1}-\ref{Rdef2}) gives
\begin{align}
R(a,b) \rightarrow R(a,b) \frac{e^{i\phi(a,b)} e^{i\theta_{12}(b)}}{e^{i\phi(b,a)}e^{i\theta_{12}(a)}}
\end{align}
We conclude that $R$ changes by a gauge transformation (\ref{gaugetransR}) with
\begin{equation}
\nu(a,b)=\phi(a,b)+\theta_{12}(b)
\label{nuphitheta2}
\end{equation}
Notice that the above expression for $\nu$ is the same as in Eq. (\ref{nuphitheta}). In other words, when we change the phases of the movement and splitting operators, the $F$ and $R$-symbols undergo gauge transformations generated by the \emph{same} $\nu(a,b)$. This is exactly what we want from our microscopic definition, as explained in Sec.~\ref{R-abs-sec}.

Next, consider the more general situation where the movement and splitting operators are changed in an arbitrary way (for fixed choice of anyon states $|a_x\>$). Denoting the new movement and splitting operators by
\begin{align}
M^a_{x'x} \rightarrow \tilde{M}^a_{x'x}, \quad \quad S(a,b) \rightarrow \tilde{S}(a,b),
\end{align}
it follows from the same arguments as in Sec.~\ref{Finvab}, that
\begin{align}
\tilde{M}^a_{x'x} |...,a_x,...\> &= e^{i\theta_{x'x}(a)} M^a_{x'x} |...,a_{x},...\> \nonumber \\
\tilde{S}(a,b)|...,ab_1,...\> &= e^{i\phi(a,b)} S(a,b)|...,ab_1,...\>
\end{align}
for some real-valued $\theta, \phi$. Substituting these relations into (\ref{Rdef1}-\ref{Rdef2}), we again see that $R(a,b)$ changes by a gauge transformation (\ref{gaugetransR}) with $\nu$ given by (\ref{nuphitheta2}).

To complete the proof of property (i), we need to check how $R$ transforms if we choose different representative anyon states $|a_x\>$. As in Sec.~\ref{Finvab}, we will assume that different choices of anyon states are related to one another by a local unitary transformation, $U$. Given this assumption, our task is to understand how $R$ changes if we replace
\begin{align}
|a_{x_1}, b_{x_2}, c_{x_3}, ...\> \rightarrow |a_{x_1}, b_{x_2}, c_{x_3}, ...\>'
\end{align}
where 
\begin{align}
|a_{x_1}, b_{x_2}, c_{x_3}, ...\>' = U |a_{x_1}, b_{x_2}, c_{x_3}, ...\>
\end{align}
for some local unitary transformation $U$. To answer this question, notice that we can choose movement and splitting operators for the states $|a_{x_1}, b_{x_2}, c_{x_3}, ...\>'$ however we like since we have already checked that this choice does not affect $R$, except by a gauge transformation. The simplest choice is
\begin{align}
(M^a_{x'x})' = U M^a_{x'x} U^\dagger, \quad \quad S'(a,b) = U S(a,b) U^\dagger
\end{align}
where $M^a_{x'x}$ and $S(a,b)$ are movement and splitting operators for the states $|a_{x_1}, b_{x_2}, c_{x_3}, ...\>$. With this choice, it is clear that $|1'\> = U|1\>$, and $|2'\> = U|2\>$. It follows that $R' = \<2'|1'\> = \<2|1\> = R$. Thus, we conclude that $R$ is invariant under a replacement of the anyon states, $|a_{x_1}, b_{x_2}, c_{x_3}, ...\> \rightarrow |a_{x_1}, b_{x_2}, c_{x_3}, ...\>'$.

\subsection{Examples}
We now illustrate our microscopic definition by computing the $R$-symbol of several different systems.

\subsubsection{Toric code}

\begin{figure}[tb]
\centering
\includegraphics[width=1.\columnwidth]{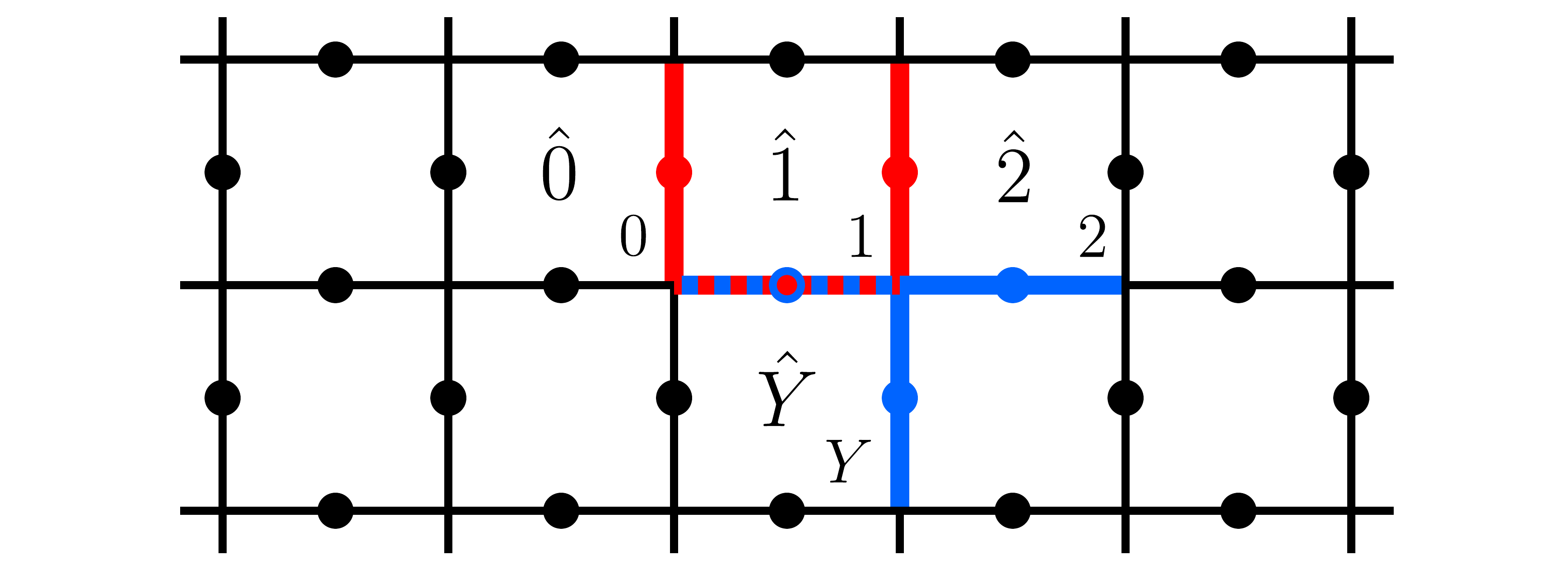}
\caption{$R$-symbol calculation for the toric code model. Vertices and plaquettes are labeled by $n$ and $\hat{n}$ respectively. The $\sigma^z$ operators applied for $M^e$ or $S(e,m)$ are colored in red. The $\sigma^z$ operators applied for $M^m, S(m,e)$ are colored in blue.  }
\label{fig:ToricR}
\end{figure}

We begin by computing the $R$-symbol $R(m,e)$ for the toric code. The first step is to label the anyon states that we will manipulate. For our calculation, we need the states corresponding to the $e$, $m$, and $em$ anyons. In the case of the $e$ anyons, we use the same notation as in Sec.~\ref{tcF}: we label the vertices along the $x$-axis by integers $n = 0, \pm 1, \pm 2,...$ and we define $|e_n\>$ to be the state with $A_{v=n} = -1$. Similarly, for the $m$ anyons, we label the plaquettes that lie just above the $x$-axis by $\hat{n}=\hat{0}, \pm \hat{1}, \pm \hat{2},...$ and we define $|m_{\hat{n}}\>$ to be the state with $B_{p= \hat{n}} = -1$ (Fig.~\ref{fig:ToricR}). Likewise, for the $em$ anyons, we define $|em_n\>$ to be the state with $A_{v=n} = B_{p = \hat{n}} = -1$. Lastly, we will need two more anyon states, which we denote by $|e_Y\>$ and $|m_{\hat{Y}}\>$. Here, $Y$ denotes a vertex, and $\hat{Y}$ denotes an adjacent plaquette, both of which are just below the $x$-axis (Fig.~\ref{fig:ToricR}).

Next, we need to choose the movement operators for the two types of anyons. For the $e$ anyons, we use the same movement operators as in Sec.~\ref{tcF}: we define $M^e_{n'n}= M^e_{nn'} = \sigma^z_{n,n'}$ where $\sigma^z_{n,n'}$ denotes the $\sigma^z$ operator on the link between adjacent vertices $n$ and $n'$. Likewise, for the $m$ anyons, we define
\beq
M^m_{\hat{n}\hat{n}'}= M^m_{\hat{n}'\hat{n}} = \sigma^x_{\hat{n},\hat{n}'}
\eeq
where $\sigma^x_{\hat{n},\hat{n}'}$ denotes the $\sigma^x$ operator on the link between adjacent plaquettes $\hat{n}$ and $\hat{n}'$.

Finally we need the splitting operators $S(e,m)$ and $S(m,e)$, which we define by
\begin{align}
S(e,m) &= \sigma^x_{\hat{1},\hat{2}} \nonumber \\
S(m,e) &= \sigma^z_{1,2} 
\end{align}

With this setup, we are now ready to compute $R(m,e)$. Using the definition (\ref{Rdef1}), we have 
\begin{align}
|1\> &= M^e_{01}  M^e_{12} M^m_{Y1}S(m,e) |em_1\> \nonumber \\
&= \sigma^z_{0,1}\sigma^z_{1,2}\sigma^x_{\hat{1},\hat{Y}}\sigma^z_{1,2}|em_1\>\nonumber\\
&=\sigma^z_{0,1}\sigma^x_{\hat{1},\hat{Y}}|em_1\>
\label{ToricRme1}
\end{align}
and 
\begin{align}
|2\> &= M^m_{Y1} M^m_{12} M^e_{01} S(e,m)|em_1\>\nonumber\\
&= \sigma^x_{\hat{1},\hat{Y}}\sigma^x_{\hat{1},\hat{2}}\sigma^z_{0,1}\sigma^x_{\hat{1},\hat{2}}|em_1\>\nonumber\\
&=\sigma^x_{\hat{1},\hat{Y}} \sigma^z_{0,1}|em_1\>
\label{ToricRme2}
\end{align}
To compare the two states, we note that the operators $\sigma^z_{0,1}$ and $\sigma^x_{\hat{1},\hat{Y}}$ act on the same link and therefore anticommute with one another. We conclude that $|1\> = -|2\>$; so
\beq
R(m,e)=-1
\eeq

Repeating the calculation for $R(e,m)$, one finds that $R(e,m) = +1$ due to the absence of anticommuting operators. More generally, with a suitable choice of conventions, it is easy to check that $R(a,b) = 1$ for all $a,b \in \{1,e,m,em\}$ except for
\begin{align*}
R(m,e) = R(em,e) = R(m,em) = R(em,em) = -1
\end{align*}

\subsubsection{Chiral boson edge theory}

\begin{figure}[tb]
\centering
\includegraphics[width=0.6\columnwidth]{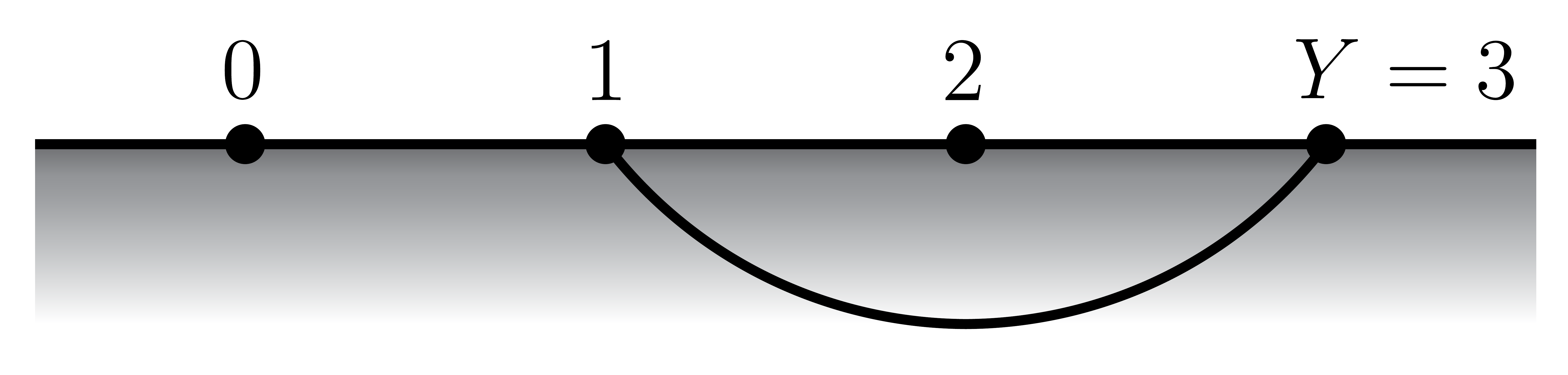}
\caption{To compute $R$ from a chiral boson edge theory, we choose $Y$ to be the point $3$ located on the edge of the topological phase. This choice is acceptable because we can think of the movement operator $M_{Y1}$ as being supported along an arc which only touches the edge at points $1$ and $3$.}
\label{fig:CEBR}
\end{figure}

Next, we compute the $R$-symbol for a general bosonic Abelian topological phase described by an $N \times N$ $K$-matrix, $K_{ij}$. As in Sec.~\ref{chibosonF}, our calculation is based on the chiral boson edge theory (\ref{phicommgen}).

The first step is to specify our conventions for anyon states and movement and splitting operators. For this step, we use exactly the same definitions as in Sec.~\ref{chibosonF}. The only new element of this calculation is that we need to define an additional set of anyon states in which the anyons are located at the special point $Y$ shown in Fig.~\ref{fig:R-micro}a. There is some subtlety in choosing the point $Y$: it is tempting to choose $Y$ to be in the \emph{bulk}, as suggested by the geometry in Fig.~\ref{fig:R-micro}a, but such a choice is not convenient since our calculation is based on an edge theory. Instead, we take $Y$ to be the point $3$ on the edge\footnote{The point $Y=3$ should not be confused with the point `$3$' that appears in the definition of the $F$-symbol: in that context it should be thought of as the point `1000' -- i.e. far from the region where anyons are manipulated.} shown in Fig.~\ref{fig:CEBR}. With this choice the movement operator $M^l_{Y1}$ is simply the operator $M^l_{31}$ defined in Eq.~\ref{Mdefchi}. This movement operator can be thought of as being supported along an arc in the bulk which meets the edge only at points $1$ and $3$ (Fig.~\ref{fig:CEBR}).

Having fixed our conventions, we are now ready to compute $R(l,m)$. Using the definition (\ref{Rdef1}) together with the expression for the splitting operator (\ref{Sdefchi}), we obtain:
\begin{align}
|1\> &= M^m_{01} M^m_{12} M^l_{31} M^m_{21}C(l,m) |[l+m]_1\> \nonumber \\
|2\> &= M^l_{31} M^l_{12} M^m_{01} M^l_{21}C(m,l) |[l+m]_1\>
\label{KRstates}
\end{align}
To find the phase difference between these two states, we first rearrange the order of the movement operators using the commutation relations (\ref{McommK}) and $M^l_{31}M^m_{21}=e^{i\alpha(l,m)}\cdot M^m_{21}M^l_{31}$: 
\begin{align}
|1\> &= M^l_{31}M^m_{01} C(l,m) |[l+m]_1\> e^{2i\alpha(l,m)}\nonumber \\
|2\> &= M^l_{31} M^m_{01} C(m,l) |[l+m]_1\>e^{i\alpha(l,m)}
\label{KRfinalstates}
\end{align}
where $\alpha(l,m)$ is given by definition (\ref{alpha}).  Then using the fact that $C(l,m)=C(m,l)$ we obtain $|1\> = e^{i \alpha(l,m)} |2\>$ so that
\beq
R(l,m)= e^{i\alpha(l,m)}
\eeq
Making the gauge transformation (\ref{gaugetransR}) with $\nu$ given as in equation (\ref{gaugetransK}), we obtain
\beq
R(l,m)=\exp[i\pi l^TK^{-1}(K-X)K^{-1}m]
\eeq
Together with the $F$-symbol in Eq. (\ref{FsymK}), this gives the complete anyon data for the bosonic Abelian topological phase with $K$-matrix $K_{ij}$. 

We can check these results in two ways. First, it is easy to verify that $F$ and $R$ obey the pentagon identity (\ref{pentid}) and hexagon equations (\ref{hexeq}). Second, we can compute the two gauge invariant quantities
\begin{align}
R(l,l)&=\exp[i\pi l^TK^{-1}l]\nonumber\\
R(l,m)R(m,l)&=\exp[2\pi i l^TK^{-1}m]
\end{align}
which describe the exchange statistics and mutual statistics of the anyons. These expressions agree with standard $K$-matrix theory~\cite{wenreview}.

\section{Non-Abelian anyons}
In this section, we explain how our microscopic definitions of $F$ and $R$ generalize to non-Abelian anyons. 
\label{sec:non-abelian}

\subsection{Review: non-Abelian anyon data}\label{sec:nonabel_review}
We begin by reviewing some basic aspects of non-Abelian anyon theories~\cite{kitaevlongpaper, preskilltqc}. These theories consist of four pieces of data:
\begin{enumerate}
\item{{\bf Set of anyon types}: a finite set of anyon types,\\$\mathcal{A} = \{a, b, c, ...\}$.}
\item{{\bf Fusion product}: an associative and commutative multiplication law on $\mathcal{A}$ of the form
\begin{align}
a \times b = \sum_c N^{ab}_c c
\end{align}
where $N^{ab}_c$ are non-negative integers called ``fusion multiplicities.''}
\item{{\bf $F$-symbol}: a complex tensor, $(F^{abc}_{def})^{\kappa\lambda}_{\mu\nu}$, where $a,b,..,f \in \mathcal{A}$, while $\mu$ runs over the set $\{1,...,N^{ab}_e\}$ with $\kappa, \lambda, \nu$ indexed similarly (see Fig.~\ref{fig:abs-non-F}).}
\item{{\bf $R$-symbol}: a complex tensor, $(R^{ab}_{c})^\mu_{\nu}$, where $a,b,c \in \mathcal{A}$, while
$\mu,\nu$ run over the set $\{1,...,N^{ab}_c\}$  (see Fig.~\ref{fig:abs-non-R}).}

\end{enumerate}

The definition of ``anyon types'' is the same as in the Abelian case, so we begin by explaining the non-Abelian fusion product. Given any pair of anyons $a, b$, we say that an anyon of type $c$ is a ``fusion product'' of $a$ and $b$ if $c$ can be converted into a pair of anyons $a, b$ by applying a unitary operator supported in a finite region around $c$. We then define the ``fusion multiplicity'' of $c$, denoted $N^{ab}_c$, as follows: $N^{ab}_c = 0$ if $c$ is not a fusion product of $a$ and $b$. Otherwise, $N^{ab}_c =m$ where $m$ is the largest integer such that there exist $m$ orthonormal states, $|a,b ; c; 1\>,...,|a,b ; c; m\>$, consisting of two anyons $a, b$, and satisfying the following properties. First, each state $|a,b; c; \mu\>$ can be obtained from a single anyon of type $c$ by applying a unitary operator supported in a finite region around $c$. Second, the $|a,b; c;\mu\>$ states are ``locally indistinguishable'': that is, they obey
\begin{align}
\<a,b; c; \mu' |O | a,b ; c ; \mu\> = \text{(const.)} \cdot \delta_{\mu \mu'}
\end{align}
for any local operator, $O$. Having defined $N^{ab}_c$, we define the fusion product of $a$ and $b$ by $a \times b = \sum_c N^{ab}_c c$.

All that remains is to explain the meaning of the $F$ and $R$ symbols. We will discuss this in detail below, but the rough picture is as follows: the $F$-symbol can be interpreted as a collection of (unitary) change of basis matrices relating multi-anyon states obtained by splitting anyons in different orders. Likewise, the $R$-symbol is a collection of (unitary) matrices that describe the non-Abelian Berry phases associated with braiding or exchanging anyons. For example, $(R^{aa}_c)^\mu_\nu$ can be interpreted as the unitary matrix associated with exchanging two $a$ particles in the fusion channel $c$. 

\subsection{Abstract definition of $F$}
\label{nonabF-abs}

\begin{figure}[tb]
\centering
\includegraphics[width=.99\columnwidth]{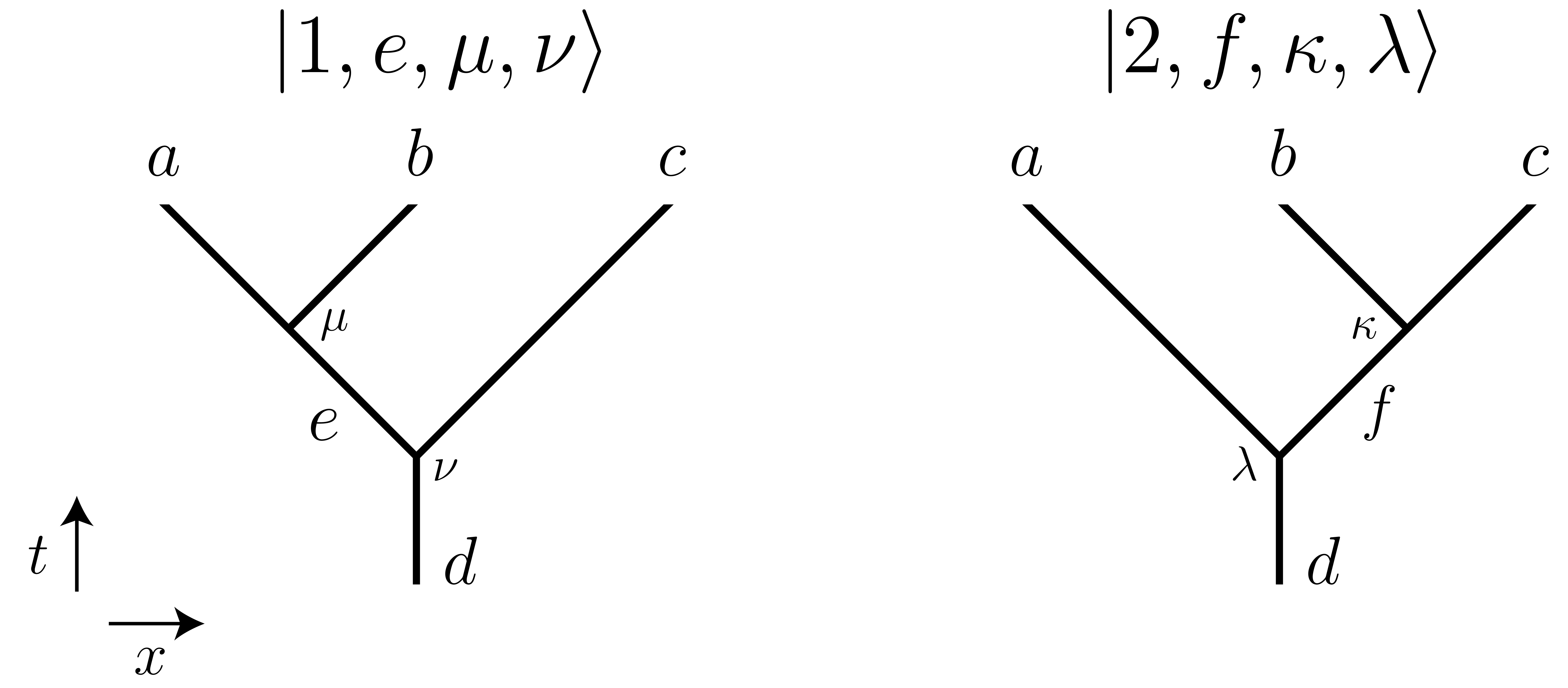}
\caption{Two processes in which an anyon $d$ splits into anyons $a, b, c$. The inner product of the final states $|1,e, \mu, \nu\>, |2,f,\kappa,\lambda\>$ defines the non-Abelian $F$-symbol.}
\label{fig:abs-non-F}
\end{figure}

We now review the abstract definition of the $F$-symbol in the non-Abelian case~\cite{kitaevlongpaper, preskilltqc}. The basic idea is to consider two different physical processes in which an anyon, $d$, splits into three anyons, $a, b, c$ (Fig.~\ref{fig:abs-non-F}). In one process, $d$ splits into anyons $e$ and $c$ and then $e$ splits into $a$ and $b$; in the other process, $d$ splits into $a$ and $f$ and then $f$ splits into $b$ and $c$. In general, there will be more than one way to split the anyons if the fusion multiplicities are larger than $1$. That is, there are multiple ``locally indistinguishable'' final states which can be obtained by each splitting. Therefore, we label the four splittings by additional indices $\mu, \nu, \kappa, \lambda$ where $\mu = 1,...,N^{ab}_e$ and the others are indexed similarly. We then denote the final states of these processes by $|1,e,\mu,\nu\>$ and $|2,f,\kappa,\lambda\>$. By construction, the set of states $\{|1,e,\mu,\nu\>\}$ and $\{|2,f,\kappa,\lambda\>\}$ are both orthonormal bases for the same subspace --- namely the subspace of states consisting of three anyons $a, b, c$ fusing to $d$. It follows that these states must be related by a unitary matrix. The $F$-symbol $(F^{abc}_{def})^{\kappa\lambda}_{\mu\nu}$ is defined to be this unitary matrix:
\begin{equation}
|1,e,\mu,\nu\> = \sum_{f, \kappa, \lambda} (F^{abc}_{def})^{\kappa\lambda}_{\mu\nu} \cdot |2,f,\kappa,\lambda\>
\end{equation}

As in the Abelian case, the above definition implies two properties of the $F$-symbol. The first property is that the $F$-symbol is only well-defined up to gauge transformations. These transformations take the form
\begin{align}
(F^{abc}_{def})^{\kappa\lambda}_{\mu\nu} \rightarrow  &\sum_{\kappa' \lambda' \mu' \nu'} (F^{abc}_{def})^{\kappa'\lambda'}_{\mu'\nu'} \nonumber \\
& \cdot (\alpha^{ec}_d)^{\nu'}_\nu (\alpha^{ab}_e)^{\mu'}_\mu (\alpha^{d}_{af})^\lambda_{\lambda'} (\alpha^{f}_{bc})^\kappa_{\kappa'}
\label{gaugetransnon}
\end{align}
where $(\alpha^{ab}_c)^{\mu'}_{\mu}$ is a family of unitary matrices of dimension $N^{ab}_c \times N^{ab}_c$, parameterized by triplets of anyons, $a, b, c$, and $\alpha^c_{ab} \equiv (\alpha^{ab}_c)^{-1}$ is the inverse matrix. To understand where these gauge transformations come from, it is useful to think of each splitting, $c \rightarrow a,b$, as being implemented by a collection of splitting operators, $S^{ab}_{c,\mu}$ , where $\mu = 1,...,N^{ab}_c$. We are free to mix different splitting operators with one another by multiplying by a unitary matrix, $(\alpha^{ab}_c)^{\mu'}_\mu$:
\begin{align}
S^{ab}_{c,\mu} \rightarrow \sum_{\mu'} (\alpha^{ab}_c)^{\mu'}_\mu S^{ab}_{c,\mu'} 
\end{align}
If we make this replacement for each of the four splittings, then $F$ changes by exactly the above gauge transformation (\ref{gaugetransnon}). 

The second property of the $F$-symbol is that it obeys the pentagon identity. In the non-Abelian case, this identity takes the form
\begin{align}
\sum_{h \lambda' \mu' \rho} (F^{abc}_{gfh})^{\lambda'\mu'}_{\lambda\mu}(F^{ahd}_{egk})^{\rho\sigma}_{\mu'\nu} &(F^{bcd}_{khl})^{\tau\rho'}_{\lambda'\rho}  = \nonumber \\
&\sum_{\kappa}  (F^{fcd}_{egl})^{\tau\kappa}_{\mu\nu} (F^{abl}_{efk})^{\rho'\sigma}_{\lambda\kappa}
\label{pentidnon}
\end{align}
This identity follows from the same reasoning as in the Abelian case.

\subsection{Microscopic definition of $F$}

\begin{figure}[tb]
\centering
\includegraphics[width=.9\columnwidth]{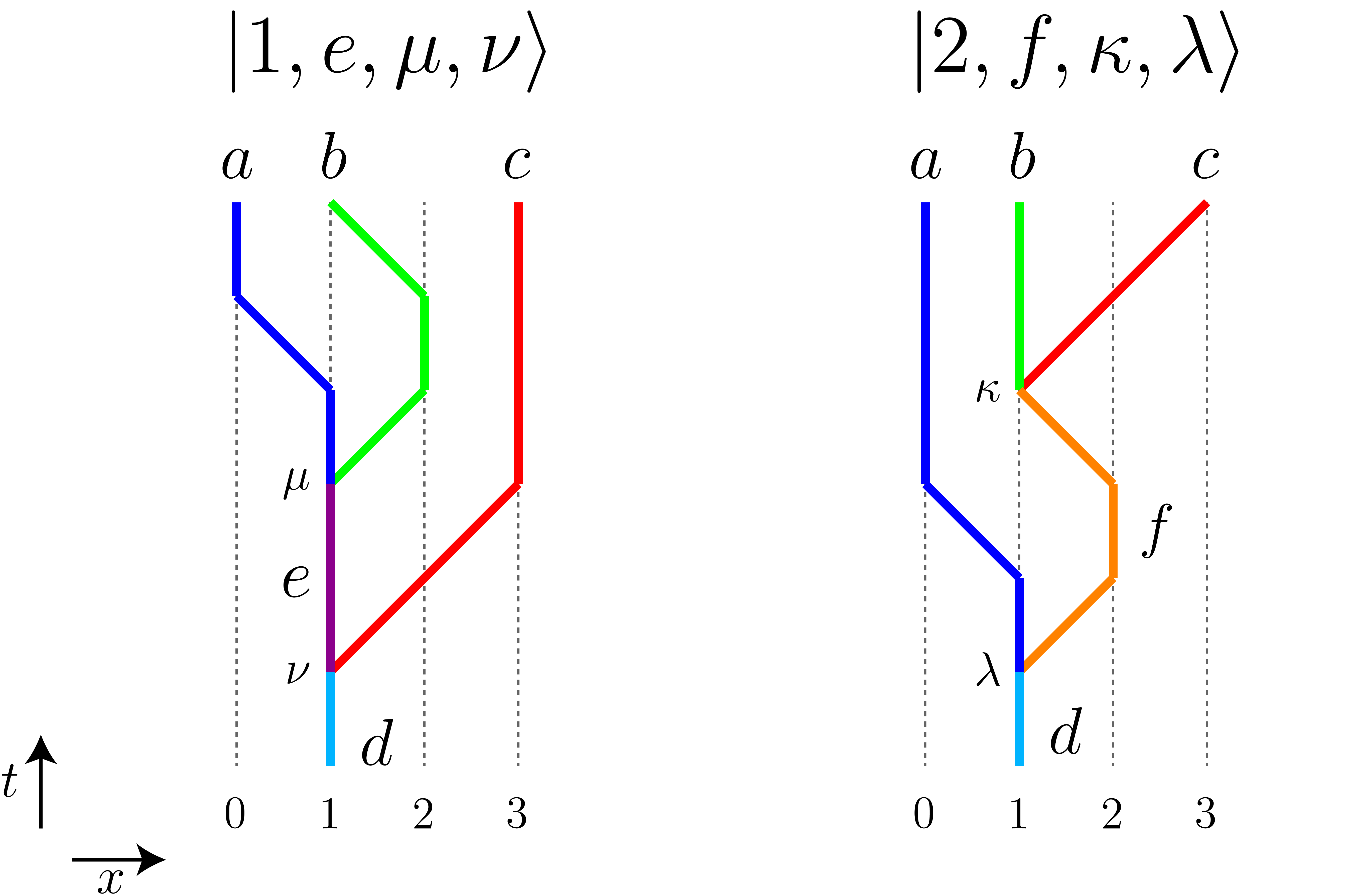}
\caption{The two microscopic processes that define the non-Abelian $F$-symbol. }
\label{fig:two_non}
\end{figure}

The microscopic definition of $F$ follows the same basic steps as is in the Abelian case. As before, we fix a line in the 2D plane and we only consider states with anyons living along this line. For each anyon $a$ and each position $x$, we let $|a_x\>$ denote a state that contains a single anyon $a$ located at point $x$. Likewise, for any finite set of well-separated points $x_1 < x_2 < x_3 < ...$ we let
\begin{align}
|a_{x_1}, b_{x_2}, c_{x_3},...; t; \mu\>, \quad \mu = 1,...,N^{abc...}_t
\label{topdegst}
\end{align}
denote an orthonormal basis of ``locally indistinguishable'' multi-anyon states that contain an anyon $a$ at point $x_1$, anyon $b$ at point $x_2$, etc., and that can be obtained from a single anyon $t$ by applying a unitary operator supported in a finite region around $t$. Here, $N^{abc...}_t$ denotes the fusion multiplicity of $t$ in the fusion product $a \times b \times c \cdots$. Also, when we say that these states are ``locally indistinguishable'' we mean that they satisfy 
\begin{equation}
\<\mu' | O | \mu\> = \text{(const.)} \cdot \delta_{\mu \mu'}
\label{topdeg}
\end{equation}
where $O$ is any local operator, and where we are using the abbreviation
\begin{align}
|\mu\> \equiv |a_{x_1}, b_{x_2}, c_{x_3},...; t; \mu\>
\end{align}

As in the Abelian case, it is useful to give a precise definition of the multi-anyon states in terms of the single anyon states: we define the multi-anyon states to be the set of \emph{all} states that have the same expectation values as the ground state for local operators supported away from all the anyons, and the same expectation values as $|a_{x_1}\>$ for local operators near $x_1$, the same expectation values as $|b_{x_2}\>$ for local operators near $x_2$, and so on. That is, the multi-anyon states are characterized by the fact that 
\begin{align}
\<...,a_x,...;t ; \mu| O | ...,a_x,...;t ; \mu\> = \<a_x |O |a_x\> 
\label{locpropna}
\end{align}
for every operator $O$ supported in the neighborhood of a single anyon $a_x$.

With this setup, we are now ready to define movement operators. For any anyon $a$ and any pair of points $x, x'$, we say that $M^a_{x'x}$ is a ``movement operator'' if it obeys two conditions: first,
\begin{equation}
M^a_{x'x} |a_x\> \propto |a_{x'}\>
\end{equation}
where the proportionality constant is a $U(1)$ phase; second, $M^a_{x'x}$ is supported in a neighborhood of the interval containing $x$ and $x'$. As in the Abelian case, the second condition, together with (\ref{locpropna}), guarantees that $M^a_{x'x}$ has a similar effect on any multi-anyon state of the form $|...,a_x,...\>$ that does not contain any other anyons between $x$ and $x'$:
\begin{equation}
M^a_{x'x} |...,a_x,...; t ; \mu\> = \sum_{\mu'} V^{\mu'}_\mu |...,a_{x'},...; t; \mu'\>
\end{equation}
where $V^{\mu'}_\mu$ is a unitary matrix. 

We now move on to define splitting operators. First, we fix two (well-separated) points on the line, `$1$' and `$2$.' Let $a, b$ be any pair of anyons and let $c$ be any anyon that appears in the product $a \times b$ with multiplicity $N^{ab}_c \geq 1$. We say that a collection of operators $\{S^{ab}_{c,\mu}\}$, where $\mu = 1,..., N^{ab}_c$, are ``splitting operators'' if $S^{ab}_{c,\mu}$ satisfies two conditions: first,
\begin{equation}
S^{ab}_{c,\mu}|c_1\> = \sum_{\mu'} Q^{\mu'}_\mu |a_1, b_2; c; \mu'\>, 
\end{equation}
where $Q^{\mu'}_\mu$ is a unitary matrix; second, $S^{ab}_{c,\mu}$ is supported in the neighborhood of the interval $[1,2]$. As in the Abelian case, the second condition guarantees that the splitting operators have a similar effect when applied to any multi-anyon state of the form $|...,c_1,...\>$, as long as the other anyons are located far from $[1,2]$: that is,
\begin{equation}
S^{ab}_{c,\mu} |...,c_1,...; t ; \nu\> = \sum_{\lambda} W^{\lambda}_{\mu \nu}|...,a_1, b_2, ...; t ; \lambda\> 
\end{equation}
for some complex coefficients, $W^{\lambda}_{\mu \nu}$. These coefficients are guaranteed to have the orthonormality property 
\begin{align}
\sum_\lambda (W^{\lambda}_{\mu \nu})^* W^{\lambda}_{\mu' \nu'} = \delta_{\mu \mu'} \delta_{\nu \nu'}
\end{align}
An equivalent way to say this is that the states $S^{ab}_{c,\mu} |...,c_1,...; t ; \nu\>$ are orthonormal.

For any choice of movement and splitting operators, we define a corresponding $F$-symbol as follows. First, starting with a single anyon $d$ at position $1$, we apply two different sequences of movement and splitting operations:
\begin{align}
|1,e,\mu,\nu\>&= M^b_{12} M^a_{01}S^{ab}_{e,\mu} M^c_{32}S^{ec}_{d,\nu}|d_1\> \nonumber \\
|2,f,\kappa,\lambda\>&=M^c_{32}S^{bc}_{f,\kappa}M^{f}_{12}M^a_{01}S^{af}_{d,\lambda}|d_1\> 
\end{align}
These two processes are shown in Figure \ref{fig:two_non}. Next, we consider the collection of states $\{|1,e,\mu,\nu\>\}$. By construction, all of these states contain anyons $a, b, c$ at positions $0, 1, 3$. In fact, the collection of states $\{|1,e,\mu,\nu\>\}$ forms an orthonormal basis for the topologically degenerate subspace spanned by $\{|a_0, b_1, c_3; d; \sigma\> : \sigma=1,...,N^{abc}_d\}$. The same is true for the second collection of states, $\{|2,f,\kappa,\lambda\>\}$. It follows that the overlaps between these two collections of states form a unitary matrix. The $F$-symbol is defined to be this unitary matrix:
\begin{equation}
(F^{abc}_{def})^{\kappa\lambda}_{\mu\nu}=\<2,f,\kappa,\lambda|1,e,\mu,\nu\>
\label{fdefnon}
\end{equation}

\subsection{Checking the definition of $F$}
As in the Abelian case, we need to check two properties of $F$ to establish that our definition is sensible: (i) different choices of anyon states and movement and splitting operators give the same $F$ up to a gauge transformation; and (ii) $F$ obeys the pentagon identity (\ref{pentidnon}). In this section, we prove property (i). As for property (ii), this follows from arguments that are almost identical to those in the Abelian case, discussed in Appendix~\ref{pentagon}.

To prove property (i), we first consider how $F$ transforms if we change the splitting and movement operators by a transformation of the form
\begin{align}
M^a_{x'x} &\rightarrow e^{i\theta^a_{x'x}} M^a_{x'x}  \nonumber \\
S^{ab}_{c,\mu} &\rightarrow \sum_{\mu'} (\beta^{ab}_c)^{\mu'}_{\mu} S^{ab}_{c, \mu'} 
\label{mstransnon}
\end{align}
where $(\beta^{ab}_c)^{\mu'}_{\mu}$ is an $N^{ab}_c \times N^{ab}_c$ unitary matrix and $\theta^a_{x'x} \in \mathbb{R}$. Substituting (\ref{mstransnon}) into the definition of $F$ (\ref{fdefnon}), it is easy to see that $F$ changes by a gauge transformation with
\begin{equation}
(\alpha^{ab}_c)^{\mu'}_{\mu} = (\beta^{ab}_c)^{\mu'}_{\mu} \cdot e^{i \theta^b_{12}}
\label{alphabetath}
\end{equation}

Next, consider the more general situation where $M$ and $S$ are changed in an arbitrary way (for a fixed choice of single\footnote{Fixing the single anyon states guarantees that the \emph{subspace} of locally indistinguishable multi-anyon states is also fixed.} anyon states $|a_x\>$). Denoting the new movement and splitting operators by
\begin{align}
M^a_{x'x} \rightarrow \tilde{M}^a_{x'x}, \quad \quad S^{ab}_{c,\mu} \rightarrow \tilde{S}^{ab}_{c,\mu},
\end{align}
it follows from the definition that
\begin{align*}
\tilde{M}^a_{x'x}|...,a_x,...; t ; \mu\> &= \sum_{\mu'}T^{\mu'}_{\mu} \cdot M^a_{x'x}|...,a_x,...; t ; \mu'\> \nonumber \\
\tilde{S}^{ab}_{c,\mu}|...,c_1,...; t ; \nu\> &=  \sum_{\mu' \nu'} U^{\mu' \nu'}_{\mu \nu} S^{ab}_{c,\mu'}|...,c_1,...; t ; \nu'\>
\end{align*}
where $T^{\mu'}_\mu$ and $U^{\mu' \nu'}_{\mu \nu}$ are unitary matrices. These matrices are highly constrained: first, the topological degeneracy property (\ref{topdeg}) implies that $T^{\mu'}_{\mu} = T \delta^{\mu'}_\mu$ for some phase factor $T$, and likewise, $U^{\mu' \nu'}_{\mu \nu} = U^{\mu'}_{\mu} \delta^{\nu'}_\nu$ for some unitary matrix $U^{\mu'}_\mu$. Next, since both movement operators are supported in a neighborhood of the interval containing $x$ and $x'$, it follows from property (\ref{locpropna}) that $T$ can only depend on $a, x, x'$. Likewise, $U^{\mu'}_{\mu}$ can only depend on $a, b, c$. Putting this all together, we conclude that
\begin{align*}
\tilde{M}^a_{x'x} |...,a_x,...; t ; \mu\> &= e^{i\theta_{x'x}(a)} M^a_{x'x} |...,a_x,...; t ; \mu\> \nonumber \\
\tilde{S}^{ab}_{c,\mu}|...,c_1,...; t ; \nu\> &=  \sum_{\mu'} (\beta^{ab}_c)^{\mu'}_{\mu} S^{ab}_{c, \mu'} |...,c_1,...; t ; \nu\>
\end{align*}
for some real-valued $\theta_{xx'}(a)$ and some unitary matrix $(\beta^{ab_c})^{\mu'}_\mu$. Substituting these relations into (\ref{fdefnon}), we again see that $F$ changes by a gauge transformation (\ref{gaugetransnon}) with $\alpha$ given by (\ref{alphabetath}).

To complete our proof of property (i), we need to check that $F$ changes at most by a gauge transformation if we choose different anyon states $|a_x\>$. This can established using nearly identical arguments to the Abelian case (see Sec.~\ref{Finvab}).

\subsection{Abstract definition of $R$}

\begin{figure}[tb]
\centering
\includegraphics[width=.5\columnwidth]{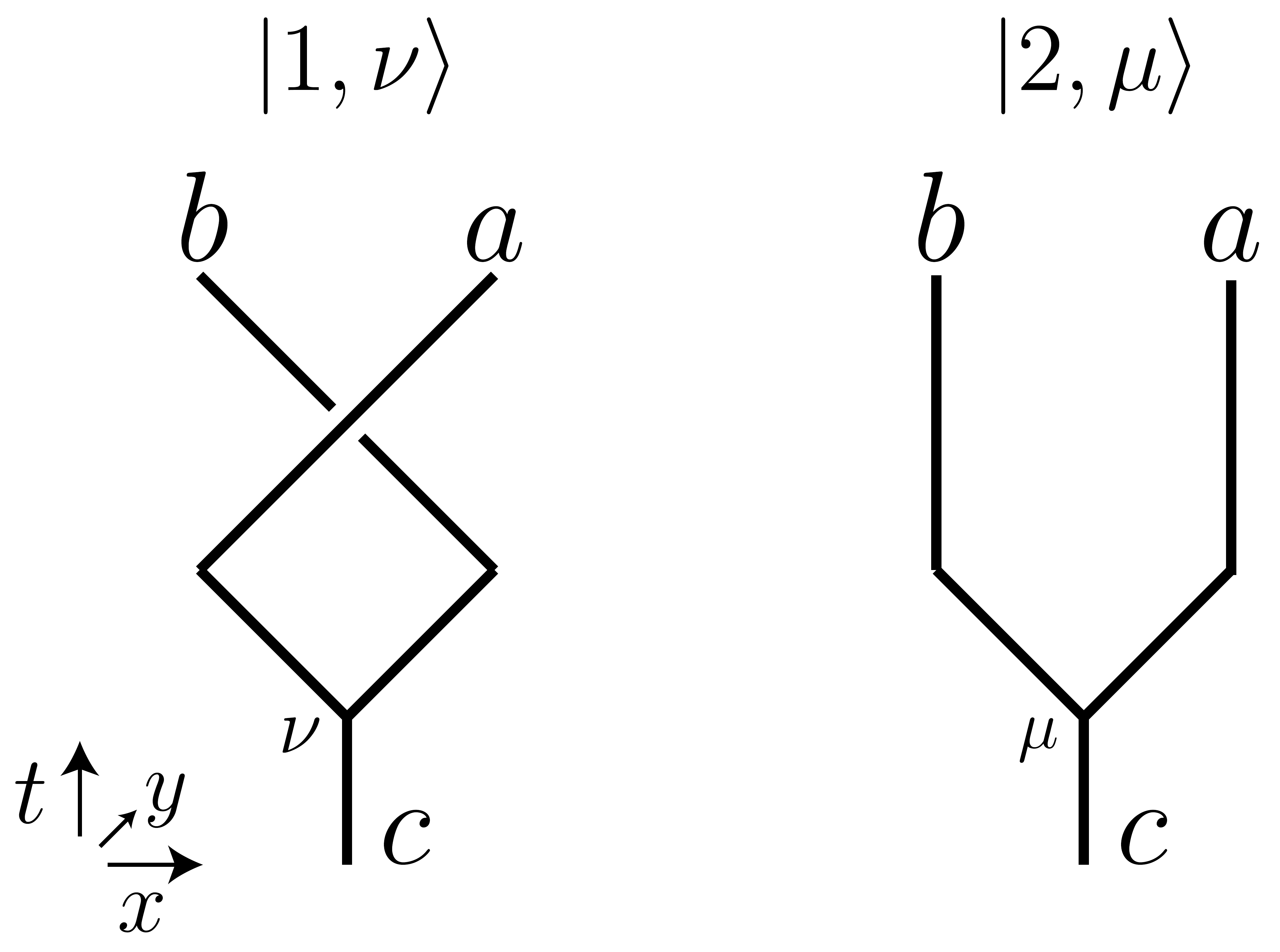}
\caption{Two processes in which an anyon $c$ splits into anyons $a,b$. The inner product of the final states $|1,\nu\>, |2, \mu\>$, defines the non-Abelian $R$-symbol.}
\label{fig:abs-non-R}
\end{figure}

We now explain the abstract definition of $R$ in the case of non-Abelian anyons~\cite{kitaevlongpaper, preskilltqc}. As always, the basic idea is to compare two different physical processes (Fig.~\ref{fig:abs-non-R}). In one process, an anyon, $c$, splits into two anyons $a, b$, with $a$ on the left and $b$ on the right, and then $a, b$ are exchanged in the counterclockwise direction; the other process, $c$ is splits into $a,b$ with $b$ on the left and $a$ on the right, without any subsequent exchange. If the fusion multiplicity $N^{ab}_c$ is larger than $1$, then are multiple ways to split $c$ into $a,b$. Therefore, we label the two splittings by additional indices $\mu, \nu$, which run from $1$ to $N^{ab}_c$. Denoting the two final states by $|1,\nu\>$ and $|2, \mu\>$, it is clear, by the same logic as in Sec.~\ref{nonabF-abs}, that these two sets of states are related by a unitary matrix. The $R$-symbol $(R^{ab}_{c})^\mu_{\nu}$ is defined to be this unitary matrix:
\begin{align}
|1,\nu\> = \sum_{\mu} (R^{ab}_c)^\mu_\nu |2,\mu\>
\label{Rdefnon}
\end{align}

Similarly to the Abelian case, two properties of the $R$-symbol follow from this definition. First, $R$ is only well-defined up to gauge transformations of the form
\begin{equation}
(R^{ab}_c)^\mu_\nu \rightarrow \sum_{\mu' \nu'}(R^{ab}_c)^{\mu'}_{\nu'} (\alpha^{ab}_c)^{\nu'}_{\nu} (\alpha^c_{ba})^\mu_{\mu'} 
\label{gaugetransRnon}
\end{equation} 
where $(\alpha^{ab}_c)^{\mu'}_{\mu}$ is the same family of unitary matrices of dimension $N^{ab}_c \times N^{ab}_c$ that appears in the gauge transformation for $F$ (\ref{gaugetransnon}). (As before, we use the notation $\alpha^c_{ab} \equiv (\alpha^{ab}_c)^{-1}$ to denote the inverse matrix). 

The second property of the $R$-symbol is that it obeys the hexagon equations. In the non-Abelian case, these take the form
\begin{align}
\sum_{\mu' \kappa} (R^{ab}_e)^{\mu'}_{\mu}  &(F^{bac}_{def})^{\kappa \lambda}_{\mu' \lambda'} (R^{ac}_f)^{\kappa'}_\kappa= \nonumber \\
&\sum_{g \sigma \tau \tau'}  (F^{abc}_{deg})^{\sigma \tau}_{\mu \lambda'}(R^{ag}_d)^{\tau'}_{\tau} (F^{bca}_{dgf})^{\kappa' \lambda}_{\sigma \tau'} \nonumber \\
\sum_{\mu' \kappa} (R^{e}_{ba})^{\mu'}_{\mu} &(F^{bac}_{def})^{\kappa \lambda}_{\mu' \lambda'} (R^{f}_{ca})^{\kappa'}_\kappa  = \nonumber \\
&\sum_{g \sigma \tau \tau'}  (F^{abc}_{deg})^{\sigma \tau}_{\mu \lambda'}(R^{d}_{ga})^{\tau'}_{\tau} (F^{bca}_{dgf})^{\kappa' \lambda}_{\sigma \tau'}
\label{hexeqnon}
\end{align}
where we use the notation $R^a_{bc} \equiv (R^{bc}_a)^{-1}$ to denote the inverse matrix. These equations follow from the same reasoning as in the Abelian case. 

\subsection{Microscopic definition of $R$}

\begin{figure}[tb]
\centering
\includegraphics[width=.9\columnwidth]{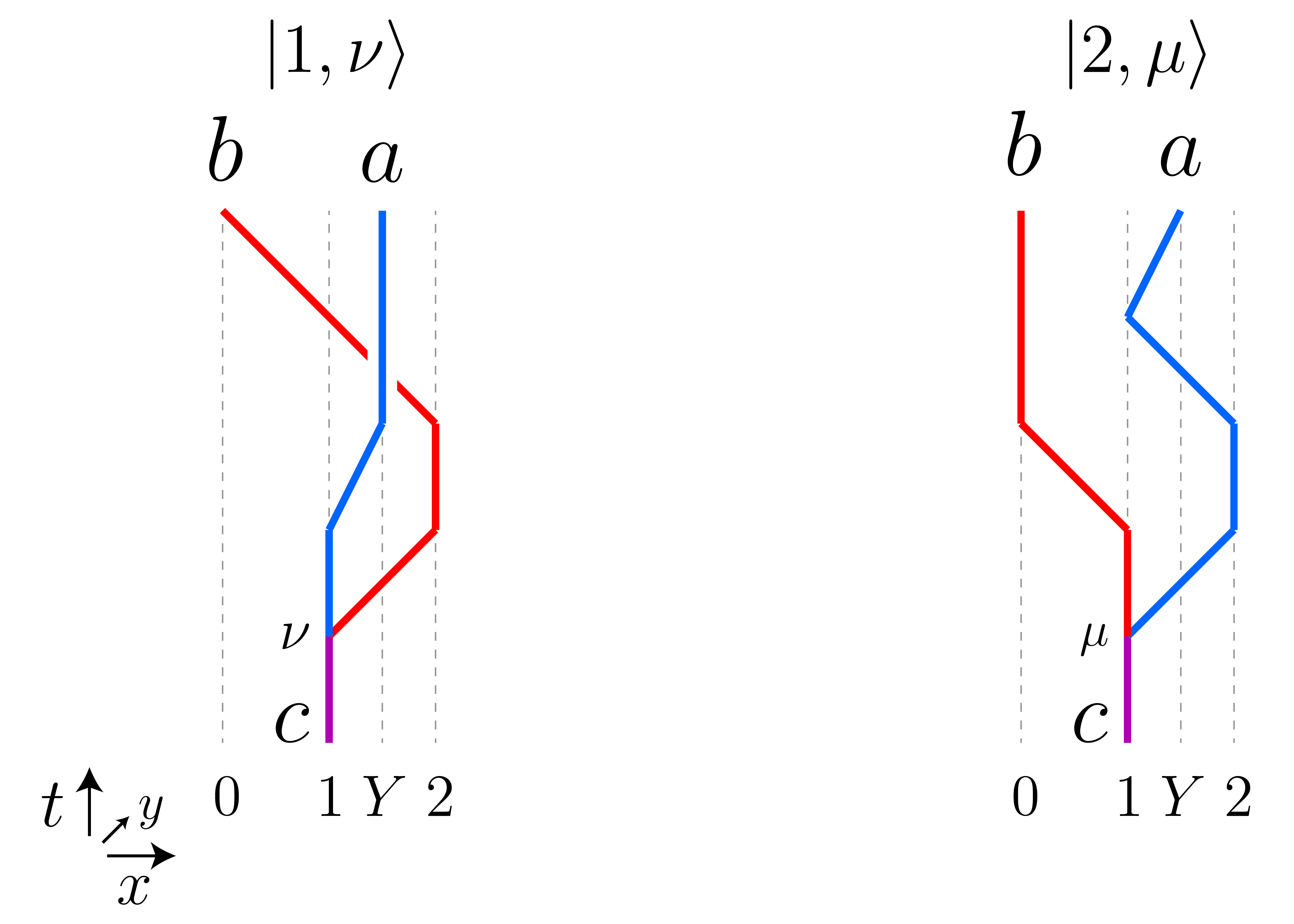}
\caption{The two processes that are compared in the microscopic definition of the non-Abelian $R$-symbol.}
\label{fig:micro-non-R}
\end{figure}

The microscopic definition of the $R$-symbol for non-Abelian anyon theories is similar to the one in the Abelian case. The first step is to choose a representative anyon state $|a_x\>$ for each anyon $a$ and each point $x$. As in the Abelian case, we include both the usual anyon states $|a_x\>$, where the position $x$ is located on the $x$-axis, as well as an additional collection of anyon states $\{|a_Y\>\}$ for some specific point $Y$ that has a negative $y$-coordinate (Fig.~\ref{fig:R-micro}a).

We then use the same notation for multi-anyon states as in Sec.~\ref{R-micro-sec}: we label them as $|a_{x_1},b_{x_2}, c_{x_3},...; t ; \mu\>$ where $a, b, c,...$ are the different anyons in the state and $x_1,x_2,x_3,...$ are their well-separated positions, ordered according to their $x$-coordinate. Also, as in the Abelian case, we introduce two additional movement operators $M^a_{Y1}$ and $M^a_{1Y}$ which move an anyon $a$ from point $1$ to point $Y$ and vice-versa. 

To define $R$, consider the initial state $|c_1\>$, i.e. the state with a single anyon $c$ at position $1$. Starting with this state, we apply two different sequences of movement and splitting operators, denoting the final states by $|1,\nu\>$ and $|2,\mu\>$:
\begin{align}
|1,\nu\> &= M^b_{01} M^b_{12} M^a_{Y1} S^{ab}_{c,\nu} |c_1\> \nonumber \\
|2,\mu\> &= M^a_{Y1} M^a_{12} M^b_{01} S^{ba}_{c,\mu} |c_1\>
\label{Rdefnon1}
\end{align}
These two processes are shown in Figure~\ref{fig:micro-non-R}. By construction, the final states, $|1, \nu\>, |2,\mu\>$, produced by these processes both contain anyons $a, b$ at positions $Y, 0$, and  form an orthonormal basis for the subspace of two anyon states at these positions. It follows that $|1, \nu\>$ and $|2, \mu\>$ are related by a unitary matrix. We define $(R^{ab}_c)^\mu_\nu$ to be this unitary matrix:
\begin{equation}
(R^{ab}_c)^{\mu}_{\nu}=\<2,\mu|1,\nu\>
\label{Rdefnon2}
\end{equation}

\subsection{Checking the definition of $R$}
We need to check two properties of $R$ to show that our definition is sensible: (i) different choices of anyon states and movement and splitting operators give the same $R$ up to a gauge transformation (\ref{gaugetransRnon}); and (ii) $R$ obeys the hexagon equations (\ref{hexeqnon}). In this section, we prove property (i). As for property (ii), this follows from arguments that are almost identical to those in the Abelian case discussed in Appendix~\ref{hexeqapp}.

The first step in proving property (i) is to consider how $R$ transforms if we replace the movement and splitting operators by
\begin{align}
M^a_{x'x} &\rightarrow e^{i\theta^a_{x'x}} M^a_{x'x}  \nonumber \\
S^{ab}_{c,\mu} &\rightarrow \sum_{\mu'} (\beta^{ab}_c)^{\mu'}_{\mu} S^{ab}_{c, \mu'} 
\end{align}
where $(\beta^{ab}_c)^{\mu'}_{\mu}$ is an $N^{ab}_c \times N^{ab}_c$ unitary matrix and $\theta^a_{x'x} \in \mathbb{R}$. Substituting this into (\ref{Rdefnon1}-\ref{Rdefnon2}) we conclude that $R$ changes by a gauge transformation (\ref{gaugetransRnon}) with
\begin{equation}
(\alpha^{ab}_c)^{\mu'}_{\mu} = (\beta^{ab}_c)^{\mu'}_{\mu} \cdot e^{i \theta^b_{12}}
\end{equation}
Importantly, the above expression for $\alpha$ is the same as in Eq. (\ref{alphabetath}): thus, the $F$ and $R$-symbols undergo gauge transformations generated by the \emph{same} $\alpha$, as desired. 

Next, consider the more general situation where the movement and splitting operators are changed in an arbitrary way (again for a fixed choice of single anyon states $|a_x\>$). Denoting the new movement and splitting operators by
\begin{align}
M^a_{x'x} \rightarrow \tilde{M}^a_{x'x}, \quad \quad S^{ab}_{c,\mu} \rightarrow \tilde{S}^{ab}_{c,\mu},
\end{align}
the same arguments as in the previous section imply that
\begin{align*}
\tilde{M}^a_{x'x} |...,a_x,...; t ; \mu\> &= e^{i\theta_{x'x}(a)} M^a_{x'x} |...,a_x,...; t ; \mu\> \nonumber \\
\tilde{S}^{ab}_{c,\mu}|...,c_1,...; t ; \nu\> &=  \sum_{\mu'} (\beta^{ab}_c)^{\mu'}_{\mu} S^{ab}_{c, \mu'} |...,c_1,...; t ; \nu\>
\end{align*}
for some real-valued $\theta_{xx'}(a)$ and some unitary matrix $(\beta^{ab_c})^{\mu'}_\mu$.
Substituting these relations into (\ref{Rdefnon1}-\ref{Rdefnon2}), we again see that $R$ changes by a gauge transformation with $\alpha$ given by (\ref{alphabetath}).

At this point we have almost proved property (i): all that remains is show that different representative anyon states $|a_x\>$ lead to the same $R$, up to a gauge transformation. This result follows from the same argument as in the Abelian case (Sec.~\ref{Rinvab}).



\section{Conclusion}
\label{conclusion}

In this paper, we have proposed microscopic definitions of both the $F$-symbol and the $R$-symbol. These definitions are quite general: they apply to arbitrary bosonic or fermionic many-body systems with local interactions and an energy gap. We have also shown that our definitions obey the pentagon and hexagon equations, thereby providing a microscopic derivation of these fundamental constraints. 

Throughout this paper, we have emphasized that our definitions provide concrete procedures for \emph{computing} anyon data from microscopic models. It is natural to wonder: can we go a step further and turn these procedures into experimental protocols? In other words, do our definitions provide a way to \emph{measure} $F$ and $R$?

The answer to this question is `yes,' at least in principle. For example, consider the Abelian $F$-symbol $F(a,b,c)$. To measure this quantity, one could first prepare an initial state of the form $\frac{1}{\sqrt{2}}|(abc)_1\> \otimes (|\up\> + |\down\>)$, where $|(abc)_1\>$ denotes an anyon state, and $|\up\>, |\down\>$ are (orthogonal) states of an ancilla qubit. Next, one could perform the sequence of movement and splitting operations in process `$1$' (\ref{fdef1}), with each operation conditioned on the ancilla qubit being in the state $|\up\>$, followed by the operations in process `$2$,' each conditioned on the ancilla qubit being in the state $|\down\>$.\footnote{Here, we assume that the movement and splitting operators are chosen so that they are \emph{unitary}.} This sequence of operations produces the final state $\frac{1}{\sqrt{2}}(|1\> \otimes |\up\> + |2\> \otimes |\down\>)$. Finally, by measuring the expectation values $\<\sigma^x\>$ and $\<\sigma^y\>$ for the ancilla qubit, one could extract the real and imaginary parts of $\<2|1\> = F(a,b,c)$. A similar scheme could be used to measure $R$; thus, in principle one can measure the \emph{complete} set of anyon data. These schemes also generalize easily to the non-Abelian case. 

That said, these schemes, both theoretical and experimental, suffer from a serious limitation: they assume a detailed microscopic understanding of anyon excitations. An important problem is to find other methods of extracting a complete set of anyon data that do not require as much information about the underlying many-body system. For example, several authors have proposed methods for computing various pieces of anyon data from a ground state wave function alone\cite{wenmodulartrans, levinwentopent, kitaevpreskilltopent, zhangvishwanath, zhangvishwanath2, zaletelpolarization, qipolarization, Haah2016}; it would be interesting if one could develop methods of this kind for the $F$-symbol or $R$-symbol.

Another interesting direction would be to consider many-body systems with \emph{symmetries}. These systems are characterized by a richer set of data that describes both the topological properties of anyon excitations as well as how they transform under the symmetries~\cite{barkeshliSET, teofradkinSET, tarantinoSET, lanwenSET}. A natural extension of this work would be to find microscopic definitions and measurement schemes for this ``symmetry-enriched''  data.


\acknowledgments

We thank R. Mong for useful discussions. K.K. and M.L. acknowledge the support of the Kadanoff Center for Theoretical Physics at the University of Chicago.
This research was supported in part by NSF DMR-1254741 as well as the Simons Foundation through the ``Ultra-Quantum Matter" Simons Collaboration.


\begin{appendix}


\section{Pentagon identity} \label{pentagon}

In this appendix, we show that the Abelian $F$-symbol, defined in Eq.~\ref{fdef2}, satisfies the pentagon identity. Here, the main reason we focus on Abelian anyons is to simplify the presentation: the generalization to the non-Abelian case is straightforward.

The first step in the proof is to pick a nice phase convention for the movement operators $M^a_{x'x}$. Specifically, we choose the phases of the movement operators so that
\begin{align}
M^a_{xx'}M^a_{x'x}|a_x\> &= |a_x\> \nonumber \\
M^a_{x''x'}M^a_{x'x}|a_x\> &= M^a_{x''x}|a_{x}\>
\label{mid}
\end{align}
With this phase convention, our space-time diagrams satisfy the following ``topological invariance'' property: consider any process, $P$, composed out of a sequence of movement and splitting operators acting on an initial state 
\begin{align} 
|i\> = |..., a_x, b_{x'}, c_{x''},...\>. 
\end{align}
For any such process, we can draw a corresponding space-time diagram. Next, consider a second process, $P'$, that acts on the same initial state, $|i\>$, and that leads to the same final configuration of anyons. Again, we can draw a corresponding space-time diagram. The topological invariance property says that if these two space-time diagrams can be continuously deformed into one another while fixing the endpoints, then the two processes produce the same final states with the same phases. That is, 
\begin{align}
P |i\> = P' |i\>
\label{topinvprop}
\end{align}
We prove this result at the end of this section.

\begin{figure}[tb]
\centering
\includegraphics[width=1\columnwidth]{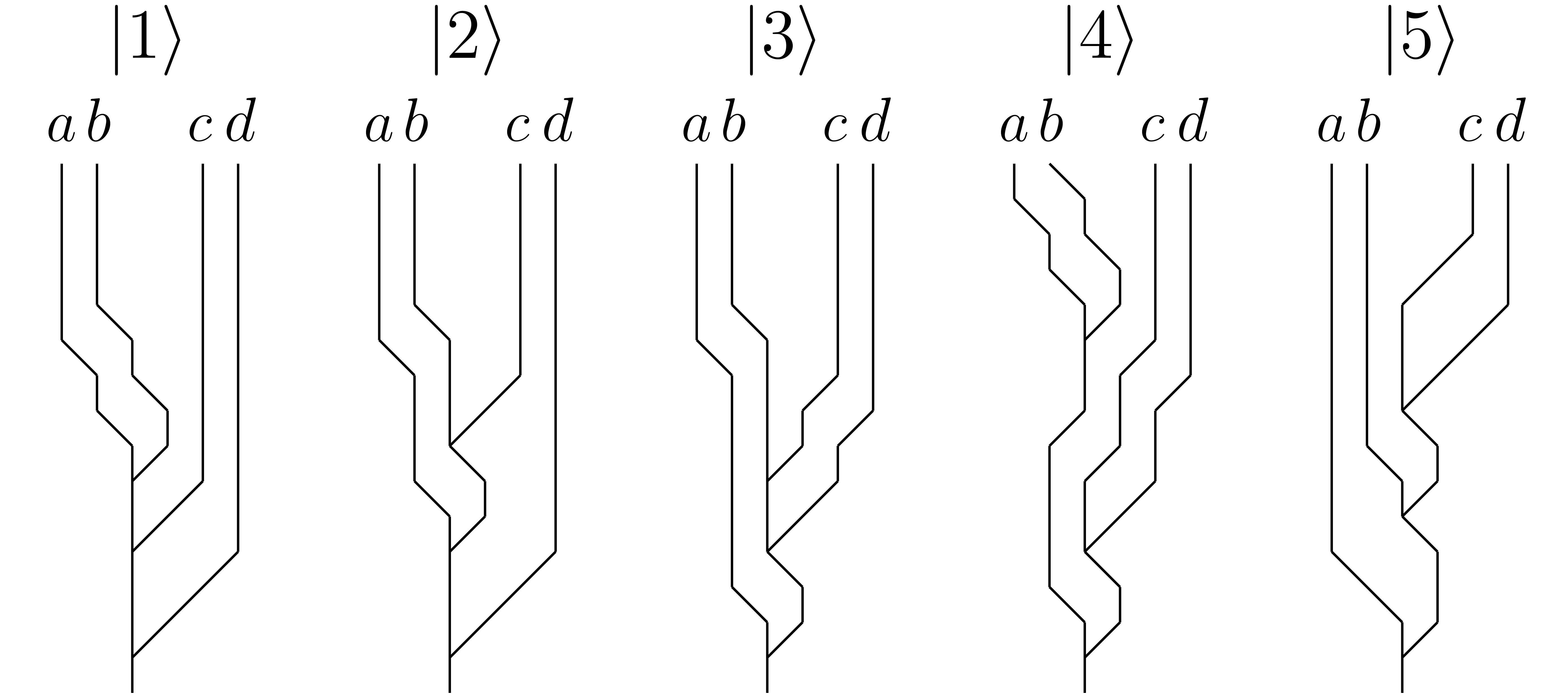}
\caption{Five microscopic processes used to derive the pentagon identity.}
\label{fig:Pentagon_states}
\end{figure}

\begin{figure}[tb]
\centering
\includegraphics[width=0.9\columnwidth]{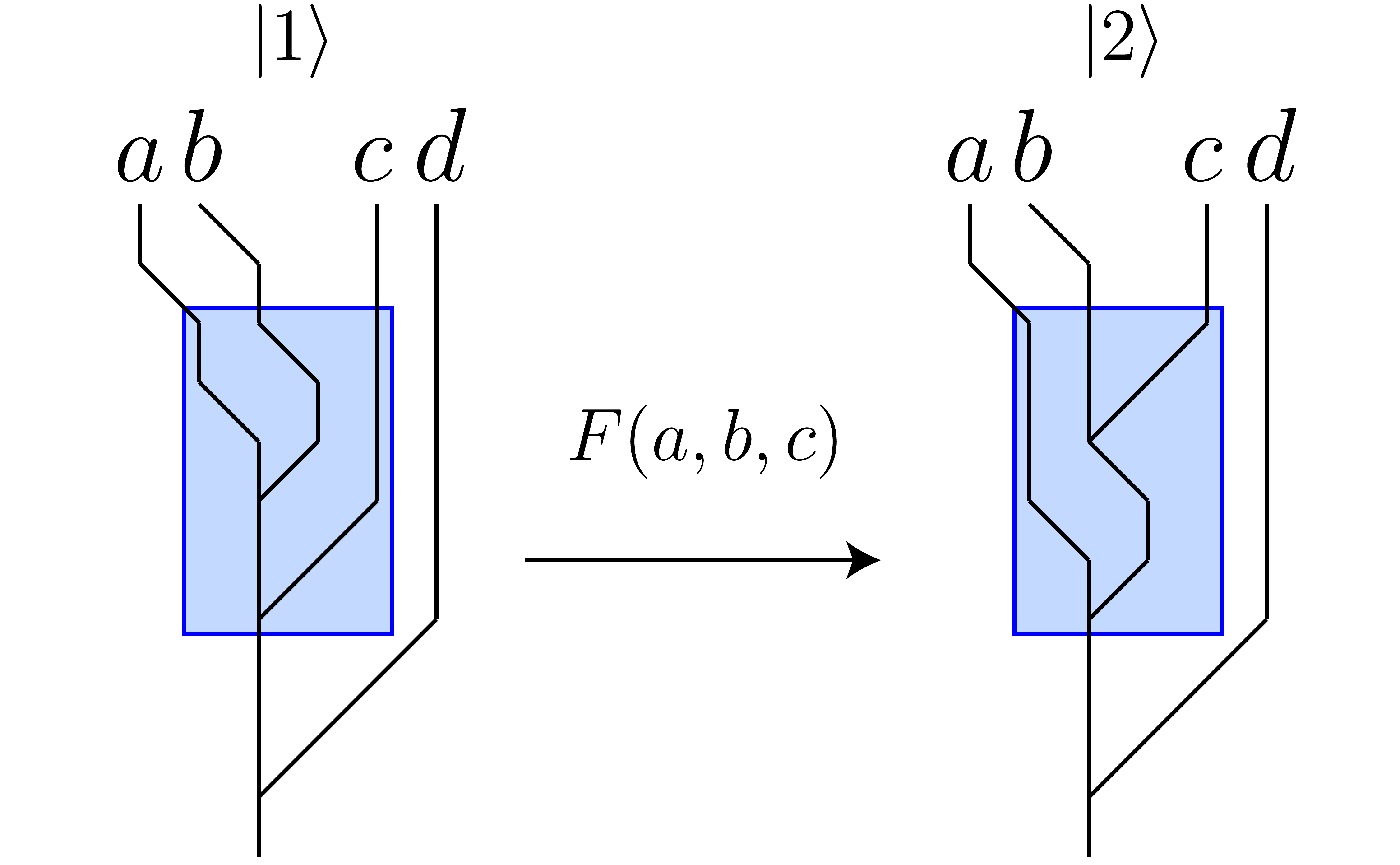}
\caption{Graphical proof of $|1\rangle=F(a,b,c)|2\rangle$: the two processes are related by a microscopic $F$-move.}
\label{fig:pent_link1}
\end{figure}

\begin{figure}[tb]
\centering
\includegraphics[width=0.9\columnwidth]{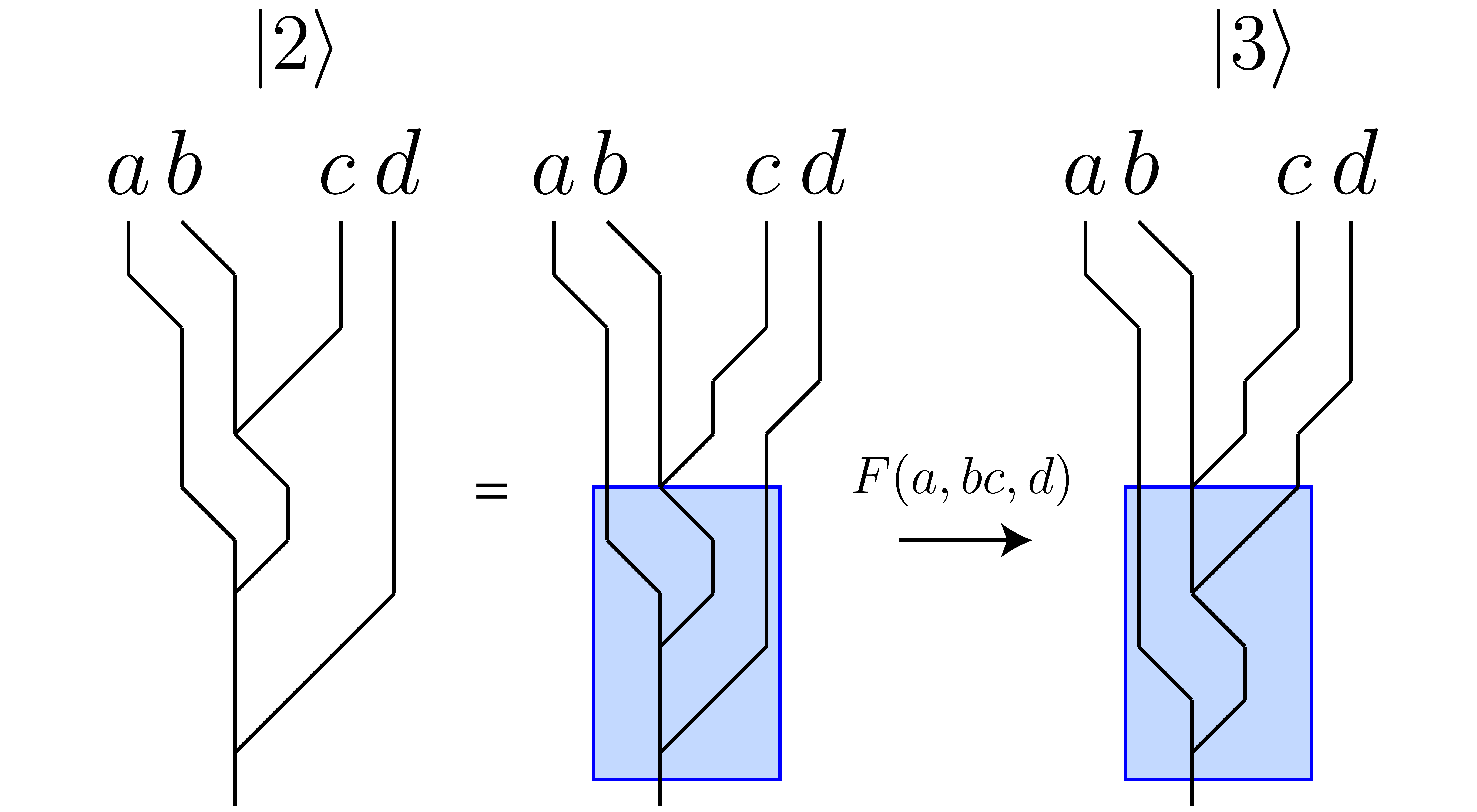}
\caption{Graphical proof of $|2\rangle=F(a,bc,d)|3\rangle$. The first equality follows from topological invariance, while the second equality follows from a microscopic $F$-move.}
\label{fig:pent_link2}
\end{figure}

\begin{figure}[tb]
\centering
\includegraphics[width=0.9\columnwidth]{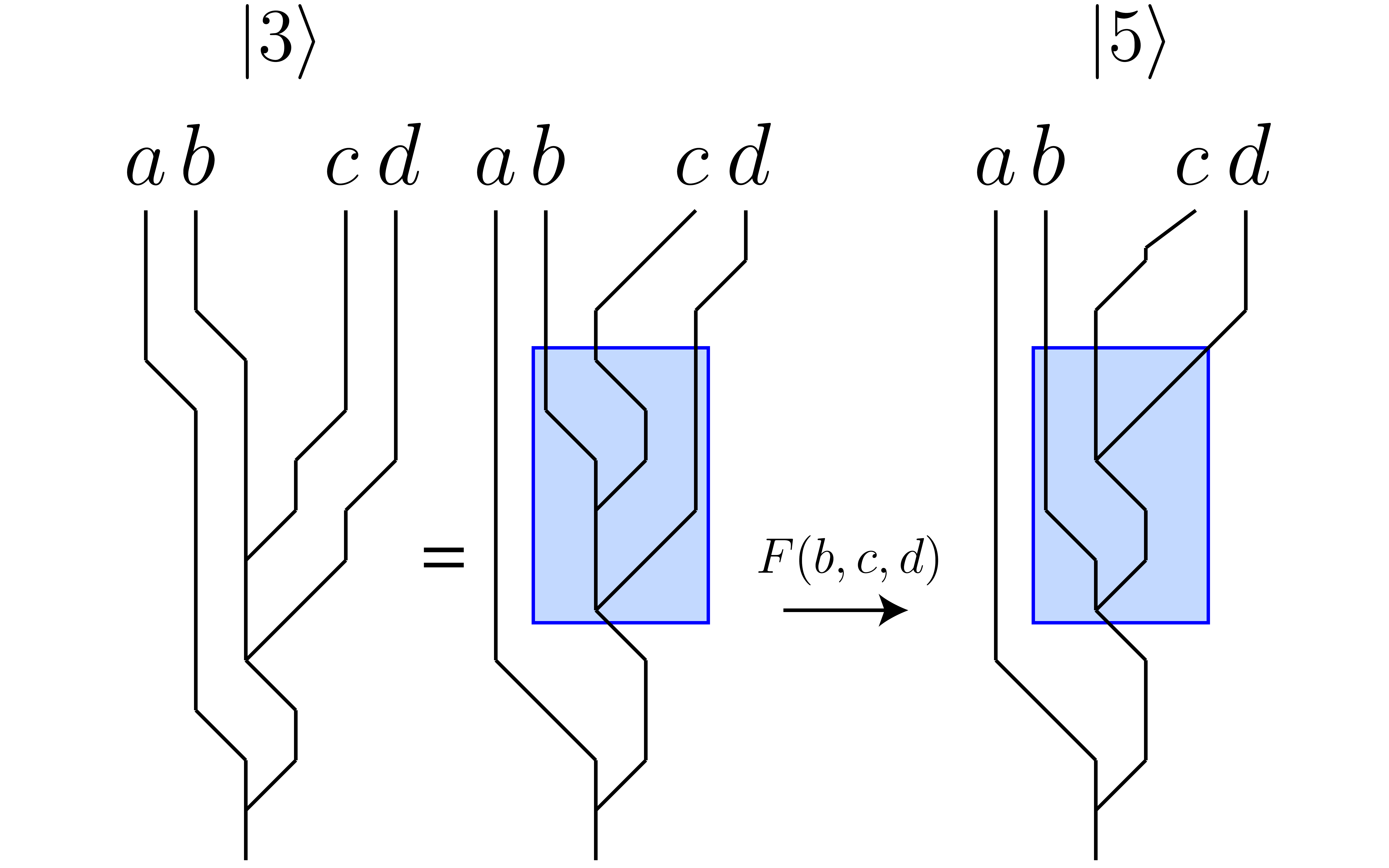}
\caption{Graphical proof of $|3\rangle=F(b,c,d)|5\rangle$. The first equality follows from topological invariance, while the second follows from a microscopic $F$-move.}
\label{fig:pent_link3}
\end{figure}

Using the above topological invariance property (\ref{topinvprop}), it is easy to show that the $F$-symbol defined in (\ref{fdef2}) satisfies the pentagon identity. The proof follows the same line of reasoning as the argument in Sec.~\ref{sec:abelian_abstract}. Consider the five processes shown in Fig. \ref{fig:Pentagon_states}. Let us denote the final states of these processes by  $|1\>, |2\>, |3\>, |4\>, |5\>$. Notice that these states are the same up to a phase since they contain the same four anyons, $a, b, c, d$, at the same four points. The idea of the proof is to compute the relative phase between states $|1\>$ and $|5\>$ in two different ways. In the first calculation, we note that
\begin{align}
|1\> &= F(a,b,c) |2\> \nonumber \\
|2\> &= F(a,bc,d) |3\> \nonumber \\
|3\> &= F(b,c,d) |5\>
\label{phase151}
\end{align}
Here, the first relation follows by noting that the diagrams corresponding to $|1\>$ and $|2\>$ differ by a microscopic $F$-move (see Fig. \ref{fig:pent_link1}). Likewise, the second relation follows in two steps (see Fig. \ref{fig:pent_link2}): in the first step, we use topological invariance to redraw the space-time diagram corresponding to $|2\>$, while in the second step we related the modified diagram to $|3\>$ using an $F$-move. The third relation follows in a similar way (Fig. \ref{fig:pent_link3}).

\begin{figure}[tb]
\centering
\includegraphics[width=0.6\columnwidth]{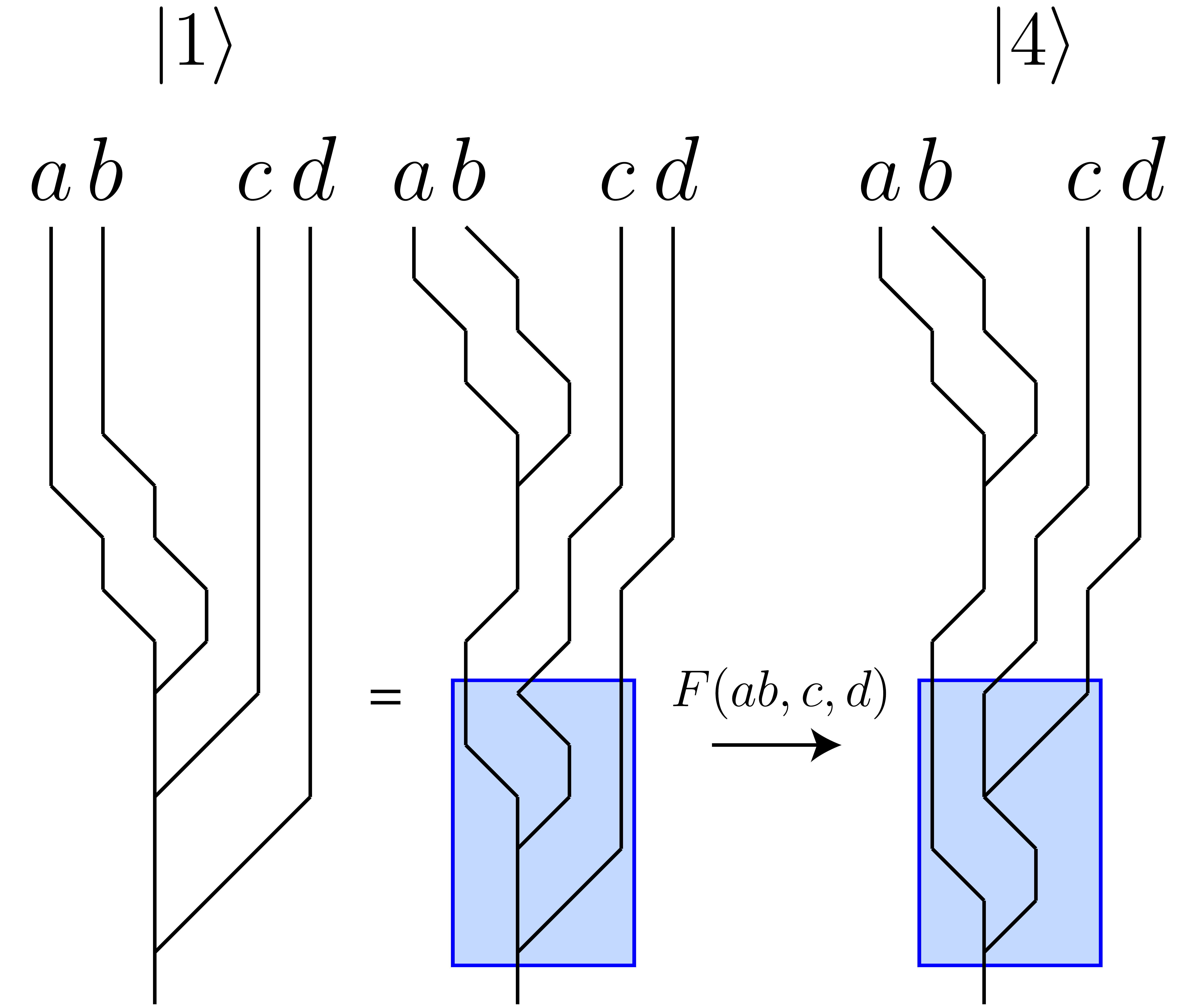}
\caption{Graphical proof of $|1\rangle=F(b,c,d)|4\rangle$. The first equality follows from topological invariance, while the second follows from a microscopic $F$-move.}
\label{fig:pent_link4}
\end{figure}

\begin{figure}[tb]
\centering
\includegraphics[width=0.9\columnwidth]{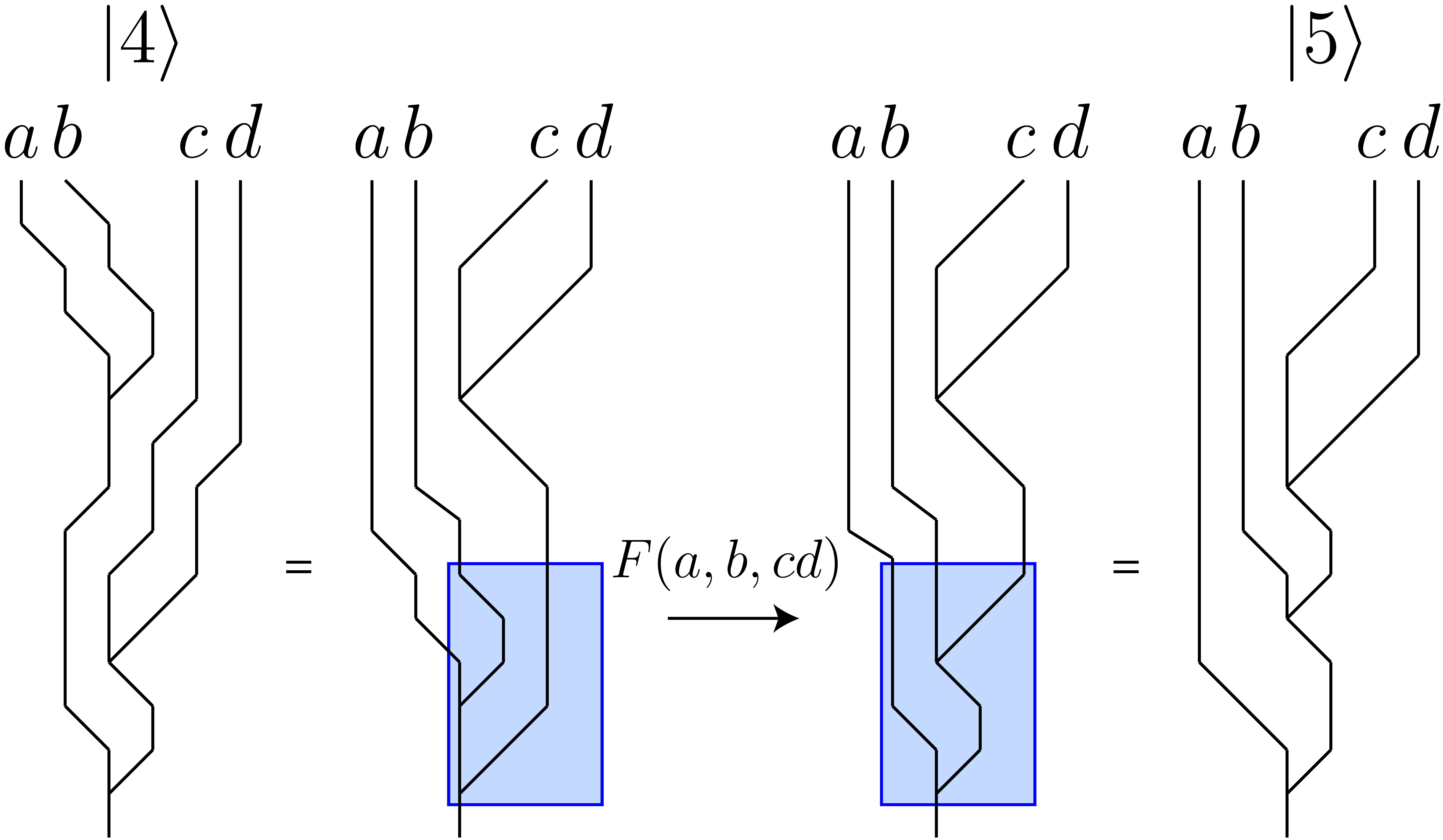}
\caption{Graphical proof of $|4\rangle=F(b,c,d)|5\rangle$. The first and third equalities follow from topological invariance, while the second follows from a microscopic $F$-move.}
\label{fig:pent_link5}
\end{figure}

For the second calculation of the relative phase between $|1\>, |5\>$, we observe that
\begin{align}
|1\> &= F(ab,c,d)|4\> \nonumber \\
|4\> &= F(a,b,cd)|5\>
\label{phase152}
\end{align}
Similarly to (\ref{phase151}), both of these relations follow from the topological invariance property (\ref{topinvprop}) together with the definition of $F$ (\ref{fdef2}). (See Figs. \ref{fig:pent_link4}-\ref{fig:pent_link5} for graphical proofs). Combining (\ref{phase151}) and (\ref{phase152}), gives the desired pentagon identity:
\begin{align*}
F(a,b,c) F(a,bc,d) F(b,c,d) = F(ab,c,d) F(a,b,cd)
\end{align*}

To complete the argument, we now prove the topological invariance property (\ref{topinvprop}). We begin by considering some special cases. First, we consider processes that (i) are composed only out of movement operators and (ii) return all anyons to their initial positions. We claim that any process, $P$, of this kind is equivalent to the identity operator in the sense that
\begin{align}
P |i\> = |i\>
\label{topinv0}
\end{align}
This claim follows from two properties of the movement operators. The first property is that movement operators commute when acting on non-overlapping intervals:
\begin{align}
[M_{x_1 x_2}, M_{x_3 x_4}] = 0
\label{mcom}
\end{align}
The second property is a multi-anyon generalization of (\ref{mid}) which follows from the locality of the movement operators:
\begin{align}
M^a_{xx'}M^a_{x'x}|...,a_x,...\> &= |...,a_x,...\> \label{mid21} \\
M^a_{x''x'}M^a_{x'x}|...,a_x,...\> &= M^a_{x''x}|...,a_{x},...\> \label{mid22}
\end{align}
Here $|...,a_x,...\>$ is any multi-anyon state that does not have any additional anyons in the interval containing $x, x'$ or $x', x''$.

To establish (\ref{topinv0}), it suffices to show that for any nontrivial process $P$, we can find a simpler process $P'$ --- i.e. a process with strictly fewer movement operators --- such that $P|i\> = P'|i\>$.  Once we establish the existence of $P'$, Eq. (\ref{topinv0}) follows immediately since we can then simplify $P$ repeatedly until we reduce it to the identity. To construct $P'$, it is convenient to make two simplifying assumptions about $P$: (i) the positions $x$ of the anyons are always integer-valued and (ii) all movements are of the form $M^a_{x'x}$, where $x' = x \pm 1$. (The second assumption can be made without loss of generality given Eq. \ref{mid22}). Next, consider the leftmost anyon, $a$, with the property that, at some point during the process, $a$ performs a movement to the right, i.e. $M^a_{x' x}$ with $x' > x$, followed later by a reverse movement $M^a_{x x'}$ with no movements of this anyon $a$ in between. Since $a$ is the \emph{leftmost} anyon with this property, it is easy to see that there cannot be any movement operators that appear between $M^a_{x' x}$ and $M^a_{x x'}$ whose region of support overlaps the interval $[x, x']$. Therefore, we can use (\ref{mcom}) to reorder the movement operators so that $M^a_{x' x}$ and $M^a_{x x'}$ appear one after the other. The two movement operators can then be removed from $P$ using the identity (\ref{mid21}). This gives the desired process $P'$. If there are \emph{no} anyons with the above property, then we follow the same reasoning as above but we consider the \emph{rightmost} anyon $a$ that performs a movement to the \emph{left}, i.e. $M^a_{x' x}$ with $x' < x$, followed later by a reverse movement $M^a_{x x'}$ with no movements of this anyon, $a$, in between. Again, this allows us to construct a simpler process $P'$.

Having established (\ref{topinv0}), the next step in the proof is to consider a slightly more general set of processes: namely, processes that are composed out of movement operators but do not necessarily return the anyons to their initial positions. We claim that $P |i\> = P' |i\>$ for any two ``movement-only'' processes $P, P'$ with the same initial state and same final configuration of anyons. To see this, consider the process $P^{inv} P'$, where $P^{inv}$ is the inverse process to $P$ --- that is, the process obtained by reversing the order of the movement operators and replacing $M^a_{x'x} \rightarrow M^a_{xx'}$. By construction, $P^{inv} P'$ has the property that it returns all anyons to their original positions so we can use (\ref{topinv0}) to deduce that $P^{inv} P' |i\> = |i\>$. It then follows that $P |i\> = P' |i\>$, proving the claim.

We are now ready to consider general processes with both splitting and movement operators. We claim that $P |i\> = P' |i\>$ for any two processes $P, P'$ whose spacetime diagrams can be continuously deformed into one another while fixing endpoints and preserving the time-ordered sequence of splittings. To prove this claim, it is useful to decompose $P$ into a product
\begin{align}
P = P_1 S_1 P_2 S_2 \cdots
\end{align}
where $S_1, S_2,...$ are splitting operators and $P_1, P_2,...$ are movement-only processes. We decompose $P'$ in a similar fashion, denoting the splitting operators by $S_1', S_2',...$ and movement-only processes by $P'_1, P'_2,\cdots$. By assumption, $S_j = S_j'$ for every $j$. On the other hand $P_j \neq P_j'$ in general. Consider the special case where $P_j$ and $P'_j$ share the same initial anyon configuration and the same final anyon configuration for each $j$. In this case, it follows immediately that $P |i\> = P'|i\>$ using the above property of movement-only processes. Next, observe that the general case -- where $P_{j}$ and $P'_{j}$ do not share the same initial and final anyon configurations -- can always be reduced to this special case: given any $P, P'$ obeying our assumption, it is not hard to construct modified processes, $\tilde{P}$ and $\tilde{P}'$ with $P|i\> = \tilde{P}|i\>$ and $P' |i\> = \tilde{P}'|i\>$ such that the corresponding movement-only processes, $\tilde{P}_{j}$ and $\tilde{P}'_{j}$, \emph{do} share the same initial and final anyon configuration for each $j$. These modified processes $\tilde{P}$ and $\tilde{P}'$ can be obtained by inserting appropriate movement operators $M^a_{xx'}$ just before each splitting operator, along with their reverse movements $M^a_{x'x}$ just after the splitting operator, and using the fact that movement operators commute with splitting operators when acting on non-overlapping intervals.

\begin{figure}[tb]
\centering
\includegraphics[width=1\columnwidth]{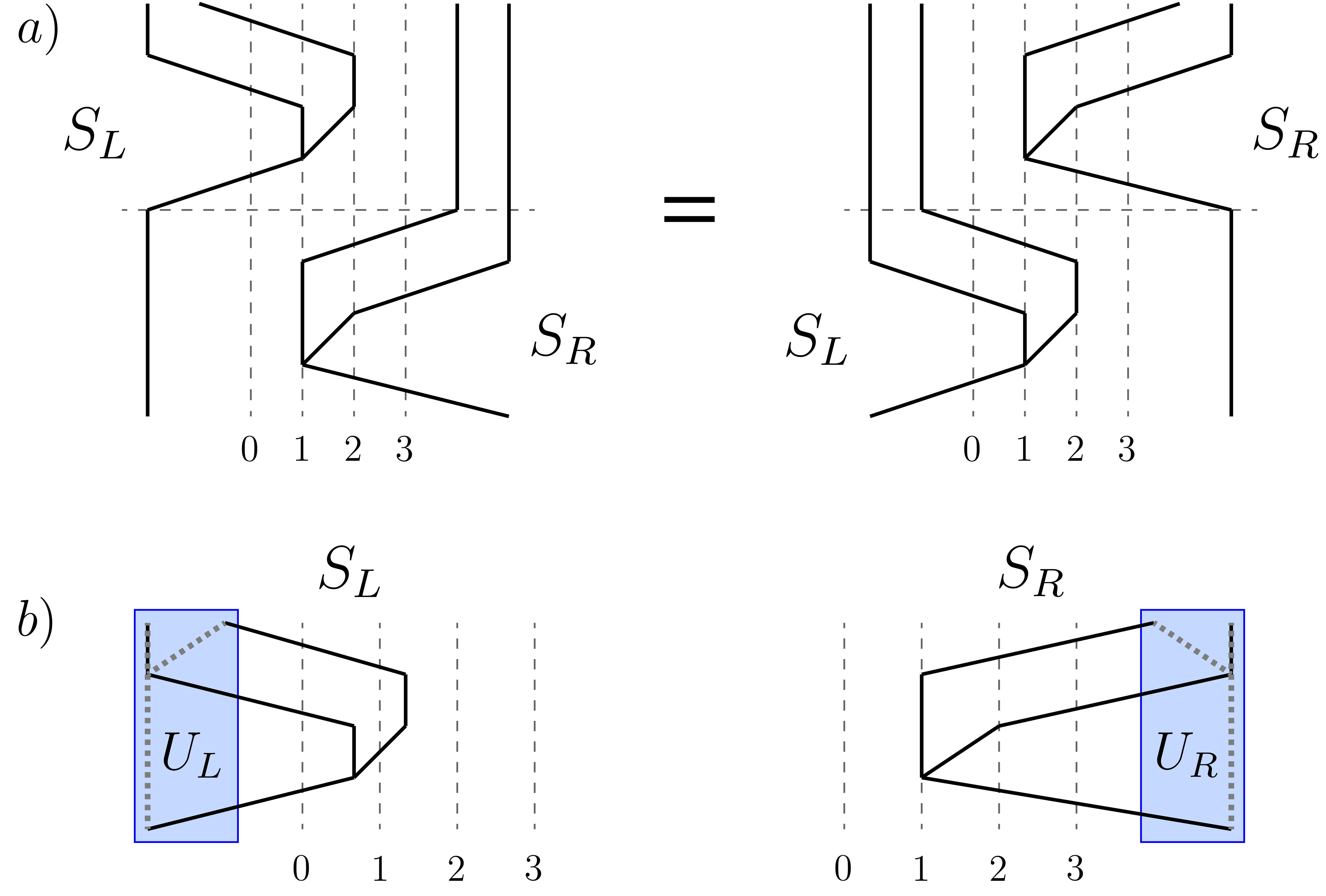}
\caption{(a) Graphical representation of the identity (\ref{scom}). (b) Schematic representation of the two unitary operators $U_L, U_R$ which are supported in the regions $x \leq 0$ and $x \geq 3$, respectively, and satisfy $U_L |i\> = S_L |i\>$ and $U_R |i\> = S_R |i\>$.}
\label{fig:SLSR}
\end{figure}

At this point, we have \emph{almost} proved the general topological invariance property (\ref{topinvprop}), but we need one more result: we need to show that the \emph{order} of splitting operators does not matter, i.e 
\begin{align}
S_L S_R |i\> = S_R S_L |i\>
\label{scom}
\end{align}
where $S_L$ and $S_R$ are the two splitting processes shown in Fig.~\ref{fig:SLSR}a. The identity (\ref{scom}), together with the claim in the previous paragraph implies topological invariance.\footnote{Strictly speaking, to establish general topological invariance, one needs to prove a slight generalization of (\ref{scom}) in which $S_L, S_R$ contain additional ``spectator'' anyon worldlines that do not participate in the splittings; however, this generalization of (\ref{scom}) can be proved using the same arguments as the special case (\ref{scom}).}

To prove the identity (\ref{scom}), we note that neither $|i\>$ nor $S_L |i\>$ contain any anyons in the region $x > -1$, which means in particular that $|i\>$ and $S_L |i\>$ share the same expectation values for any operator supported in $x > 0$. It then follows from the properties of the Schmidt decomposition that there exists a unitary operator, $U_L$, supported in the complementary region $x \leq 0$ such that
\begin{align}
U_L |i\> = S_L |i\> 
\end{align}
By the same reasoning, there exists a unitary operator, $U_R$, supported in the region $x \geq 3$ such that
\begin{align}
U_R |i\> = S_R |i\>
\end{align}
Next, we observe that
\begin{align}
[U_L, U_R] = [S_L, U_R] = [U_L, S_R] = 0
\label{UScom}
\end{align}
since each pair of operators is supported on non-overlapping regions. Combining these identities, we derive
\begin{align}
S_L S_R |i\> = S_L U_R |i\> = U_R S_L |i\> = U_R U_L |i\> = U_L U_R |i\> \nonumber \\
S_R S_L |i\> = S_R U_L |i\> = U_L S_R |i\> = U_L U_R |i\>
\end{align}
This proves the reordering identity (\ref{scom}) and completes our proof of the topological invariance property (\ref{topinvprop}).


\section{Hexagon equations}
\label{hexeqapp}
\begin{figure}[tb]
\centering
\includegraphics[width=1\columnwidth]{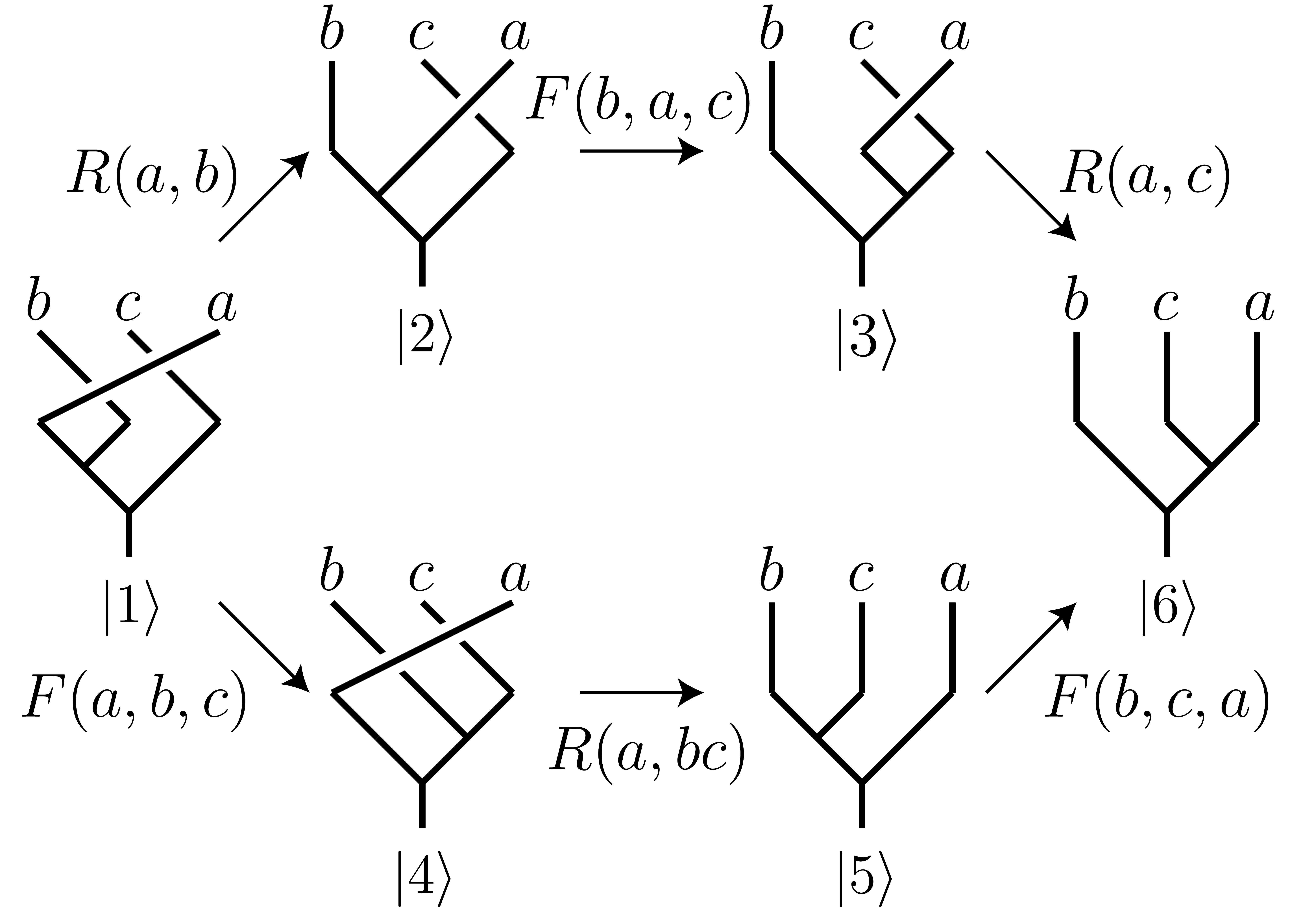}
\caption{Graphical derivation of the first hexagon equation (\ref{hexeq}).}
\label{fig:hex1}
\end{figure}

In this appendix, we show that our definitions of the Abelian $F$ and $R$ symbols obey the hexagon equations (\ref{hexeq}). Our proof closely follows the derivation of the pentagon identity given in Appendix~\ref{pentagon} and like that derivation it is straightforward to generalize it to non-Abelian anyons. 

\begin{figure}[tb]
\centering
\includegraphics[width=1\columnwidth]{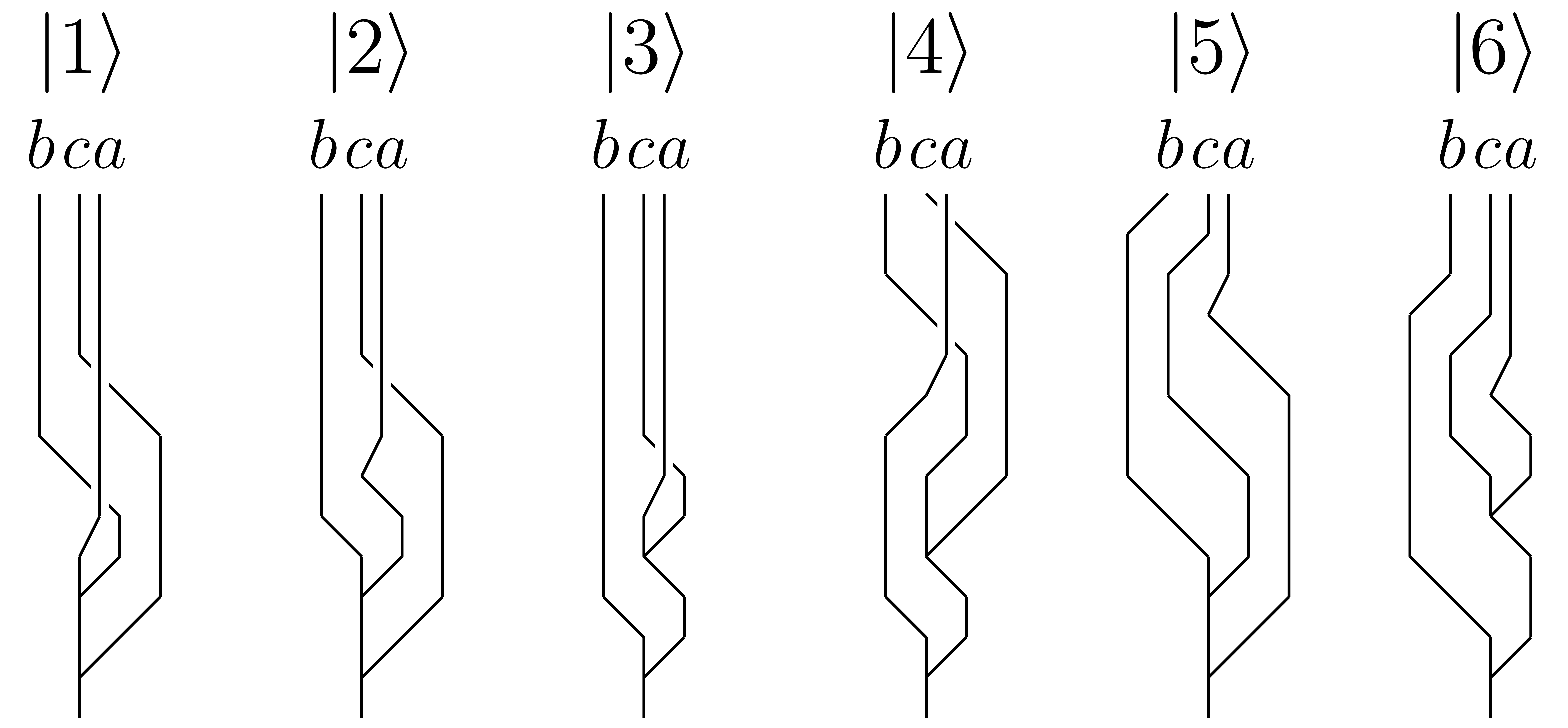}
\caption{Six microscopic processes used to derive the first hexagon equation.}
\label{fig:hex_states}
\end{figure}

We will focus on the first hexagon equation since it is slightly easier to prove with our conventions. The first step of the proof is to construct \emph{microscopic} processes that have the same structure as the six abstract processes shown in Fig.~\ref{fig:hex1}. This step is analogous to the first step of the proof in Appendix~\ref{pentagon}, in which we constructed the microscopic processes shown in Fig.~\ref{fig:Pentagon_states}. Just as in Fig.~\ref{fig:Pentagon_states}, we draw out explicit diagrams for these microscopic processes in Fig.~\ref{fig:hex_states}. These processes are constructed so that they satisfy the following properties: (i) all six processes start in the same initial state and end with the same final particle positions; (ii) all anyon movements occur between either nearest neighbor integer points $n$ and $n\pm1$ or between $1$ and $Y$; and (iii) each crossing in Fig.~\ref{fig:hex1} is implemented, microscopically in Fig.~\ref{fig:hex_states}, by a particle sitting at $Y$ and another particle moving between $2$ and $1$. 

Let us denote the final states produced by these six microscopic processes by $|1\>,...,|6\>$. By construction, $|1\>,...,|6\>$ have the same anyons in the same positions, so they only differ by a phase. The main idea of the proof is to compute the phase difference between $|1\>$ and $|6\>$ in two different ways. The first calculation proceeds along the top path of Figure~\ref{fig:hex1}: we claim that
\begin{align}
|1\> &= R(a,b) |2\> \nonumber \\
|2\> &= F(b,a,c)|3\> \nonumber \\
|3\> &= R(a,c) |6\>
\label{123}
\end{align}
Each of these three equalities follow from a property that we call ``weak topological invariance.'' To state this property, let $P, P'$ be any two processes composed out of movement and splitting operators, and suppose that $P, P'$ start in the same initial state, $|i\>$, and end in the same anyon configuration. The weak topological invariance property states that $P |i\> = P' |i\>$ as long as the spacetime diagrams corresponding to $P, P'$ -- which we regard as decorated planar graphs -- can be continuously deformed into one another while fixing the endpoints and preserving the time-ordered sequence of splittings and crossings. 

Assuming weak topological invariance, which we will prove below, it is easy to derive the first equality in (\ref{123}). The argument is similar to that in Fig.~\ref{fig:pent_link5}: one starts with the two microscopic processes $P_1, P_2$ that produce states $|1\>, |2\>$. One then constructs two modified processes $P'_1, P'_2$ whose spacetime diagrams are equivalent to $P_1, P_2$ according to weak topological invariance, and which are related to one another by a microscopic $R$-move. By weak topological invariance, we know that $P'_1, P'_2$ must produce the same final states as $P_1, P_2$. At the same time, since the $P_1', P_2'$ are related to one another by a microscopic $R$-move, we know that their final states differ by $R(a,b)$; hence we deduce $|1\> = R(a,b)|2\>$. The other two equalities in (\ref{123}) are derived similarly.

\begin{figure}[tb]
\centering
\includegraphics[width=1\columnwidth]{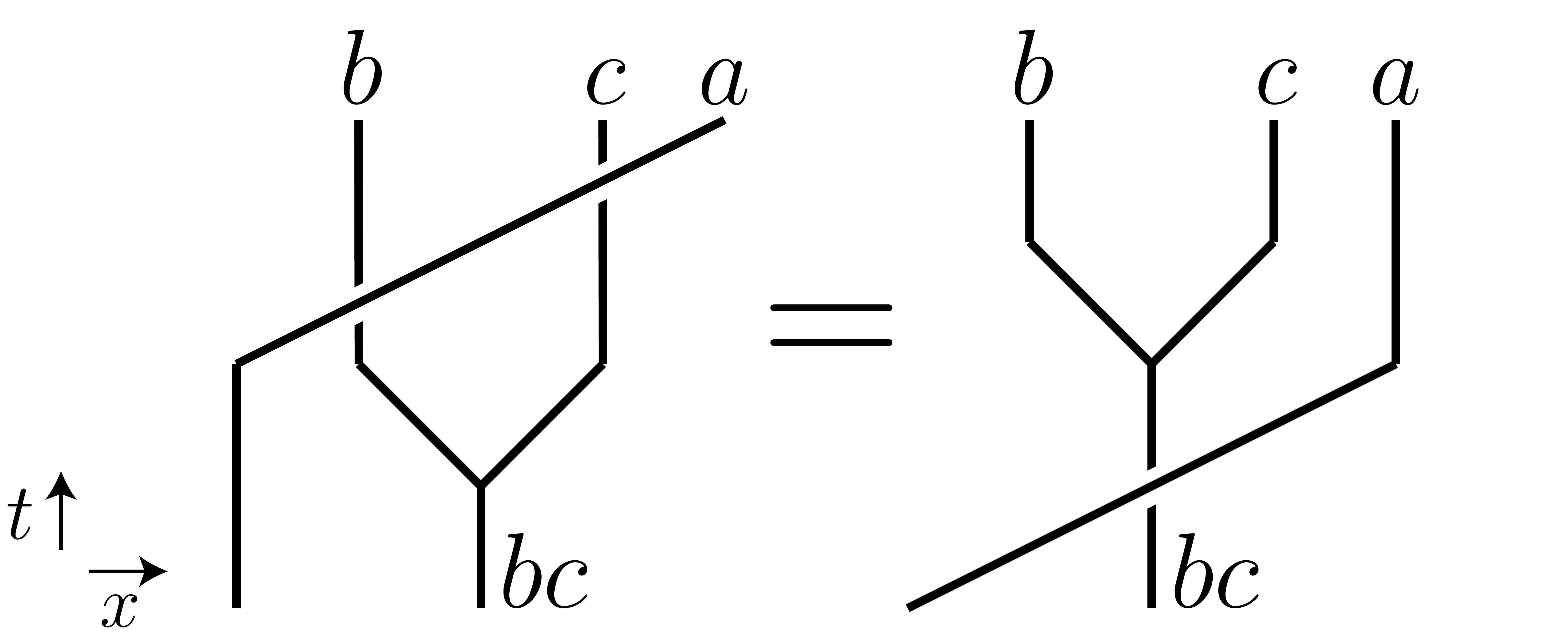}
\caption{Graphical representation of the identity that is needed to relate states $|4\>$ and $|5\>$ in Fig.~\ref{fig:hex1}.}
\label{fig:hex-id-a}
\end{figure}

Next, we calculate the phase difference between $|1\>$ and $|6\>$ by going along the \emph{bottom} path of figure \ref{fig:hex1}. We claim that
\begin{align}
|1\> &= F(a,b,c) |4\> \nonumber \\
|4\> &= R(a,bc)|5\> \nonumber \\
|5\> &= F(b,c,a) |6\>
\label{145}
\end{align}
As in the previous calculation, the first and third equalities in (\ref{145}) follow immediately from weak topological invariance. The second equality however requires an additional identity beyond weak topological invariance, namely the identity shown schematically in Figure~\ref{fig:hex-id-a}. This identity, together with weak topological invariance, is sufficient to show $|4\> = R(a,bc)|5\>$. 

Combining (\ref{123}) and (\ref{145}) gives the desired hexagon equation:
\begin{align*}
R(a,b)F(b,a,c)R(a,c)= F(a,b,c) R(a,bc) F(b,c,a)
\end{align*}

To complete the argument we need to establish two properties that we used above: (1) weak topological invariance and (2) the identity in Fig.~\ref{fig:hex-id-a}. We begin by proving weak topological invariance. 

As in Appendix~\ref{pentagon}, the first step of the proof is to consider the special class of processes that (i) are composed only out of movement operators; (ii) have spacetime diagrams that do not contain any crossings; and (iii) return all anyons to their initial positions. We claim that any process, $P$, of this kind obeys $P|i\> = |i\>$. The proof is virtually identical to the corresponding claim in Appendix~\ref{pentagon}, so we will not repeat it here.

The next step is to consider a slightly more general set of processes, namely processes that are composed out of movement operators and have spacetime diagrams without crossings, but do not necessarily return the anyons to their initial positions. We claim that $P|i\> = P'|i\>$ for any two processes of this kind. Again, the proof is the same as the corresponding claim in Appendix~\ref{pentagon}.

We are now ready to prove weak topological invariance. Suppose that $P, P'$ are two processes that can be deformed into one another while fixing the endpoints and preserving the time-ordered sequence of splittings and crossings. We wish to show $P|i\> = P'|i\>$. Similarly to Appendix~\ref{pentagon}, we decompose $P$ into a product
\begin{align}
P = P_1 V_1 P_2 V_2 \cdots
\label{Pdecomp}
\end{align}
where $P_1, P_2,...$ are movement-only processes without any crossings and where $V_1, V_2,...$ are `vertex operators' consisting of either a splitting operator $S(a,b)$ or a crossing operator (i.e. a movement operator $M^a_{12}$ or $M^a_{21}$ in the presence of a particle at position $Y$). We decompose $P'$ in a similar fashion, denoting the movement-only processes by $P'_1, P'_2,...$ and vertex operators by $V_1', V_2',\cdots$. By assumption, $V_j = V_j'$ for every $j$, but $P_j$ and $P'_j$ may be different. Consider the special case where $P_j$ and $P'_j$ share the same initial anyon configuration and the same final anyon configuration for each $j$. In this case, it follows immediately that $P |i\> = P'|i\>$ using the above property of movement-only processes. Next note that the general case -- where $P_{j}$ and $P'_{j}$ do not share the same initial and final anyon configurations -- can always be reduced to this special case: given any $P, P'$ it is not hard to construct modified processes, $\tilde{P}$ and $\tilde{P}'$, with $P|i\> = \tilde{P}|i\>$ and $P' |i\> = \tilde{P}'|i\>$ such that the corresponding movement-only processes $\tilde{P}_{j}$ and $\tilde{P}'_{j}$ \emph{do} share the same initial and final anyon configuration for each $j$. These modified processes $\tilde{P},\tilde{P}'$ can be obtained in the same way as in Appendix~\ref{pentagon}, i.e. by inserting appropriate movement operators $M^a_{xx'}$ just before each splitting operator, along with their reverse movements $M^a_{x'x}$ just after the splitting operator, and using the fact that movement operators commute with splitting operators when acting on non-overlapping intervals.

\begin{figure}[tb]
\centering
\includegraphics[width=1\columnwidth]{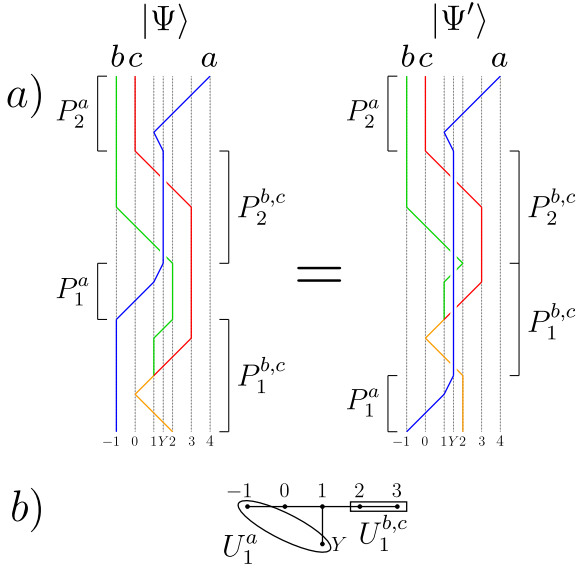}
\caption{(a) The relation $|\Psi\> = |\Psi'\>$ is the first step in our proof of the identity in Fig.~\ref{fig:hex-id-a}. (b) Regions of support of the operators $U_1^a, U_1^{b,c}$ used in proof of $|\Psi\> = |\Psi'\>$.}
\label{fig:hex-id1-micro}
\end{figure}

\begin{figure}[tb]
\centering
\includegraphics[width=0.85\columnwidth]{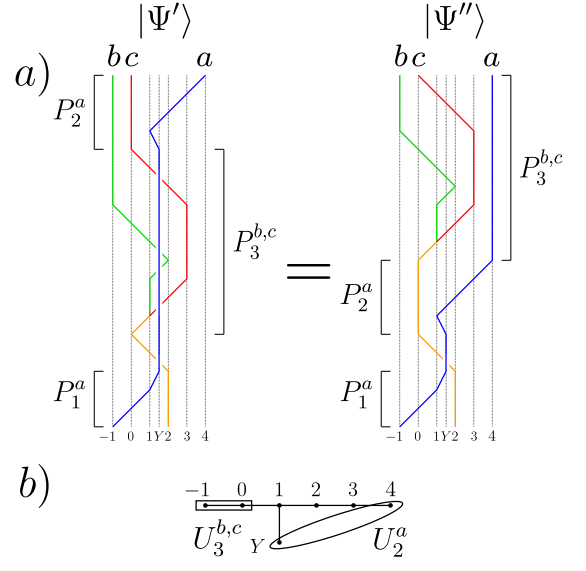}
\caption{(a) The relation $|\Psi'\> = |\Psi''\>$ is the second step in our proof of the identity in Fig.~\ref{fig:hex-id-a}. (b) Regions of support of the operators $U_2^a, U_3^{b,c}$ used in proof of $|\Psi'\> = |\Psi''\>$.}
\label{fig:hex-id2-micro}
\end{figure}

All that remains is to prove the identity shown in Fig.~\ref{fig:hex-id-a}. We will do this in two steps. First we will show that $|\Psi\> = |\Psi'\>$ where $|\Psi\>, |\Psi\>'$ are the two final states produced by the processes shown in Fig.~\ref{fig:hex-id1-micro}a. Next we will show that $|\Psi'\> = |\Psi''\>$ where $|\Psi''\>$ is the final state produced by the second process in Fig.~\ref{fig:hex-id2-micro}a. Taken together, these two equalities imply that $|\Psi\> = |\Psi''\>$. This last result is the microscopic analog of the identity in Fig.~\ref{fig:hex-id-a}.

To prove $|\Psi\> = |\Psi'\>$, we note that
\begin{align}
|\Psi\> = P^a_2 P^{b,c}_2 P^a_1 P^{b,c}_1  |i\> \nonumber \\
|\Psi'\> = P^a_2 P^{b,c}_2 P^{b,c}_1 P^a_1 |i\> 
\end{align}
where $|i\>$ is the initial state and $P^{b,c}_1, P^a_1, P^{b,c}_2, P^a_2$ are the processes shown in Fig.~\ref{fig:hex-id1-micro}a. Given these two expressions, our problem reduces to showing that $P^a_1 P^{b,c}_1  |i\> = P^{b,c}_1 P^a_1 |i\>$. We prove this in the same way that we proved the identity (\ref{scom}) in Appendix~\ref{pentagon}. The first step is to note that there exists a (unitary) operator, $U^a_1$, with a region of support, shown in Fig.~\ref{fig:hex-id1-micro}b, such that $U^a_1|i\>=P^a_1|i\>$. Likewise, there exists a (unitary) operator, $U^{b,c}_1$, with a region of support shown in Fig.~\ref{fig:hex-id1-micro}b, such that $U^{b,c}_1|i\>=P^{b,c}_1|i\>$. Given that these operators have non-overlapping regions of support we know that:
\begin{align}
[U^a_1, U^{b,c}_1] = [P^a_1, U^{b,c}_1] = [U^a_1, P^{b,c}_1] = 0
\end{align}
Combining these identities, the claim follows easily:
\begin{align*}
&P^a_1P^{b,c}_1|i\>=P^a_1U^{b,c}_1|i\>=U^{b,c}_1P^a_1|i\>=U^{b,c}_1U^a_1|i\>\nonumber\\
&=U^a_1U^{b,c}_1|i\>=U^a_1P^{b,c}_1|i\>=P^{b,c}_1U^a_1|i\>=P^{b,c}_1P^a_1|i\>,
\end{align*}

To prove that  $|\Psi'\> = |\Psi''\>$, we use similar reasoning. First we note that
\begin{align}
|\Psi'\> = P^a_2 P^{b,c}_3 (M^{bc}_{01}M^{bc}_{12}P^a_1) |i\> \nonumber \\
|\Psi''\> =  P^{b,c}_3 P^a_2 (M^{bc}_{01}M^{bc}_{12}P^a_1) |i\> 
\end{align}
as one can see in Fig.~\ref{fig:hex-id2-micro}a. Given these expressions, our problem is equivalent to showing that $P^a_2, P^{b,c}_3$ commute with each other when acting on the state $|m\>=M^{bc}_{01}M^{bc}_{12}P^a_1|i\>$. To show this commutation relation, we note that there exists a unitary operator, $U^{b,c}_3$, with a region of support shown in Fig.~\ref{fig:hex-id2-micro}b, such that $U^{b,c}_3|m\>=P^{b,c}_3|m\>$. Likewise, there exists a unitary operator, $U^a_2$, with a region of support shown in  Fig.~\ref{fig:hex-id2-micro}b such that $U^a_2|m\>=P^a_2|m\>$. Using the same manipulations as above, the claim follows immediately:
\begin{align*}
&P^a_2P^{b,c}_3|m\>=P^a_2U^{b,c}_3|m\>=U^{b,c}_1P^a_2|m\>=U^{b,c}_3U^a_2|m\>\nonumber\\
&=U^a_2U^{b,c}_3|m\>=U^a_2P^{b,c}_3|m\>=P^{b,c}_3U^a_2|m\>=P^{b,c}_3P^a_2|m\>,
\end{align*}
This completes the proof of the first hexagon equation.

\section{Derivation of Eq.~\ref{Sidentity_dsem}}
\label{semapp}
In this appendix, we derive the identity (\ref{Sidentity_dsem}) that we used in our analysis of the doubled semion model, that is:
\begin{align}
S(s,s)|0\> \propto a^\dagger_1 a^\dagger_2 |0\>
\end{align}
where $S(s,s)$ is the splitting operator in the doubled semion model and $a_n^\dagger$ is the semion creation operator.

The derivation follows from straightforward algebra. First we write the semion creation operator $a_n^\dagger$ as a product
\begin{align}
a^\dagger_n = W_n T_n U_n
\end{align}
where
\begin{align}
W_n &= \prod_{m<n}\sigma^z_{m,L} \sigma^z_{m,R}, \nonumber \\
T_n &= \prod_{m < n}(-1)^{\frac{1}{4}(1+\sigma^x_{m,L})(1-\sigma^x_{m,R})} \nonumber \\
U_n &= \prod_{m < n}i^{\frac{1-\sigma^x_{m,U}}{2}}
\end{align}
It follows that
\begin{align}
a^\dagger_1 a^\dagger_2 &= (W_1 T_1 U_1) (W_2 T_2 U_2) \nonumber \\
&= (W_1 T_1 W_2 T_2) (U_1 U_2) \nonumber \\
&= (W_1 W_2) (W_2 T_1 W_2 T_2) (U_1 U_2)
\label{a1a2app}
\end{align}
where the second equality follows from the fact that the $U_i$ operators commute with $W_j, T_j$. 

Next, we note that
\begin{align}
W_1 W_2 &= \sigma^z_{1,L}\sigma^z_{1,R} \nonumber \\
U_1 U_2 &= i^{\frac{1-\sigma^x_{1,U}}{2}} \prod_{m < 1} \sigma^x_{m,U}
\label{w1w2}
\end{align} 
Also,
\begin{align*}
W_2 T_1 W_2 &= \prod_{m < 1}(-1)^{\frac{1}{4}(1-\sigma^x_{m,L})(1+\sigma^x_{m,R})} 
\end{align*}
so
\begin{align}
W_2 T_1 W_2 T_2 = (-1)^{\frac{1}{4}(1+\sigma^x_{1,L})(1-\sigma^x_{1,R})} \prod_{m < 1}\sigma^x_{m,L}\sigma^x_{m,R} 
\label{t1t2}
\end{align}

Substituting (\ref{w1w2}) and (\ref{t1t2}) into (\ref{a1a2app}) gives
\begin{align}
a^\dagger_1 a^\dagger_2 &= \sigma^x_{0,R} \sigma^z_{1,L}\sigma^z_{1,R} (-1)^{\frac{1}{4}(1+\sigma^x_{1,L})(1-\sigma^x_{1,R})} i^{\frac{1-\sigma^x_{1,U}}{2}} \prod_{m < 1} Q_m
\label{a1a22app}
\end{align}
where 
\begin{align}
Q_m \equiv \sigma^x_{m,L} \sigma^x_{m,U} \sigma^x_{m-1,R}.
\end{align}
denotes the $Q_v$ operator on vertex $m$.

Acting both sides of (\ref{a1a22app}) on the ground state $|0\>$ and using the fact that $Q_m|0\> = |0\>$, we obtain
\beq
a^\dagger_1a^\dagger_2|0\>=S(a,a)|0\>
\eeq
as we wished to show.

\end{appendix}

\bibliography{bosonic_procedure}

\end{document}